\pdfoutput=1
\documentclass[AMA,STIX1COL]{WileyNJD-v2}

\usepackage{bm}

\DeclareFontEncoding{FMS}{}{}
  \DeclareFontSubstitution{FMS}{futm}{m}{n}
\DeclareSymbolFont{symbols}{FMS}{futm}{m}{n}
\DeclareFontEncoding{FMX}{}{}
  \DeclareFontSubstitution{FMX}{futm}{m}{n}
\DeclareSymbolFont{largesymbols}{FMX}{futm}{m}{n}
\DeclareMathDelimiter{\llbracket}{\mathopen}{symbols}{153}{largesymbols}{133}
\DeclareMathDelimiter{\rrbracket}{\mathclose}{symbols}{154}{largesymbols}{134}
\DeclareSymbolFont{largesymbols}{LS2}{stixex}{m}{n}
\DeclareMathSymbol{\sumop}{\mathop}{largesymbols}{"B3}

\articletype{Research Article}%

\begin{document}

\title{A multiscale approach to hybrid RANS/LES wall modeling within a high-order discontinuous Galerkin scheme using function enrichment}

\author[]{Benjamin Krank*}

\author[]{Martin Kronbichler}

\author[]{Wolfgang A. Wall}

\authormark{BENJAMIN KRANK \textsc{et al}}

\address[]{\orgdiv{Institute for Computational Mechanics}, \orgname{Technical University of Munich}, \orgaddress{\country{Germany}}}

\corres{*Benjamin Krank, Institute for Computational Mechanics, Technical University of Munich, Boltzmannstr. 15, 85748 Garching, Germany. \email{krank@lnm.mw.tum.de}}

\abstract[Summary]{We present a novel approach to hybrid RANS/LES wall modeling based on function enrichment, which overcomes the common problem of the RANS--LES transition and enables coarse meshes near the boundary. While the concept of function enrichment as an efficient discretization technique for turbulent boundary layers has been proposed in an earlier article by Krank \& Wall ({\it J. Comput. Phys.} 316 (2016) 94--116), the contribution of this work is a rigorous derivation of a new multiscale turbulence modeling approach and a corresponding discontinuous Galerkin discretization scheme. In the near-wall area, the Navier--Stokes equations are explicitly solved for an LES and a RANS component in one single equation. This is done by providing the Galerkin method with an independent set of shape functions for each of these two methods; the standard high-order polynomial basis resolves turbulent eddies where the mesh is sufficiently fine and the enrichment automatically computes the ensemble-averaged flow if the LES mesh is too coarse. As a result of the derivation, the RANS model is consistently applied solely to the RANS degrees of freedom, which effectively prevents the typical issue of a log-layer mismatch in attached boundary layers. As the full Navier--Stokes equations are solved in the boundary layer, spatial refinement gradually yields wall-resolved LES with exact boundary conditions. Numerical tests show the outstanding characteristics of the wall model regarding grid independence, superiority compared to equilibrium wall models in separated flows, and achieve a speed-up by two orders of magnitude compared to wall-resolved LES.}


\jnlcitation{\cname{%
\author{B. Krank}, 
\author{M. Kronbichler}, and 
\author{W.A. Wall}}, 
\ctitle{A multiscale approach to hybrid RANS/LES wall modeling within a high-order discontinuous Galerkin scheme using function enrichment}, \cjournal{Int. J. Numer. Meth. Fluids}, \cvol{2017;00:1--32}.}

\maketitle

\section{Introduction}
The simulation of turbulent boundary layers poses two fundamental challenges: (i) the mean velocity gradient in the viscous sublayer can become very high in engineering applications; this gradient has to be resolved or modeled for the prediction of the skin friction, and (ii) eddies of various spatial and temporal scales have to be resolved or modeled for the computation of the mean velocity profile (and thus the skin friction). Classical hybrid RANS/LES wall modeling is popular in research and industry as it promises a vast reduction in computational cost compared to wall-resolved LES through modeling of the near-wall turbulence (cf. (ii)), while being potentially much more accurate than RANS in separated flows. Yet the velocity gradient is usually resolved by the scheme (cf. (i)). A complete review of wall modeling approaches and RANS/LES methods would be beyond the scope of this article, so we refer to a series of review articles in this field\cite{Spalart00,Piomelli08,Frohlich08,Sagaut13,Larsson16}. Within the high-order DG method, detached-eddy simulation models\cite{Krank17c,Wurst13}, a hybrid RANS/ILES model\cite{Zhu16}, and wall-stress models based on equilibrium wall functions\cite{Wiart17,Frere16,Frere17} have been developed.

The major challenge in hybrid RANS/LES methods is the transition region from RANS to LES\cite{Piomelli08b}. The problem is that ``LES content has to be generated in the outer part of the boundary layer''\cite{Sagaut13} since the momentum transfer is fully modeled in RANS but relies on the turbulent structures in LES. If the model parameters are carefully tuned within the transition region, the results may be quite convincing\cite{Shur08}; an error in the sum of the resolved and modeled Reynolds stress would otherwise produce a so-called log-layer mismatch in equilibrium boundary layers\cite{Piomelli08b}. However, existing zonal RANS/LES models often rely heavily on empiricism, and grid-dependence can be an issue. Enhancements, including artificial forcing methods or synthetic turbulence, may improve the results obtained for attached boundary layers\cite{Sagaut13}.

\subsection{A multiscale approach to wall modeling via function enrichment}

In this work, we propose a novel approach to hybrid RANS/LES wall modeling, which resolves the velocity gradient at the wall with coarser meshes (cf. (i)), models the near-wall turbulence (cf. (ii)), and additionally solves the problem of the RANS--LES transition. The basic idea is that, in the vicinity of the wall, the solution is decomposed into a RANS and an LES component according to the ancient strategy \emph{divide et impera}. The two components overlap in this near-wall layer, and together they represent the velocity solution. The decoupling is achieved by applying the RANS eddy viscosity on the RANS degrees of freedom only and allows the energy-carrying turbulent structures to fully develop inside the near-wall layer, which entirely circumvents the aforementioned problem of transition at the RANS/LES interface. This structure of the eddy viscosity term is a direct result of applying a new three-level hybrid RANS/LES filter, which is defined based on Germano's framework of additive filters\cite{Germano04}. Despite the fact that the two components are expressed and modeled separately, the full Navier--Stokes equations are computed in the whole boundary layer in a single equation, with exact boundary conditions. This ensures robustness in challenging flow conditions such as separated flows featuring a high adverse pressure gradient. The velocity gradient is resolved, although the first off-wall node for the LES may be placed in a range $y_1^+\sim 1\ldots100$, which is possible due to the interaction of the RANS and LES scales and due to the use of a tailored spatial discretization via function enrichment.

Function enrichment in the context of wall modeling was recently proposed by Krank and Wall for the standard (continuous) FEM\cite{Krank16} and for the discontinuous Galerkin method (DG)\cite{Krank17c,Krank16c}; see also the monograph~\cite{Krank18} for further details. The concept of function enrichment utilizes the basic characteristic of the Galerkin method: The user chooses a set of shape functions, and the method automatically finds the best match among those available shape functions. For the LES solution, we take classical polynomial shape functions. However, the RANS solution is based on a discrete solution space tailored for this specific application. The key idea is to construct the shape functions by using a wall-law as an enrichment function, which enables the resolution of the high velocity gradient at the no-slip boundary with very few degrees of freedom. The general principle of function enrichment is universal and has been successfully employed in a broad spectrum of other computational disciplines\cite{Farhat01,Belytschko09,Fries10}.

Once the solution is decomposed into a RANS and an LES component, each with different shape functions, we employ a second beneficial feature of the Galerkin method: The weighting functions have the same structure as the solution functions, allowing an explicit decomposition of the governing equations into separate equations for the RANS and the LES components -- although the problem is still solved in one single equation -- according to the variational multiscale methodology\cite{Hughes95,Hughes98}. The scale separation allows appropriate turbulence modeling for each of the scales, a RANS turbulence model for the RANS equation and an LES turbulence model for the LES scale, exactly as they are derived through the three-level hybrid RANS/LES filter. We note that this is a fundamental difference to the application of function enrichment in detached-eddy simulation (DES)~\cite{Krank17c}, where the domain is clearly separated into the inner RANS region and the outer LES region, with a `grey area' in between, since the RANS eddy viscosity acts on the polynomial velocity component as well in that study; these differences are detailed in the subsequent section.

\subsection{Comparison to other hybrid RANS/LES methodologies}

This composition of the solution relates the present method to the nonlinear disturbance equations (NLDE)\cite{Morris97,Labourasse02,Labourasse04}, which have been proposed in order to reconstruct the energy-carrying turbulent structures (LES) around an averaged mean flow (RANS), with the primary application in the field of aeroacoustics. In the NLDE, the RANS and LES components overlap in a part of the domain, as they do in the present method. Several features of our approach distinguish between the methods:
\begin{itemize}
\item In NLDE, the RANS equations are first solved for the mean flow and the turbulent fluctuations are reconstructed in a subsequent step, whereas the RANS and LES solutions are computed simultaneously in a single equation in the present model.
\item In NLDE, the LES represents fluctuations around a mean flow, while the LES and RANS are each a variable fraction of one solution in the present methodology.
\item The NLDE usually require the LES to be wall-resolved, whereas in our model the RANS provides the LES with the necessary dissipation if the LES is underresolved, ie, not all energetic scales have to be computed in the LES.
\end{itemize}

The multiscale wall model is further compared to two of the classical hybrid RANS/LES wall models, the original DES approach by Spalart et al\cite{Spalart97}, which was recently applied with wall modeling via function enrichment in Krank et al\cite{Krank17c} (in the delayed DES\cite{Spalart06} (DDES) variant), and the method of blending subgrid models through a weighted sum by Baggett\cite{Baggett98}. Both models follow the idea of two spatially separated zones, a RANS layer and an LES bulk flow, with a transition region in between. 
They compare to the proposed method as follows:
\begin{itemize}
\item DES achieves the blending by limiting the wall distance function in the Spalart--Allmaras model by a characteristic grid length scale such that the RANS model acts as a one-equation LES model in the bulk flow, see, eg\cite{Wurst13} for an application with DG. The present method applies the idea of limiting the wall distance of the RANS model by a characteristic cell length scale in order to account for the resolved Reynolds stresses.
\item The method of weighted sum blends the subgrid stress tensor from a RANS model at the wall into an LES model in the bulk flow via an explicitly defined blending function. This approach has been refined by Germano\cite{Germano04} and Rajamani and Kim\cite{Rajamani10}, who suggested to blend LES and RANS filtering operators instead of subgrid models, and it is this filter that will be extended to a three-level filter in the present work.
\item The major difference of the proposed approach to both the DES and the weighted sum method is, however, that a marginally/underresolved LES,  which is not directly affected by the RANS eddy viscosity, extends to the no-slip boundary. It is this difference that avoids nonphysical velocity fluctuations, such as described by Baggett\cite{Baggett98}, within the present method. A direct qualitative and quantitative comparison of DES using function enrichment\cite{Krank17c} and the present multiscale wall model will be presented in Section~\ref{sec:channel}.
\end{itemize}

\subsection{Outline}
The wall model has been implemented in a high-performance discontinuous Galerkin incompressible flow solver, which was previously applied for computation of DNS and LES\cite{Krank16b,Krank17b}, DES\cite{Krank17c}, and RANS\cite{Krank16c}. The wall model may be implemented in a continuous Galerkin (standard FEM) framework as well, but the discontinuous Galerkin method provides the simplest formulation for enrichments including adaptation. The flow solver is implemented using fast state-of-the-art matrix-free algorithms.

The remainder of this paper is organized as follows. In Section~\ref{sec:multiscale_approach}, we derive consistent governing equations for the RANS and LES components in one equation and show how these equations may be solved in the framework of the variational multiscale method. In Section~\ref{sec:spaces}, the concept of function enrichment for turbulent boundary layers is revisited. Section~\ref{sec:dgsolver} gives details on the solution procedure and variational forms of the present discontinuous Galerkin methodology. In Section~\ref{sec:examples}, numerical examples demonstrate the excellent mesh independence of the present wall model and show how the full consistency of the model enables accurate results under separated flow conditions. Conclusions close the article in Section~\ref{sec:conclusion}.

\section{A multiscale approach to wall modeling}
\label{sec:multiscale_approach}
The primary innovation of this work is the derivation of a consistent multiscale RANS/LES wall modeling framework for application to wall modeling via function enrichment. Wall modeling via function enrichment has been introduced by Krank and Wall\cite{Krank16} and represents a highly efficient discretization method problem-tailored for turbulent boundary layers. The topic of near-wall subgrid modeling in the hybrid RANS/LES context has recently been addressed by extending the wall-function-enriched RANS solver based on the Spalart--Allmaras model\cite{Krank16c} to DES\cite{Spalart97}. While that work allows the use of comparably coarse grids in wall-normal direction in the inner layer as compared to classical DES, the issue of the RANS--LES transition persisted. However, the Galerkin method in conjunction with function enrichment enables a much more general turbulence model, which avoids the common problem of the RANS--LES transition entirely; the derivation of this model is the topic of this section.

We derive a consistent multiscale framework for application to wall modeling via function enrichment and define a suitable subgrid model. Further essential ingredients to the wall modeling approach are the discrete function spaces including function enrichment, which are recited in Section~\ref{sec:spaces}, as well as crucial modifications to the discontinuous Galerkin formulation of the spatial discretization, which are presented in Section~\ref{sec:dgsolver}. 

In the following, the solution of the incompressible Navier--Stokes equations in a layer near the wall is split into two parts, an eddy-resolving component (LES) and an ensemble-averaged component (RANS). The equations are derived by applying a new three-level filter to the Navier--Stokes equations, which is defined based on Germano's framework for additive filtering. The implications of this filter are that both the RANS eddy viscosity term and the definition of the eddy viscosity variable depend only on the RANS component, thus the LES component can evolve independently and is limited solely by its own resolution. Further, the RANS and the LES components are not spatially separated, but overlap in the near-wall layer.

In this section, we begin with a presentation of the governing equations in Section~\ref{sec:ins}. Subsequently, the scale separation is introduced by a three-level filter and the differences to the classical hybrid RANS/LES filter are discussed in Section~\ref{sec:scalesep}. Finally, Section~\ref{sec:vms} shows how the variational multiscale method is employed to solve for the RANS and LES solutions using weak forms and appropriate solution as well as weighting functions.

\subsection{Incompressible Navier--Stokes Equations}
\label{sec:ins}
We consider the incompressible Navier--Stokes equations in conservation form:
\begin{alignat}{5}
&\nabla \cdotp \bm{u} &&= 0  &&\text{ \hspace{0.2cm} in } \Omega \times [0,\mathcal{T}], \label{eq:conti} \\
\frac{\partial \bm{u}}{\partial t} + &\nabla \cdotp \left(\bm{\mathcal{F}}^c(\bm{u}) + p \bm{\mathcal{I}} - \bm{\mathcal{F}}^{\nu}(\bm{u})\right) &&= \bm{f} &&\text{ \hspace{0.2cm} in } \Omega \times [0,\mathcal{T}],
\label{eq:mom}
\end{alignat}
with the velocity vector $\bm{u}=\left(u_1,u_2,u_3\right)^T$, the pressure $p$, the body force $\bm{f}=\left(f_1,f_2,f_3\right)^T$, the three-dimensional spatial domain $\Omega$, and the simulation time $\mathcal{T}$. The convective and viscous fluxes are:
\[
\bm{\mathcal{F}}^c(\bm{u}) = \bm{u} \otimes \bm{u}, \hspace{1cm} 
\bm{\mathcal{F}}^{\nu}(\bm{u}) = 2 \nu \bm{\epsilon}(\bm{u}),
\]
where the kinematic viscosity is denoted $\nu$ and the rate-of-deformation tensor $\bm{\epsilon}=1/2\left(\nabla \bm{u}+(\nabla \bm{u})^T\right)$.

At $t=0$, an initial velocity field $\bm{u}_0$ is specified as
\begin{equation}
\bm{u} = \bm{u}_0 \text{ \hspace{0.2cm} in } \Omega \times \{0\}.
\end{equation}
Periodic as well as no-slip Dirichlet boundary conditions on solid walls $\partial \Omega^D = \partial \Omega$ close the problem with
\begin{equation}
\bm{u} = \bm{g}_{\bm{u}} = \bm{0} \text{ \hspace{0.2cm} on } \partial \Omega^D \times [0,\mathcal{T}].
\end{equation}

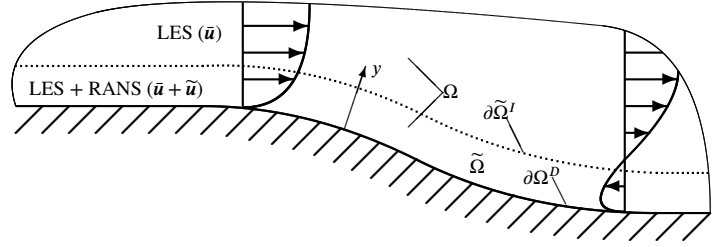
\begin{figure}[t]
\centering
\begin{picture}(220,100)
\thicklines
\put(-20,55){\line(1,0){70}}
\qbezier(50,55)(90,55)(130,35)
\qbezier(130,35)(170,15)(220,15)
\put(220,15){\line(1,0){20}}
\put(-10,55){\line(-1,-1){10}}
\put(0,55){\line(-1,-1){10}}
\put(10,55){\line(-1,-1){10}}
\put(20,55){\line(-1,-1){10}}
\put(30,55){\line(-1,-1){10}}
\put(40,55){\line(-1,-1){10}}
\put(50,55){\line(-1,-1){10}}
\put(59.5,54.5){\line(-1,-1){10}}
\put(69,54){\line(-1,-1){10}}
\put(77.5,52.5){\line(-1,-1){10}}
\put(86,51){\line(-1,-1){10}}
\put(94,49){\line(-1,-1){10}}
\put(101.8,46.8){\line(-1,-1){10}}
\put(109.2,44.2){\line(-1,-1){10}}
\put(116,41){\line(-1,-1){10}}
\put(123,38){\line(-1,-1){10}}
\put(130,35){\line(-1,-1){10}}
\put(136.8,31.8){\line(-1,-1){10}}
\put(143.8,28.8){\line(-1,-1){10}}
\put(151,26){\line(-1,-1){10}}
\put(158.5,23.5){\line(-1,-1){10}}
\put(166.5,21.5){\line(-1,-1){10}}
\put(174.5,19.5){\line(-1,-1){10}}
\put(183,18){\line(-1,-1){10}}
\put(191.7,16.7){\line(-1,-1){10}}
\put(200.5,15.5){\line(-1,-1){10}}
\put(210.2,15.2){\line(-1,-1){10}}
\put(220,15){\line(-1,-1){10}}
\put(230,15){\line(-1,-1){10}}
\put(240,15){\line(-1,-1){10}}
\qbezier[35](-20,70)(15,70)(50,70)
\qbezier[43](50,70)(91,72)(132,52)
\qbezier[43](132,52)(171,32)(220,30)
\qbezier[10](220,30)(230,30)(239,30)
\thinlines
\footnotesize
\put(-15,59){LES + RANS ($\bar{\bm{u}}+\widetilde{\bm{u}}$)}
\put(33,80){LES ($\bar{\bm{u}}$)}
\put(150,30){$\widetilde{\Omega}$}
\put(140,58){${\Omega}$}
\put(140,60){\line(-1,1){13}}
\put(140,60){\line(-1,-1){13}}
\put(170,26){$\partial\Omega^D$}
\put(181,28.5){\line(1,-2){5}}
\put(155,50){$\partial\widetilde{\Omega}^I$}
\put(164,50){\line(1,-2){6}}
\qbezier(-20,55)(-28,80)(10,90)
\qbezier[800](10,90)(50,100)(200,85)
\qbezier(200,85)(240,80)(240,15)
\put(103,46){\vector(1,3){8}}
\put(113,67){$y$}
\thicklines
\qbezier(65,54.5)(90,54.5)(90,93.0)
\put(65,55){\line(0,1){38.8}}
\put(65,65){\vector(1,0){19.5}}
\put(65,75){\vector(1,0){23.3}}
\put(65,85){\vector(1,0){25}}
\qbezier(208,15.5)(190,15.5)(208,35)
\qbezier(208,35)(228,55)(228,67.8)
\put(208,15){\line(0,1){68.3}}
\put(208,75){\line(1,0){14.3}}
\put(208,65){\vector(1,0){20.1}}
\put(208,55){\vector(1,0){16.4}}
\put(208,45){\vector(1,0){9.6}}
\put(208,25){\vector(-1,0){8}}
\end{picture}
\caption{Zoom-in on the near-wall area. The full domain is denoted $\Omega$ and a subset in a layer near the wall $\widetilde{\Omega}\subset\Omega$. In $\widetilde{\Omega}$, the solution is composed of a RANS and an LES component, both in variable composition, while the solution outside this area is represented by an LES component only. The interface in between is denoted $\partial\widetilde{\Omega}^I$.}
\label{fig:sketch}
\end{figure}

\subsection{Three-level hybrid RANS/LES filter}
\label{sec:scalesep} 

We propose a novel hybrid RANS/LES filter based on Germano's framework of additive filtering\cite{Germano04}. The filter is constructed such that an underresolved LES solution overlaps with the RANS solution in the near-wall area. In this subsection, we first recite the classical two-level additive hybrid RANS/LES filter\cite{Germano04} and then use that framework to define a new three-level filter. In applying the new filter to the incompressible Navier--Stokes equations, we obtain a set of equations for an eddy-resolving velocity variable and a statistical velocity variable, and their sum represents the final velocity solution. These equations have formally seven unknowns, three LES velocity components, three RANS velocity components, and one pressure variable. In practice, the LES and RANS components will be solved by taking a different set of shape functions tailored for each of the two scales (see Section~\ref{sec:vms} and~\ref{sec:spaces}).

The derivations address the near-wall area of a computational domain. Figure~\ref{fig:sketch} gives an overview of the general concept and the variables defining the wall model. The near-wall region is denoted $\widetilde{\Omega}\subset \Omega$; the baseline LES solver is used outside this area. Coupling conditions defining the transition from the near-wall region to the outer region at the interface $\partial \widetilde{\Omega}^I$ are also discussed in this section.

\subsubsection{The classical two-level additive hybrid RANS/LES filter}
Germano's hybrid RANS/LES filter\cite{Germano04} defines the velocity and pressure as a weighted sum of the LES-filtered and RANS-averaged quantities, yielding
\begin{align}
\bm{u}_H &= a_1 \bar{\bm{u}} + a_2 \langle \bm{u} \rangle,\\
p_H &= a_1 \bar{p} + a_2 \langle p \rangle.
\end{align}
Herein, $\bar{(\cdotp)}$ denotes an LES-filtering operator and $\langle \cdotp \rangle$ a statistical (ensemble-averaging) operator. As an LES filter, implicit grid filtering is commonly considered, which may also be interpreted as a variational projection in the context of the Galerkin method\cite{Hughes00}.
The continuous blending functions $0 \leq a_1 \leq 1$ and $a_2=1-a_1$ may be spatially dependent and are predefined. The filter fulfills two basic requirements, namely
\begin{align}
a_1 + a_2 &= 1,\\
\langle \bm{u}_H \rangle &= \langle \bm{u} \rangle,
\end{align}
and equivalently for the pressure.
Applying this filter to the governing equations~\eqref{eq:conti}--\eqref{eq:mom} and using the rule of permutation of the derivative and the filtering operator\cite{Germano04}, we obtain
\begin{alignat}{3}
&\nabla \cdotp \bm{u}_H &&= 0  , \label{eq:conti_H} \\
\frac{\partial \bm{u}_H}{\partial t} + &\nabla \cdotp \left(\bm{\mathcal{F}}^c\left(\bm{u}_H\right) + p_H \bm{\mathcal{I}} - \bm{\mathcal{F}}^{\nu}\left(\bm{u}_H\right)\right) &&= \bm{f}- \nabla \cdotp \bm{\tau}_H,
\label{eq:mom_H}
\end{alignat}
where $\bm{\tau}_H$ is the corresponding subgrid stress tensor, which is given by Germano\cite{Germano04}. It is noted that the chain rule applies in all spatial derivatives if $a_1$ is spatially dependent.

We argue that the construction of this filter promotes some of the issues associated to hybrid RANS/LES methods, such as in the transition region. For any blending factor $a_1 < 1$, the resolved LES scales in $\bm{u}_H$ are not allowed to evolve with their full magnitude as they are multiplied by a factor smaller than one, which induces a damping of the resolved turbulence. Therefore, the smallest resolved scales would be larger than the given grid filter size and the method does not exhibit optimal efficiency in resolving turbulent scales. It is one of the primary incentives of this work to construct a numerical method which allows the LES scales to evolve independently, only limited by the coarseness of the LES filter size. This is achieved by proposing a new formulation of the hybrid RANS/LES filter in the following, which allows the use of $a_1=1$ throughout.

\subsubsection{A new three-level additive hybrid RANS/LES filter}
We modify the classical additive hybrid RANS/LES filter by adding a third term, yielding
\begin{align}
\bm{u}_H &= a_1 \bar{\bm{u}} + a_2 \langle \bm{u} \rangle + a_3 \langle \bar{\bm{u}} \rangle,\\
p_H &= a_1 \bar{p} + a_2 \langle p \rangle + a_3 \langle \bar{p} \rangle,
\end{align}
for the filtered velocity and pressure. Herein, the third filtering operator $\langle \bar{(\cdotp)} \rangle$ defines a hierarchical filter by successively applying an LES filter and a statistical filter. This particular construction enables the choice of the constant factors $a_1=1$, $a_2=1$, and $a_3=-1$ throughout the near-wall region. The third component subtracts the statistical average of the resolved variables from the balance. Due to $a_1=1$, resolved turbulent structures are not damped artificially by the method. The eddy resolving component $\bar{\bm{u}}$ may be significantly underresolved as we place the first off-wall node in the region of $y_1^+\sim 1...100$, so the energy-carrying scales do not have to be resolved in the near-wall area. 
The resulting filtering operator fulfills the requirements:
\begin{align}
a_1 + a_2 + a_3 &= 1,\\
\langle \bm{u}_H \rangle &= \langle \bm{u} \rangle, \label{eq:req2}
\end{align}
and equivalently for the pressure.

The present hybrid filter may be related to the NLDE, where full LES and RANS solutions overlap in a part of the domain\cite{Morris97,Labourasse02,Labourasse04}. The NLDE is formally derived using a hierarchical scale separation operator, applying both an LES filter and a RANS operator, such that $\langle \bar{\bm{u}}\rangle = \langle {\bm{u}}\rangle$ holds by definition. This characteristic stands in contrast to the present approach, since the LES filter size is usually so coarse that the mean flow is not captured sufficiently by the LES scale and we have $\langle \bar{\bm{u}}\rangle \neq \langle {\bm{u}}\rangle$.

When this approach is applied to the incompressible Navier--Stokes equations, the derivatives of $a_i$ with respect to the spatial coordinates vanish and the subgrid tensor becomes\cite{Germano04}:
\begin{align}
\bm{\tau}_H =& a_1 \bm{\tau}_{\mathrm{LES}} + a_2 \bm{\tau}_{\mathrm{RANS}} + a_3  \bm{\tau}_3 \nonumber\\
&+ a_1 a_2 \left(\bar{\bm{u}}-\langle \bm{u}\rangle \right)\otimes\left(\bar{\bm{u}}-\langle \bm{u}\rangle\right) + a_1 a_3 \left(\bar{\bm{u}} - \langle \bar{\bm{u}} \rangle\right)\otimes\left(\bar{\bm{u}} - \langle \bar{\bm{u}} \rangle\right) + a_2 a_3 \left(\langle \bm{u}\rangle - \langle \bar{\bm{u}} \rangle\right)\otimes\left(\langle \bm{u}\rangle - \langle \bar{\bm{u}} \rangle\right),
\label{eq:tauH}
\end{align}
with the LES and RANS subgrid tensors
\begin{align}
\bm{\tau}_{\mathrm{LES}} &= \overline{\bm{u}\otimes \bm{u}} - \bar{\bm{u}}\otimes \bar{\bm{u}}, \\
\bm{\tau}_{\mathrm{RANS}} &= \langle \bm{u} \otimes \bm{u} \rangle - \langle \bm{u} \rangle \otimes \langle \bm{u} \rangle = \langle \bm{u}' \otimes \bm{u}' \rangle, \\
\bm{\tau}_{3} &= \langle \overline{\bm{u}\otimes \bm{u}}\rangle - \langle \bar{\bm{u}}\rangle\otimes \langle \bar{\bm{u}}\rangle = \langle \bm{\tau}_{\mathrm{LES}}\rangle + \langle \bar{\bm{u}}' \otimes \bar{\bm{u}}' \rangle,
\end{align}
where $(\cdotp)'$ denotes the fluctuating component of the respective quantity, ie $\bm{u} = \langle \bm{u}\rangle +\bm{u}'$.
In order to simplify the subgrid terms in the following, we define a new velocity variable summarizing the statistical velocity contributions, $\widetilde{\bm{u}} = \langle \bm{u} \rangle - \langle \bar{\bm{u}} \rangle$. 

{\bf Remark:} The quantity $\widetilde{\bm{u}}$ plays a major role in the remainder of this article, as we will explicitly solve for this component in the final numerical method and the velocity solution is considered in the form $\bm{u}_H=\bar{\bm{u}}+\widetilde{\bm{u}}$, as a sum of an LES and a RANS component.

Inserting the constants $a_i$ in Equation~\eqref{eq:tauH} and application of the relation~\eqref{eq:req2}, $\langle \bm{u} \rangle = \langle \bar{\bm{u}} \rangle + \widetilde{\bm{u}}$, yields
\begin{align}
\bm{\tau}_H =& \bm{\tau}_{\mathrm{LES}}+ \bm{\tau}_{\mathrm{RANS}} - \bm{\tau}_{3} 
+ \left(\bar{\bm{u}}-\langle \bm{u}\rangle \right)\otimes\left(\bar{\bm{u}}-\langle \bm{u}\rangle\right) - \left(\bar{\bm{u}} - \langle \bar{\bm{u}} \rangle\right)\otimes\left(\bar{\bm{u}} - \langle \bar{\bm{u}} \rangle\right) - \underbrace{\left(\langle \bm{u}\rangle - \langle \bar{\bm{u}} \rangle\right)}_{\widetilde{\bm{u}}}\otimes \underbrace{\left(\langle \bm{u}\rangle - \langle \bar{\bm{u}} \rangle\right)}_{\widetilde{\bm{u}}}\\
=& \bm{\tau}_{\mathrm{LES}} - \langle \bm{\tau}_{\mathrm{LES}}\rangle + \langle \bm{u}' \otimes \bm{u}' \rangle - \langle \bar{\bm{u}}' \otimes \bar{\bm{u}}' \rangle - \bar{\bm{u}}' \otimes \widetilde{\bm{u}} - \widetilde{\bm{u}} \otimes \bar{\bm{u}}'.
\end{align}

In the near-wall area, we assume that Reynolds stresses dominate, allowing the simplification
\begin{equation}
\bm{\tau}_H \approx  \langle \bm{u}' \otimes \bm{u}' \rangle - \langle \bar{\bm{u}}' \otimes \bar{\bm{u}}' \rangle \label{eq:approx_subgrid_stress}.
\end{equation}
Herein, the LES subgrid stresses play a minor role in the vicinity of the wall, as barely any turbulence is resolved due to an overly coarse LES mesh and usually almost the entire solution is represented by $\widetilde{\bm{u}}$. In the present work, we consider the implicit LES capabilities of the numerical method\cite{Krank16b,Krank17b}, where the numerical upwind fluxes introduce an appropriate amount of dissipation in the high-frequency range, so any LES subgrid tensor is omitted in the following. Further remarks on how to include an explicit LES model are given in Section~\ref{sec:vms}. As a model for the remaining subgrid terms, we employ an eddy viscosity closure
\begin{align}
\bm{\tau}_H &\approx  \langle \bm{u}' \otimes \bm{u}' \rangle -  \langle \bar{\bm{u}}' \otimes \bar{\bm{u}}' \rangle \nonumber \\
&\approx -2 \nu_t \left(\bm{\epsilon}\left(\langle {\bm{u}}\rangle \right) -  \bm{\epsilon}\left(\langle \bar{\bm{u}}\rangle\right)\right) \label{eq:eddyvisc}\\
&= -2 \nu_t \bm{\epsilon}\left(\widetilde{\bm{u}}\right) \nonumber 
\end{align}
with the RANS eddy viscosity $\nu_t$. The relevant velocity component, on which the eddy viscosity model acts, is the RANS component $\widetilde{\bm{u}}$, which is a direct result of the formal definition of the three-level filter, as indicated in Equation~\eqref{eq:eddyvisc}.

{\bf Remark:} The earlier elaborations on using a constant blending factor of $a_1=1$ manifest themselves in the numerical method: the LES degrees of freedom are not directly impacted by the RANS eddy viscosity, so they are not damped.

For simplicity and due to the limitation to the inner layer in the present application, we employ Prandtl's algebraic mixing length eddy viscosity model including van Driest's damping function\cite{vanDriest56}
\begin{equation}
\nu_t = l_{\mathrm{mix}}^2 |\bm{\epsilon}\left(\widetilde{\bm{u}}\right)|
\label{eq:mix}
\end{equation}
with the mixing length
\begin{equation}
l_{\mathrm{mix}} = \kappa y_{\mathrm{DES}} \left(1-\mathrm{exp}\left({-y^+/A^+}\right)\right),
\label{eq:lmix}
\end{equation}
where $A^+ = 26$, the norm $|\bm{\epsilon}\left(\widetilde{\bm{u}}\right)|=\sqrt{2\bm{\epsilon}\left(\widetilde{\bm{u}}\right):\bm{\epsilon}\left(\widetilde{\bm{u}}\right)}$, and a modified wall distance $y_{\mathrm{DES}}$. We further note that the Reynolds stresses in Equation~\eqref{eq:approx_subgrid_stress} include the difference between the full and resolved stress tensor $\langle{\bm{u}}' \otimes {\bm{u}}'\rangle - \langle\bar{\bm{u}}' \otimes \bar{\bm{u}}'\rangle$. It would be possible to explicitly compute all terms involving the resolved scales in Equation~\eqref{eq:approx_subgrid_stress} $\left( - \langle\bar{\bm{u}}' \otimes \bar{\bm{u}}'\rangle\right)$ by averaging over homogeneous directions and subtracting the result from the eddy viscosity tensor in the philosophy of Medic et al\cite{Medic05}. In this work we account for the resolved Reynolds stress tensor by considering the classical DES approach of Spalart et al\cite{Spalart97}, which limits the wall distance $y$ with the characteristic grid length scale $h$ times a model constant $C_{\mathrm{DES}}$:
\begin{equation}
y_{\mathrm{DES}} = \min\left(y,C_{\mathrm{DES}} h\right) \approx \left(\frac{1}{y^\beta}+\frac{1}{\left(C_{\mathrm{DES}} h\right)^\beta}\right)^{-1/\beta},
\label{eq:ydes}
\end{equation}
and we use the latter modification in order to avoid kinks in the residual. For $C_{\mathrm{DES}}$, Spalart et al\cite{Spalart97} specify a value of the order of unity and a careful calibration using turbulent channel flow yields $C_{\mathrm{DES}}=0.85$ with the solver presented in Section~\ref{sec:dgsolver}. We further employ the constant $\beta=6$ and take $h$ as the wall-normal width of the element $\Delta y_e$ divided by $k+1$, $h=\Delta y_e/(k+1)$, with $k$ the polynomial degree of the discrete function space of $\bar{\bm{u}}$ (see Section~\ref{sec:enrichment}) and $k+1$ nodes in each spatial direction. We note that a factor of $k$ instead of $k+1$ is sometimes used in the DG context, see, eg, Wurst et al\cite{Wurst13}, but the factor $k+1$ seems to be more consistent with the analysis carried out Moura et al\cite{Moura15}.

{\bf Remark:} Taking the wall-normal cell width in Equation~\eqref{eq:ydes} stands in contrast to the recommendation in DES of using $\max(\Delta x,\Delta y, \Delta z)$\cite{Shur99} as a characteristic cell length. The wall-normal grid size is chosen in this work as we require the flow to be fully turbulent at the interface $\partial\widetilde{\Omega}^I$, where the RANS component ends, and the resolved LES stresses are assumed to behave approximately universal in wall-normal direction. The requirement that turbulent eddies have to be resolved by the first off-wall cell limit the cell aspect ratio $\max(\Delta x,\Delta z)/\Delta y_1$ in the same way as for wall-stress models. The specific cell aspect ratio requirements are analyzed in detail in Section~\ref{sec:channel}.

As the final result we get for the hybrid-filtered Navier--Stokes equations
\begin{alignat}{3}
&\nabla \cdotp \left(\bar{\bm{u}}+\widetilde{\bm{u}}\right) &&= 0,\label{eq:finalconti}\\
\frac{\partial \left(\bar{\bm{u}}+\widetilde{\bm{u}}\right)}{\partial t} + & \nabla \cdotp \left(\bm{\mathcal{F}}^c\left(\bar{\bm{u}}+\widetilde{\bm{u}}\right)
+ p_H \bm{\mathcal{I}} - \bm{\mathcal{F}}^{\nu}\left(\bar{\bm{u}}+\widetilde{\bm{u}}\right) - \bm{\mathcal{F}}^{\nu_t}\left(\widetilde{\bm{u}}\right)\right) &&=  \bm{f}, \label{eq:finalmom}
\end{alignat}
where the components of the velocity solution have been expanded in $\bm{u}_H = \bar{\bm{u}} + \widetilde{\bm{u}}$. We solve for these two velocity components, an eddy-resolving velocity component $\bar{\bm{u}}$ and a statistical velocity component $\widetilde{\bm{u}}$, explicitly in our numerical method. The idea is that the method computes turbulent eddies where the LES mesh is sufficiently fine. If the mesh is too coarse to resolve the near-wall flow, the method automatically promotes the RANS modes and computes the flow, or parts of it, in a statistical sense. Since the resolution of the pressure is of much less relevance in turbulent boundary layers, the pressure is kept as a single variable filtered with the hybrid RANS/LES filter.

{\bf Remark:} A straightforward discretization of Equations~\eqref{eq:finalconti} and~\eqref{eq:finalmom} with the same function space for $\bar{\bm{u}}$ and $\widetilde{\bm{u}}$ would not be possible, as there are formally seven unknowns (three components each in $\bar{\bm{u}}$ and $\widetilde{\bm{u}}$ as well as $p_H$) whereas solely four equations are available. This underdetermination requires the user to make further assumptions on the composition of the solution, in particular that the two components of the velocity field are described with finite numerical resolution and using linearly independent shape functions. An efficient description of this basis should take further knowledge about turbulent boundary layers into account and our choice is presented in Section~\ref{sec:spaces}. The general framework of weak forms in a multiscale context, where different shape functions are used for particular scales, is discussed in the subsequent section.

Outside of the near-wall layer $\widetilde{\Omega}$, only the LES scale $\bar{\bm{u}}$ is considered, and the coupling condition on the interface $\partial \widetilde{\Omega}^I$ in Figure~\ref{fig:sketch} is $ \bm{u}_H = \bar{\bm{u}}$. Additionally, we require that the viscous flux by the eddy viscosity term $\bm{\mathcal{F}}^{\nu_t}\left(\widetilde{\bm{u}}\right)$ becomes zero in normal direction to avoid a kink in the solution, giving the interface condition
\begin{equation}
\bm{\mathcal{F}}^{\nu_t}\left(\widetilde{\bm{u}}\right) \cdotp \bm{n} = \bm{0} \text{ \hspace{0.2cm} on } \partial \widetilde{\Omega}^I \times [0,\mathcal{T}].
\label{eq:interface_condition}
\end{equation}
Since $\nu_t$ is in general not zero, this condition is equivalent to $\bm{\epsilon}\left(\widetilde{\bm{u}}\right)\cdotp \bm{n}=\bm{0}$.

\subsection{Variational multiscale formulation}
\label{sec:vms}
The goal of the present subsection is twofold: We explain, on an abstract level, how the solvability of Equations~\eqref{eq:finalconti} and~\eqref{eq:finalmom} is enabled despite the seemingly underdetermination of the equation system. This is done by choosing weighting functions for the weak form which are of the same structure as the velocity components and by taking a linearly independent functions for the two scales $\bar{\bm{u}}$ and $\widetilde{\bm{u}}$ with finite dimension each. An efficient set of discrete linearly independent function spaces via function enrichment is presented in Section~\ref{sec:spaces}. Furthermore, we explain how the viscous operator in Equation~\eqref{eq:finalmom} has to be modified in order to enable a physically suitable RANS solution to develop.

For this purpose, we derive an abstract weak form of Equations~\eqref{eq:finalconti} and~\eqref{eq:finalmom} in the standard procedure by multiplication of these equations with appropriate weighting functions $\bm{v}\in\mathcal{V}^{\bm{u}}$ as well as $q\in\mathcal{V}^{{p}}$ and integration over the whole spatial domain $\Omega$. As a result we obtain the variational form of the incompressible Navier--Stokes equations as
\begin{eqnarray}
\mathscr{C}\left(q,\bar{\bm{u}}+\widetilde{\bm{u}}\right) =& 0, \label{eq:generic_weak_q} \\
\mathscr{M}\left(\bm{v},\frac{\partial \left(\bar{\bm{u}}+\widetilde{\bm{u}}\right)}{\partial t}\right) + \mathscr{F}^c\left(\bm{v},\bar{\bm{u}}+\widetilde{\bm{u}}\right) + \mathscr{P}\left(\bm{v}, p_H\right) - \mathscr{F}^{\nu}\left(\bm{v},\bar{\bm{u}}\right) - \mathscr{F}^{\nu+\nu_t}\left(\bm{v},\widetilde{\bm{u}}\right) =& l\left(\bm{v}\right).\label{eq:generic_weak_v}
\end{eqnarray}
Herein, the terms correspond to the transient (mass) term $\mathscr{M}$, convective term $\mathscr{F}^c$, pressure term $\mathscr{P}$, the rearranged viscous terms $\mathscr{F}^{\nu}$ and $\mathscr{F}^{\nu+\nu_t}$, and the right-hand-side term $l$, as well as the velocity-divergence term of the continuity equation $\mathscr{C}$. All terms are bilinear regarding $\bm{v}$ as well as $\bar{\bm{u}}$ and $\widetilde{\bm{u}}$ except for the convective term, which is nonlinear in the second slot. The detailed definitions of these operators will be elaborated in the context of the solver description in Section~\ref{sec:dgsolver}.

The classical Bubnov--Galerkin method suggests to take weighting functions of the same space as the solution functions. We choose appropriate solution functions $\bar{\bm{u}}\in\mathcal{V}^{\bar{\bm{u}}}$ and $\widetilde{\bm{u}}\in\mathcal{V}^{\widetilde{\bm{u}}}$, assume direct sum decomposition of the underlying spaces $\mathcal{V}^{{\bm{u}}} = \mathcal{V}^{\bar{\bm{u}}} \oplus \mathcal{V}^{\widetilde{\bm{u}}}$, ie, $\left(\bar{\bm{u}} + \widetilde{\bm{u}}\right)\in\mathcal{V}^{{\bm{u}}}$, and we require $\mathcal{V}^{\bar{\bm{u}}} \cap \mathcal{V}^{\widetilde{\bm{u}}}=\{\bm{0}\}$ for linear independence. Employing the same basis for the weighting functions, we obtain two components for $\bm{v}$, $\bar{\bm{v}}=\mathcal{V}^{\bar{\bm{u}}}$ and $\widetilde{\bm{v}}=\mathcal{V}^{\widetilde{\bm{u}}}$, and we have $\bm{v} = \bar{\bm{v}} + \widetilde{\bm{v}}$. By inserting this ansatz into the weak forms \eqref{eq:generic_weak_q} and \eqref{eq:generic_weak_v}, the momentum equation may be decomposed into two equations resembling the two scales of the solution according to the classical variational multiscale paradigm\cite{Hughes95,Hughes98}: an LES scale, weighted with $\bar{\bm{v}}$, and a RANS scale, weighted with $\widetilde{\bm{v}}$, yielding
\begin{eqnarray}
\mathscr{C}\left(q,\bar{\bm{u}}+\widetilde{\bm{u}}\right) =& 0,\label{eq:generic_vms_q}\\
\mathscr{M}\left(\bar{\bm{v}},\frac{\partial \left(\bar{\bm{u}}+\widetilde{\bm{u}}\right)}{\partial t}\right) + \mathscr{F}^c\left(\bar{\bm{v}},\bar{\bm{u}}+\widetilde{\bm{u}}\right) + \mathscr{P}\left(\bar{\bm{v}}, p_H\right) - \mathscr{F}^{\nu}\left(\bar{\bm{v}},\bar{\bm{u}}\right) - \mathscr{F}^{\nu+\nu_t}\left(\bar{\bm{\bm{v}}},\widetilde{\bm{u}}\right)=& l\left(\bar{\bm{v}}\right),\label{eq:generic_vms_vbar}\\
\mathscr{M}\left(\widetilde{\bm{v}},\frac{\partial \left(\bar{\bm{u}}+\widetilde{\bm{u}}\right)}{\partial t}\right) + \mathscr{F}^c\left(\widetilde{\bm{v}},\bar{\bm{u}}+\widetilde{\bm{u}}\right) + \mathscr{P}\left(\widetilde{\bm{v}}, p_H\right) - { \mathscr{F}^{\nu}\left(\widetilde{\bm{v}},\bar{\bm{u}}\right)} - \mathscr{F}^{\nu+\nu_t}\left(\widetilde{\bm{v}},\widetilde{\bm{u}}\right) =& l\left(\widetilde{\bm{v}}\right).\label{eq:generic_vms_vtilde}
\end{eqnarray}
The equation for the continuity equation remains unchanged. Through the weighting of Equations~\eqref{eq:generic_vms_vbar} and \eqref{eq:generic_vms_vtilde} with linearly independent function spaces, the equation system is well posed and resolves the underdetermination discussed in the previous section.

The equation for the LES scale, Equation~\eqref{eq:generic_vms_vbar}, is essentially a standard LES approach on a background convective velocity $\widetilde{\bm{u}}$, similar to NLDE\cite{Labourasse02}, except for an additional eddy viscosity term based on $\widetilde{\bm{u}}$, providing the LES scale with physical dissipation in regions where the energy-carrying scales are not sufficiently resolved. Potential LES subgrid models contained in $\bm{\tau}_{\mathrm{LES}}$ would also be added to this equation.

Equation~\eqref{eq:generic_vms_vtilde} represents the RANS scale, which resolves a variable fraction of the averaged velocity. It may be observed that the LES solution couples into the RANS equation in~\eqref{eq:generic_vms_vtilde} and this coupling should be investigated in more detail. Considering again the NLDE methodology as a reference, the RANS equations are well defined without the additional LES source terms. However, the definition of the eddy viscosity in this work (Equation~\eqref{eq:mix}) takes the resolved Reynolds stresses of the LES arising in the convective term into account, such that the full convective flux should be included, and thus the transient term. In contrast, there is no physical justification for the viscous LES part. Indeed, our numerical tests revealed that the method tends to promote an unphysical RANS mode, such as a RANS solution directed adverse to the primary flow direction even in attached boundary layers, if the viscous LES term is included in the RANS equation. Further arguments to be considered are that we expect the viscous RANS flux to be dominant in the underresolved LES region $\left(|\bm{\mathcal{F}}^{\nu}\left(\bar{\bm{u}}\right)| \ll |\bm{\mathcal{F}}^{\nu+\nu_t}\left(\widetilde{\bm{u}}\right)|\right)$. Additionally, numerical stability issues according to a coercivity analysis (in Section~\ref{sec:coercivity}) suggest that this term would significantly restrict the range of application if taken into account. We conclude that the viscous LES term should be canceled from the RANS equation according to
\begin{equation}
\mathscr{M}\left(\widetilde{\bm{v}},\frac{\partial \left(\bar{\bm{u}}+\widetilde{\bm{u}}\right)}{\partial t}\right) + \mathscr{F}^c\left(\widetilde{\bm{v}},\bar{\bm{u}}+\widetilde{\bm{u}}\right) + \mathscr{P}\left(\widetilde{\bm{v}}, p_H\right) - \mathscr{F}^{\nu+\nu_t}\left(\widetilde{\bm{v}},\widetilde{\bm{u}}\right) = l\left(\widetilde{\bm{v}}\right) \label{eq:generic_vms_vtilde_mod}
\end{equation}
in replacement of Equation~\eqref{eq:generic_vms_vtilde}. It is noted that this modification solely has an impact on one component of the equations and therefore primarily represents a measure for helping the method to find a physically suitable composition of $\bar{\bm{u}}$ and $\widetilde{\bm{u}}$ within the solution space. Similar modifications in the framework of the variational multiscale method are frequently considered regarding scale separation into large and small resolved/unresolved scales in LES turbulence modeling, see eg the review article\cite{Gravemeier06}.

In this section, a consistent set of governing equations has been derived, which decomposes the incompressible Navier--Stokes equations into a RANS scale and an LES scale. We have argued that a linearly independent function space is required for these two components in order to guarantee solvability of the equation system. In the next section, we show how the concept of function enrichment may be used to construct a highly efficient function space for $\bar{\bm{u}}$ and $\widetilde{\bm{u}}$.

\section{RANS and LES velocity components via function enrichment}
\label{sec:spaces}
The concept of function enrichment was introduced by Krank and Wall\cite{Krank16} within the standard FEM as a way to economize the computation of turbulent boundary layers. This approach has since been applied to the discontinuous Galerkin method for DES\cite{Krank17c} and the RANS equations by Krank et al\cite{Krank16c}; the present section is closely related to the latter articles. The basic idea is to construct additional shape functions using a wall-law as enrichment function, which enable the use of very coarse meshes in the near-wall region along with exact boundary conditions since the function space is capable of resolving the sharp boundary layer gradient present in the mean velocity. In the current work, this capability is used to resolve the RANS scale according to Section~\ref{sec:multiscale_approach} in a layer near the wall with very few degrees of freedom. The high-order polynomial component additionally represents eddies where the mesh is sufficiently fine. Since these basis functions are linearly independent, they may be used to approximate the RANS and LES scales in the variational multiscale method described in Section~\ref{sec:vms}. The major idea is summarized in Section~\ref{sec:enrichment} and the enrichment function employed in this article is presented in Section~\ref{sec:vandriest}.

\subsection{Function enrichment}
\label{sec:enrichment}
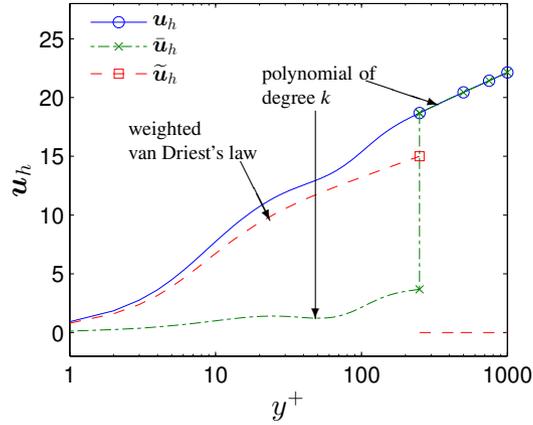
\begin{figure}[t]
\centering
\includegraphics[trim= 0mm 0mm 0mm 0mm,clip,scale=0.7]{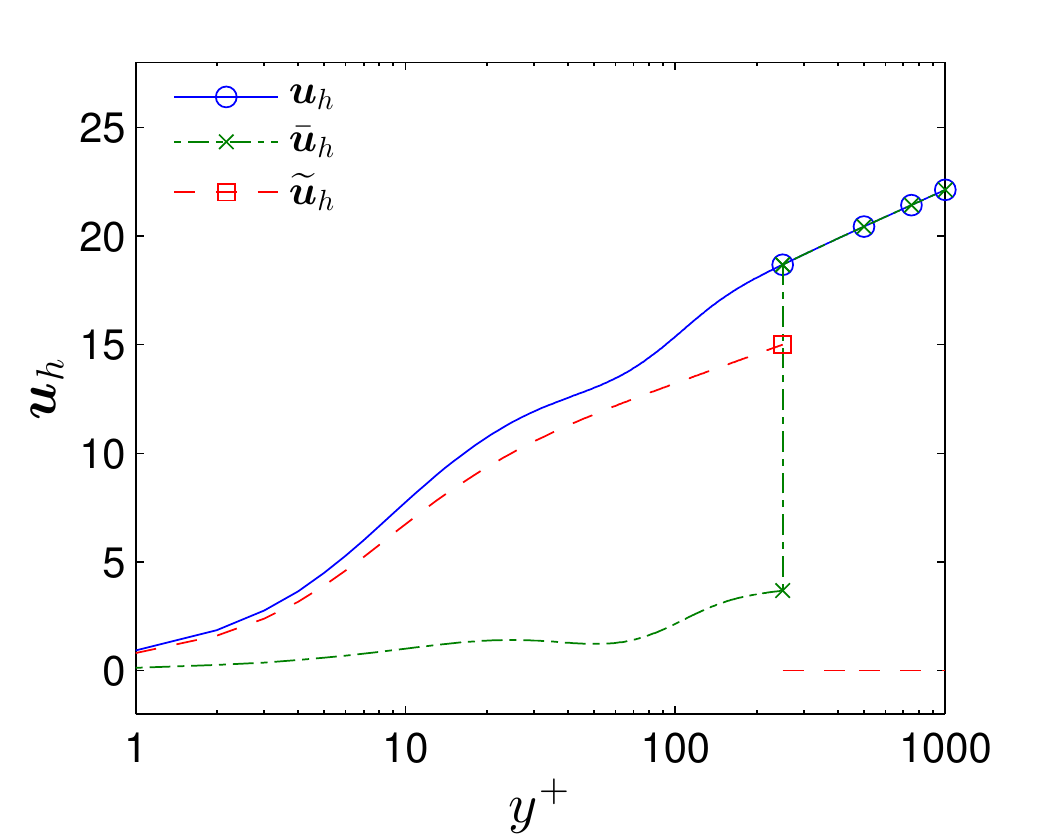}
\begin{picture}(300,0)
\put(136,110){\vector(1,-2){11}}
\put(94,116){\footnotesize \parbox{2.1cm}{weighted\\van Driest's law}}
\put(183,140){\vector(3,-1){27}}
\put(164,129){\vector(0,-1){78}}
\put(144,136){\footnotesize \parbox{1.5cm}{polynomial of degree $k$}}
\end{picture}
\caption{Decomposition of a boundary layer into a high-order polynomial component and a weighted wall function. The decomposition applies solely within the first off-wall cell.}
\label{fig:decomp}
\end{figure}

We consider a tessellation of the three-dimensional spatial computational domain $\Omega_h \subset \mathbb{R}^3$ into $N_e$ nonoverlapping hexahedral finite elements $\Omega_e$, given as $\Omega_h = \bigcup_{e=1}^{N_e} \Omega_e$. The subscript $(\cdotp)_h$ relates the respective variable to a characteristic element size $h$. The discrete velocity solution is denoted as $\bm{u}_h$ in the following and, within the first element layer near the no-slip wall, it consists of two parts, a high-order polynomial component $\bar{\bm{u}}_h\left(\bm{x},t\right)\in\mathcal{V}_h^{\bar{\bm{u}}}=\left(\mathcal{V}_h^p\right)^3$ and an enrichment component $\widetilde{\bm{u}}_h\left(\bm{x},t\right)\in \mathcal{V}_h^{\widetilde{\bm{u}}}$, yielding
\begin{equation}
\bm{u}_h\left(\bm{x},t\right)=\bar{\bm{u}}_h\left(\bm{x},t\right)+\widetilde{\bm{u}}_h\left(\bm{x},t\right).
\label{eq:space}
\end{equation}
These two components are of course related to the respective RANS and LES contributions of the solution as introduced in the previous section. In each element, we use a polynomial of degree $k$ to represent the LES component of both $\bar{\bm{u}}_h$ and $p_h$, for example for the pressure:
\begin{equation*}
\mathcal{V}_h^p = \{p_h \in L^2 (\Omega) : p_h |_{\Omega_e} \in P_k(\Omega_e), \forall e \in \Omega_h \},
\end{equation*}
where $P_k$ are all polynomials up to the tensor degree $k$ and the solution allows for discontinuities at the element boundaries. Written in terms of a finite element expansion, the standard polynomial component in each element becomes
\begin{equation}
\bar{\bm{u}}_h(\bm{x},t)=\sumop_{B \in N^{k}} N_B^{k}(\bm{x}) \bar{\bm{u}}_B(t),
\label{eq:std_fe}
\end{equation}
where $N_B^{k}$ is the $B^{\mathrm{th}}$ shape function of degree $k$ and $\bar{\bm{u}}_B$ is the corresponding nodal value. The enrichment may analogously be written in terms of an FE expansion of tensor degree $l$, multiplied by an enrichment function $\psi$ according to
\begin{equation}
\widetilde{\bm{u}}_h(\bm{x},t)=\psi(\bm{x},t)\sumop_{B \in N^{l}} N_{B}^{l}(\bm{x})\widetilde{\bm{u}}_{B}(t);
\label{eq:enrichment}
\end{equation}
this expansion corresponds to the general definition of $\widetilde{\bm{u}}_h$ in $\widetilde{\Omega}_h \subset {\Omega}_h$:
\begin{equation}
\mathcal{V}_h^{\widetilde{\bm{u}}} = \{\widetilde{\bm{u}}_h \in \left(L^2(\Omega)\right)^3 : \widetilde{\bm{u}}_h |_{\Omega_e} \in \left(\psi P_l(\Omega_e)\right), \forall e \in \widetilde{\Omega}_h \},
\end{equation}
and we set $\widetilde{\bm{u}}_h=\bm{0}$ outside $\widetilde{\Omega}_h$. As indicated, $\widetilde{\Omega}_h$ corresponds to the RANS/LES zone in Figure~\ref{fig:sketch}, and, in the present higher order context (eg $k=4$), this layer is chosen equivalent to the width of the first off-wall cell. If $k\leq2$, two cell layers may be considered.
This construction allows the numerical method to resolve solutions similar to the enrichment function with very coarse meshes. At the same time, full consistency is maintained and the method has sufficient flexibility in flow situations where the solution differs from the wall-law through the high-order polynomial component. In the context of boundary-layer enrichments, we therefore employ a wall function as enrichment function, which will be introduced in the subsequent section. It is noted that the blending of function spaces is not an issue within the discontinuous Galerkin method since neighboring elements may have nonconforming shape functions by default, due to the weak element coupling. As a consequence, ramp functions used within the standard FEM\cite{Krank16} are not necessary. The resulting decomposition of $\bm{u}_h$ into the two components is illustrated in Figure~\ref{fig:decomp}. The discrete pressure variable is adequately represented by a high-order polynomial as high gradients are not expected in the pressure distribution.

\subsection{Enrichment function}
\label{sec:vandriest}
In our previous work\cite{Krank16,Krank16c,Krank17c}, we employed Spalding's law as the enrichment function, which has proven to be suitable both for RANS, DES, and LES. Preliminary investigations of the present wall model have indicated the significant potential of using the particular wall function which is consistent with the eddy viscosity model presented in Section~\ref{sec:scalesep} in the viscous sublayer and the buffer layer ($y^+<30$). For example, the predictions of the wall shear stress could be enhanced in accuracy and robustness. This wall function has been proposed by van Driest\cite{vanDriest56} and fulfills the boundary conditions $u^+(y=0)=0$ and $\frac{\partial u^+}{\partial y^+}|_{y=0}=1$. The enrichment function is defined as
\begin{equation}
\psi = \int_0^{y^+}\frac{2\ \mathrm{d}y^+}{1+\sqrt{1+\left(2\kappa y^+(1-\mathrm{exp}(-y^+/A^+))\right)^2}}
\label{eq:psi}
\end{equation}
with $u^+ = \psi$, and we take $\kappa=0.41$ and $A^+=26$. Since the enrichment function has to be evaluated on every quadrature point of the numerical method in a layer of elements near the wall, and an efficient evaluation may not be obvious, we describe a possible approach in Appendix~\ref{sec:vdriest_num}.

As the enrichment function is universal with respect to the wall coordinate $y^+=y \sqrt{\tau_w/\rho}/\nu$, given the wall shear stress $\tau_w$ and the density $\rho$, the wall coordinate scales the enrichment function in wall-normal direction to match the local wall shear stress. We have developed an adaptation algorithm\cite{Krank16,Krank16c} which allows for spatial and temporal variations of the flow in the wall model by discretizing the wall shear stress $\tau_w$ based on linear continuous shape functions $N_B^{c,m}$ ($m=1$): The wall shear stress is computed on each node $B$ in $N^{c,m}$ similar to\cite{Krank16c} with
\begin{equation}
\tau_{w,B}=\frac{\lVert\int_{\partial \Omega^D} N_B^{c,m}(\bm{x})\bm{\tau}_w \ \mathrm{d}A \rVert}{\int_{\partial \Omega^D} N_B^{c,m}(\bm{x}) \ \mathrm{d}A},
\end{equation}
where $\bm{\tau}_w$ is the wall shear stress vector, which may be computed as $\bm{\tau}_w=\rho \nu \left(\nabla \bm{u}_h\right) \cdotp \bm{n}$.
Note that we compute the integral for each component of the numerator and subsequently take the vector norm. The nodal values are interpolated using
\begin{equation}
\tau_{w,h}=\sumop_{B \in N^{c,m}}N_B^{c,m} \tau_{w,B}.
\end{equation}
The quantity $\tau_{w,h}$ includes a spatial coarsening of the wall shear stress distribution similar to\cite{Krank16}. This coarsening is physically motivated since the wall function represents a relation for the mean quantities, ie, the mean wall shear stress gives a mean velocity profile. Without this coarsening, one would observe a statistical overprediction of $\tau_w$. The wall distance is likewise defined using a finite element expansion where the weights $y_{B}$ are given as the shortest distance between the current and the closest wall node,
\begin{equation}
y_{h}=\sumop_{B \in N^{c,m}}N_B^{c,m} y_{B},
\end{equation}
which facilitates the evaluation of the enrichment shape functions and their spatial gradients. The $y^+$-variable in the mixing length eddy viscosity model in Equation~\eqref{eq:lmix} is also computed based on these definitions of $\tau_{w,h}$ and $y_{h}$.

The wall shear stress is updated at the beginning of every time step using the velocity solution of the previous time step for $\tau_{w,h}$, resulting in a new FE space in every time step. A cell-wise $L^2$-projection is employed to project the solution vector onto the new function space with
\begin{equation}
\int_{\Omega_e}\bm{v}_h^{n,\mathrm{new}} \cdotp \bm{u}_h^{n,\mathrm{new}} \ \mathrm{d}\Omega = \int_{\Omega_e} \bm{v}_h^{n, \mathrm{new}} \cdotp \bm{u}_h^{n, \mathrm{old}} \ \mathrm{d}\Omega,
\label{eq:l2}
\end{equation}
where $n$ is the current time step number.
Supplementary solution vectors required by the time integration scheme are projected as well or recomputed from the velocity solution $\bm{u}_h^{n,\mathrm{new}}$, if possible. Instead of using a wall shear stress based on the solution of the previous time step, it would also be possible to use a higher order estimate such as an extrapolated wall shear stress based on a series of previous time steps for improved accuracy. Since the `lag' in $\tau_{w,h}$ is small in this work due to the semi-explicit formulation of the scheme, such an extrapolation is not considered.

Finally, the present adaptation algorithm also allows to switch off the wall model when not needed, ie, when the standard polynomial component is sufficient to resolve the necessary scales according to\cite{Krank16c}. In particular, one observes problems in conditioning when the enriched element spans a $y^+$-range smaller than approximately 15 wall units, depending on the respective choice of $k$ and $l$, as the standard and enrichment shape functions become close to linearly dependent when $\tau_w \to 0$, yielding $\psi\to 0$. This issue is circumvented by switching off the enrichment and the complete multiscale wall model if all quadrature points of the wall-nearest cell lie in $y^+<30$. A parameter study regarding the latter quantity, using turbulent channel flow with the setup applied in Figure~\ref{fig:ch_wallres_u}, yielded almost indistinguishable results for values in the range of 20 to 30. In case $y^+$ becomes larger than 30 at any quadrature point of this cell in a subsequent time step, the wall model is switched on again by taking the enrichment in the solution of the new time step into account and an appropriate RANS solution develops automatically within a few time steps.

\section{Application to a high-order discontinuous Galerkin solver}
\label{sec:dgsolver}
The present multiscale wall model may in principle be implemented in any discontinuous Galerkin flow solver. In order to exemplify the approach, we employ our high-performance matrix-free incompressible code INDEXA\cite{Krank16b}. An extension to compressible flow, where the energy equation has to be discretized in addition, is straightforward since, in adiabatic boundary layers, high gradients are not expected in the energy variable, allowing it to be treated in the same way as the pressure variable by the high-order polynomial space. The present section gives an overview of the temporal discretization (Section~\ref{sec:temp}) and the spatial operators of the variational form (Section~\ref{sec:spatial}), and presents a numerical stability analysis of the viscous multiscale term (Section~\ref{sec:coercivity}). Finally, we discuss possibilities for the efficient implementation of the enriched elements (Section~\ref{sec:implementation}).

\subsection{Temporal dual splitting scheme}
\label{sec:temp}
The incompressible Navier--Stokes equations are integrated in time using the high-order semi-explicit dual-splitting scheme by Karniadakis et al\cite{Karniadakis91}, which decomposes the governing equations into three sub-steps. In the first step, the convective term is treated explicitly as
\begin{equation}
\frac{\gamma_0 \hat{\bm{u}}-\sumop_{i=0}^{J-1}\left(\alpha_i \bm{u}^{n-i}\right)}{\Delta t} = - \sumop_{i=0}^{J-1}\beta_i \nabla \cdotp \bm{\mathcal{F}}^c\left(\bm{u}^{n-i}\right)+\bm{f}^{n+1},
\label{eq:convstep}
\end{equation}
yielding the first intermediate velocity $\hat{\bm{u}}$. Herein, $\Delta t$ denotes the time increment, $n$ the time step, $J$ the order of accuracy, and $\alpha_i$, $\beta_i$, as well as $\gamma_0$ time integration constants of the extrapolation and backward-differentiation (BDF) schemes. The explicit character of this step restricts the time step size according to the Courant--Friedrichs--Lewy (CFL) condition. The time step size is chosen adaptively in order to maximize the time increments in each step with the accurate CFL condition\cite{Krank16c}, and the time integration parameters are calculated to match variable time increments. We take a value for the Courant number of $\mathrm{Cr}=0.14$ and a second order accurate scheme ($J=2$) for all examples considered in this article. Unlike $k$-adaptivity, we have not observed time step restrictions through the enrichment.

In the second step of the time integration scheme, a Poisson equation is solved for the pressure $p^{n+1}$ at the time $t^{n+1}$,
\begin{equation}
-\nabla^2 p^{n+1} = -\frac{\gamma_0}{\Delta t} \nabla \cdotp \hat{\bm{u}}.
\label{eq:poisson}
\end{equation}
We adopt the high-order boundary conditions for the Poisson problem\cite{Karniadakis91},
\begin{equation}
\nabla p^{n+1} \cdotp \bm{n}= -\left(\sumop_{i=0}^{J-1}{\beta_i\left(\nabla \cdotp \bm{\mathcal{F}}^c\left(\bm{u}_h^{n-i}\right) + \nu\nabla \times \bm{\omega}^{n-i}\right)-\bm{f}^{n+1}}\right)\cdotp \bm{n}
\label{eq:bc_d_pres}
\end{equation}
on $\partial \Omega^D$, where we have made use of the boundary conditions $\bm{g}_{\bm{u}}=\bm{0}$ and $\nu_t=0$. Only the solenoidal viscous part is taken into consideration\cite{Karniadakis91}, and the vorticity $\bm{\omega}$ is computed as in Krank et al\cite{Krank16c,Krank16b}. We further note that the present dual-splitting scheme allows equal-order interpolation of velocity and pressure\cite{Karniadakis13}. With the pressure available, the velocity field is made divergence-free in a local projection step, given as
\begin{equation}
\hat{\hat{\bm{u}}}=\hat{\bm{u}}-  \frac{\Delta t}{ \gamma_0} \nabla p^{n+1},
\end{equation}
resulting in the second intermediate velocity $\hat{\hat{\bm{u}}}$. Finally, the viscous effects are taken into account in a Helmholtz-like equation,
\begin{equation}
\frac{\gamma_0}{\Delta t}\left(\bm{u}^{n+1}-\hat{\hat{\bm{u}}}\right) = \nabla \cdotp \bm{\mathcal{F}}^{\nu}\left(\bm{u}^{n+1}\right),
\label{eq:visc}
\end{equation}
yielding the velocity solution $\bm{u}^{n+1}$ at time $t^{n+1}$.
\subsection{Galerkin formulation}
\label{sec:spatial}

Variational formulations are derived for all steps of the time integration scheme (Equations~\eqref{eq:convstep}--\eqref{eq:visc}) and the weak forms are related to the symbolic variational formulations discussed in Section~\ref{sec:vms}. These weak forms are adapted from\cite{Krank16c,Krank16b} with major modifications of the viscous term summarized in the following. They consist of element-wise volume integrals and integrals over interior faces $\partial \Omega_e^{\Gamma}=\partial \Omega_e^-\cap \partial \Omega_e^+$ between two adjacent elements $\Omega_e^-$ and $\Omega_e^+$. Unit normal vectors point outwards of the respective cells and are denoted $\bm{n}_{\Gamma}^-$ and $\bm{n}_{\Gamma}^+$ such that $\bm{n}_{\Gamma}^-=-\bm{n}_{\Gamma}^+$ holds. Jumps in the solution at the interfaces between two neighboring elements are abbreviated with $[\phi]=\phi^--\phi^+$ and $\llbracket \phi \rrbracket = \phi^- \otimes \bm{n}_{\Gamma}^- + \phi^+ \otimes \bm{n}_{\Gamma}^+$. Likewise, an averaging operator defines the mean value $\{\!\{\phi\}\!\}=w^-\phi^-+w^+\phi^+$ with the weights $w^-=w^+=1/2$ unless specified otherwise. Boundary conditions are imposed by prescribing external values for $\phi^+$ and $\nabla \phi^+$ on $\partial \Omega^D$ and interface conditions on $\partial \widetilde{\Omega}^I$. The weak form of the multiscale model further makes use of the definitions $\widetilde{\Omega}_e=\Omega_e\cap\widetilde{\Omega}$ (enriched cell interior), $\partial \widetilde{\Omega}_e^I=\partial\Omega_e \cap \partial \widetilde{\Omega}^I$ (face on $\partial \widetilde{\Omega}^I$, exactly one face enriched), and $\partial \widetilde{\Omega}_e=\partial \Omega_e\cap \widetilde{\Omega} \backslash \partial \widetilde{\Omega}_e^I$ (both adjacent faces enriched). The superscripts $(\cdotp)^-$ and $(\cdotp)^+$ are omitted in the remainder of this section if possible without ambiguity. The subsequent weak forms employ the common notation for $L^2$ inner products abbreviating volume integrals with $(a,b)_{\Omega_e}=\int_{\Omega_e}a b \ \mathrm{d}\Omega$ for scalars, $(\bm{a},\bm{b} )_{\Omega_e}=\int_{\Omega_e}\bm{a} \cdotp \bm{b} \ \mathrm{d}\Omega$ for vectors, and $(\bm{a},\bm{b} )_{\Omega_e}=\int_{\Omega_e}\bm{a} : \bm{b} \ \mathrm{d}\Omega$ for tensors. The notation for face integrals is analogous.

We derive element-wise weak forms for each term of the Navier--Stokes equations by multiplying the respective term with a weighting function and integrating over one element volume:
\begin{itemize}
\item {\it Mass term:} The mass term is obtained without further modification as
\begin{equation}
\mathscr{M}_e({\bm{v}},\bm{u}) = \left(\bm{v},\bm{u}\right)_{\Omega_e}.
\end{equation}
Neighboring cells are not connected due to the absence of face terms.
\item {\it Convective term:}
The weak form is integrated by parts, yielding
\begin{equation}
\mathscr{F}^c_e({\bm{v}},\bm{u}) = -\left(\nabla\bm{v}, \bm{\mathcal{F}}^c\left(\bm{u}\right)\right)_{\Omega_e}+\left(\bm{v},\bm{\mathcal{F}}^{c*}\left(\bm{u}\right)\cdotp\bm{n}_{\Gamma}\right)_{\partial \Omega_e}
\end{equation}
for the convective term. The convective flux is defined via the local Lax--Friedrichs flux
\begin{equation}
    \bm{\mathcal{F}}^{c*}\left(\bm{u}\right)=
                  \{\!\{\bm{\mathcal{F}}^{c}\left(\bm{u}\right)\}\!\}+\Lambda/2\llbracket \bm{u}\rrbracket,
\end{equation}
and we take the maximum of $\Lambda = \max(\lambda^-,\lambda^+)$ across the interface of the largest eigenvalue of the flux Jacobian (see Reference\cite{Krank16b})
\begin{equation}
\begin{array}{ll}
\lambda^{\mp} = 2|\bm{u}^{\mp} \cdotp \bm{n}_{\Gamma}|.
\end{array}
\end{equation}

\item{\it Pressure gradient:}
The pressure gradient is integrated by parts as suggested in Krank et al\cite{Krank16b}, yielding
\begin{equation}
\mathscr{P}_e(\bm{v}, p) = - \left(\nabla \cdotp \bm{v}, p\right)_{\Omega_e} + \left(\bm{v}, \{\!\{p\}\!\} \bm{n}_{\Gamma}\right)_{\partial \Omega_e}.
\end{equation}
\item {\it Velocity divergence:}
The velocity divergence is integrated by parts according to\cite{Krank16b} as well and results in
\begin{equation}
\mathscr{C}_e(q,\bm{u}) = -\left(\nabla q, \bm{u}\right)_{\Omega_e} + \left(q,\{\!\{{\bm{u}}\}\!\}\cdotp \bm{n}_{\Gamma}\right)_{\partial \Omega_e}.
\end{equation}

\item {\it Viscous multiscale term:}
We discuss a suitable viscous term in view of the modifications introduced in Section~\ref{sec:vms}. The baseline viscous implementation of our solver is an interior penalty method in symmetric form in the standard incompressible flow solver\cite{Krank16b} and in nonsymmetric form in the enriched solver\cite{Krank16c}. The nonsymmetric variant of the interior penalty method has the advantage of being stable with very low requirements on the penalty parameter\cite{Riviere99}, which is beneficial in the case of nonpolynomial shape functions, such as the present boundary layer enrichment, where the derivation of inverse estimates on the fly is not economical.

The viscous term considered in this work follows Reference\cite{Krank16c} and the impact of the LES solution $\bar{\bm{u}}$ on the equations for the RANS scale weighted with $\widetilde{\bm{v}}$ are canceled as suggested in Section~\ref{sec:vms}. We propose the following formulation of the viscous multiscale term:
\begin{alignat}{9}
&&&\mathrm{volume \ terms:} && \mathrm{adjoint \ terms:} && \mathrm{std. \ consist. \ terms:} && \mathrm{penalty \ terms:} \nonumber \\
&\mathscr{F}^{\nu}_e(\bar{\bm{v}},\widetilde{\bm{v}},\bar{\bm{u}},\widetilde{\bm{u}})=&& & &  \nonumber\\
&\widetilde{\bm{v}}, \widetilde{\bm{u}}: \hspace{0.2cm}&-&\left(\bm{\epsilon}\left(\widetilde{\bm{v}}\right),\bm{\mathcal{F}}^{\nu+\nu_t}\left(\widetilde{\bm{u}}\right)\right)_{\widetilde{\Omega}_e} &-& \left(w\bm{\mathcal{F}}^{\nu+\nu_t}\left(\widetilde{\bm{v}}\right),\llbracket\widetilde{\bm{u}}\rrbracket\right)_{\partial \widetilde{\Omega}_e} &+& \left(\widetilde{\bm{v}},\{\!\{\bm{\mathcal{F}}^{\nu+\nu_t}\left(\widetilde{\bm{u}}\right)\}\!\}\cdotp \bm{n}_{\Gamma}\right)_{\partial \widetilde{\Omega}_e} &-& \left(\widetilde{ \bm{v}},\tau_{\mathrm{IP}} \nu \llbracket \widetilde{\bm{u}}\rrbracket \cdotp \bm{n}_{\Gamma}\right)_{\partial {\Omega}_e} \nonumber \\
&\widetilde{\bm{v}}, \bar{\bm{u}}: &&    & -& \left(w\bm{\mathcal{F}}^{\nu+\nu_t}\left(\widetilde{\bm{v}}\right),\llbracket\bar{\bm{u}}\rrbracket\right)_{\partial \widetilde{\Omega}_e} &+& \left(\widetilde{ \bm{v}},\{\!\{\bm{\mathcal{F}}^{\nu}\left(\bar{\bm{u}}\right)\}\!\}\cdotp \bm{n}_{\Gamma}\right)_{\partial \widetilde{\Omega}_e^I} &-& \left(\widetilde{ \bm{v}},\tau_{\mathrm{IP}} \nu \llbracket \bar{ \bm{u}}\rrbracket \cdotp \bm{n}_{\Gamma}\right)_{\partial {\Omega}_e} \label{eq:viscous_multiscale}\\
&\bar{\bm{v}}, \widetilde{\bm{u}}:&-&\left(\bm{\epsilon}\left(\bar{ \bm{v}}\right),\bm{\mathcal{F}}^{\widehat{\nu+\nu_t}}\left(\widetilde{\bm{u}}\right)\right)_{\widetilde{\Omega}_e} &-& \left(\frac{1}{2}\bm{\mathcal{F}}^{\nu}\left(\bar{ \bm{v}}\right),\llbracket\widetilde{\bm{u}}\rrbracket\right)_{\partial \widetilde{\Omega}_e^I}  &+& \left(\bar{ \bm{v}},\{\!\{\bm{\mathcal{F}}^{\nu+\nu_t}\left(\widetilde{\bm{u}}\right)\}\!\} \cdotp \bm{n}_{\Gamma}\right)_{\partial \widetilde{\Omega}_e} &-& \left(\bar{ \bm{v}},\tau_{\mathrm{IP}} \nu \llbracket \widetilde{\bm{u}}\rrbracket \cdotp \bm{n}_{\Gamma}\right)_{\partial {\Omega}_e} \nonumber\\
&\bar{\bm{v}}, \bar{\bm{u}}:&-&\left(\bm{\epsilon}\left(\bar{ \bm{v}}\right),\bm{\mathcal{F}}^{\nu}\left(\bar{\bm{u}}\right)\right)_{\Omega_e} &-& \left(\frac{1}{2}\bm{\mathcal{F}}^{\nu}\left(\bar{ \bm{v}}\right),\llbracket\bar{\bm{u}}\rrbracket\right)_{\partial \Omega_e} &+& \left(\bar{ \bm{v}},\{\!\{\bm{\mathcal{F}}^{\nu}\left(\bar{\bm{u}}\right)\}\!\}\cdotp \bm{n}_{\Gamma}\right)_{\partial \Omega_e} &-& \left(\bar{ \bm{v}},\tau_{\mathrm{IP}} \nu \llbracket \bar{ \bm{u}}\rrbracket \cdotp \bm{n}_{\Gamma}\right)_{\partial {\Omega}_e} \nonumber
\end{alignat}
where all terms are expanded into the individual contributions from the RANS and LES scale. Herein, the row ($\widetilde{\bm{v}}, \widetilde{\bm{u}}$) represents the RANS contribution to the RANS equation, the row ($\widetilde{\bm{v}}, \bar{\bm{u}}$) the LES contribution to the RANS equation, the row ($\bar{\bm{v}}, \widetilde{\bm{u}}$) the RANS contribution to the LES equations and finally the last row the LES contributions to the LES scale. The columns correspond to the respective volume terms, the adjoint face terms, the standard consistency face terms, and the penalty terms. We note the skew-symmetry of the standard consistency and adjoint face terms as well as the symmetry of the penalty terms.

In Equation~\eqref{eq:viscous_multiscale} we do not entirely cancel the LES impact on the RANS equations but shift the adjoint term at the inner enriched faces $\partial \widetilde{\Omega}_e$ from the row ($\bar{\bm{v}}, \widetilde{\bm{u}}$) to the row ($\widetilde{\bm{v}}, \bar{\bm{u}}$). This modification is consistent, since it relies on the discontinuity of the velocity solution at element boundaries, and is required in order to guarantee a stable scheme through skew-symmetric face terms. The skew-symmetry is beneficial in the context of the coercivity analysis in the subsequent section. In contrast, on the interface between enriched and nonenriched cells $\partial \widetilde{\Omega}^I_e$, the opposite face terms are considered, namely the adjoint term in row ($\bar{\bm{v}}, \widetilde{\bm{u}}$) and the skew-symmetric standard consistency term in row ($\widetilde{\bm{v}}, \bar{\bm{u}}$) again in order to allow the fulfillment of the coercivity argument. The resulting Galerkin formulation is in agreement with the interface conditions $\bar{\bm{u}}^+ = \bar{\bm{u}}^- + \widetilde{\bm{u}}^-$ and $\widetilde{\bm{u}}^+ = \bm{0}$ on $\partial \widetilde{\Omega}_e^I$ (ie, $\llbracket{\bm{u}}\rrbracket=\left(\bar{\bm{u}}^- + \widetilde{\bm{u}}^- - \bar{\bm{u}}^+\right)\otimes\bm{n}_{\Gamma}$ and $\llbracket\widetilde{\bm{u}}\rrbracket=\widetilde{\bm{u}}^-\otimes\bm{n}_{\Gamma}$), where the left face $(\cdotp)^-$ is enriched and the right face $(\cdotp)^+$ is not enriched, without loss of generality. However, the condition $\bm{\epsilon}\left(\widetilde{\bm{u}}^-\right)\cdotp \bm{n}=\bm{0}$ is not exactly fulfilled as we have $\bm{\mathcal{F}}^{\nu+\nu_t}\left(\widetilde{\bm{u}}^-\right)\cdotp \bm{n}_{\Gamma}^- =\{\!\{ \bm{\mathcal{F}}^{\nu}\left(\bar{\bm{u}}\right) \}\!\}\cdotp \bm{n}_{\Gamma}$ on $\partial \widetilde{\Omega}_e^I$, but the error is considered to be small since kinks have not been observed in the solution at that boundary. As an alternative, our numerical tests have shown that the standard consistency term in row ($\widetilde{\bm{v}}, \bar{\bm{u}}$) may be neglected, which yields a better fulfillment of the latter interface condition while problems in stability have not been observed with this modified formulation. The numerical examples shown in Section~\ref{sec:examples} yet employ the variant with proven stability.

We further note that the material parameter $\nu$ is applied to the interior penalty term in the whole domain. As interior penalty stabilization parameter $\tau_{\mathrm{IP}}$ we use the definition by Hillewaert\cite{Hillewaert13} (see also References\cite{Krank16c,Krank16b}), since this choice has yielded favorable results in underresolved turbulent flows (ILES) in Krank et al\cite{Krank16b}. On the Dirichlet boundary, we have taken measures to enhance the accuracy of the weakly enforced no-slip condition\cite{Krank16c}. In the present work, we increase the interior penalty parameter $\tau_{\mathrm{IP}}$ on all Dirichlet boundary faces by a factor of $10$, which has a similar effect while being more solver-friendly in 3D. The weights of the averaging operators with spatially varying material parameter $\nu_t$ included in the $\bm{\mathcal{F}}^{\nu+\nu_t}$-terms are given through harmonic weighting\cite{Krank16c,Burman12}:
\begin{equation}
w^-=\frac{\nu+\nu_t^+}{2\nu+\nu_t^-+\nu_t^+}, \hspace{1cm}
w^+=\frac{\nu+\nu_t^-}{2\nu+\nu_t^-+\nu_t^+}.
\label{eq:harmweights}
\end{equation}
On $\partial \widetilde{\Omega}^I$, all terms including the eddy viscosity vanish, making a consideration of the varying material law unnecessary.

\item {\it Velocity div-div penalty:}
The present scheme requires a div-div penalty operator for stabilization of mass conservation, see Reference\cite{Krank16b} for a detailed discussion. The corresponding operator reads
\begin{equation}
\mathscr{D}_e\left({\bm{v}},{\bm{u}}\right)= \left(\nabla \cdotp \bm{v}, \tau_{\mathrm{D}}\nabla \cdotp {{\bm{u}}}\right)_{\Omega_e}.
\label{eq:div-div}
\end{equation}
We note that the contribution of this term vanishes if the velocity field is exactly divergence-free, as it relies on the continuity residual $\nabla \cdotp {{\bm{u}}}$. The stabilization parameter $\tau_{\mathrm{D}}$ is given as\cite{Krank16b}
\begin{equation}
\tau_{\mathrm{D}}=\frac{ \Vert {\bm{u}}^n\Vert h \Delta t}{\mathrm{Cr}},
\label{eq:tauD}
\end{equation}
where $ \Vert{\bm{u}}^n\Vert$ is the volume-averaged velocity magnitude of the previous time step, $h$ the element length based on the cube root of the element volume and Cr the global Courant number.

\item {\it Pressure Laplace term:} The pressure Poisson equation further requires the discretization of a Laplace term. We use the symmetric interior penalty method\cite{Arnold82}, yielding
\begin{equation}
\mathscr{L}_e\left({q}, p^{n+1}\right) = - \left(\nabla q, \nabla p^{n+1} \right)_{\Omega_e} + \frac{1}{2}\left(\nabla q,\llbracket p^{n+1}\rrbracket\right)_{\partial \Omega_e} + \left(q,\bm{\mathcal{P}}^*\cdotp \bm{n}_{\Gamma}\right)_{\partial \Omega_e}.
\end{equation}
For the numerical flux $\bm{\mathcal{P}}^*$, we include an interior penalty term on internal faces and use the pressure boundary conditions on the Dirichlet boundary~\eqref{eq:bc_d_pres} as
\begin{equation}
    \bm{\mathcal{P}}^*=\left\{
                \begin{array}{ll}
                  \{\!\{\nabla p^{n+1}\}\!\}-\tau_{\mathrm{IP}}\llbracket p^{n+1} \rrbracket  &\text{ \hspace{0.1cm}  on }\partial \Omega_e^{\Gamma} \text{ and}\\
                  -\left(\sumop_{i=0}^{J-1}{\beta_i\left(\nabla \cdotp \bm{\mathcal{F}}^c\left(\bm{u}_h^{n-i}\right)+\nu\nabla \times \bm{\omega}_h^{n-i}\right)-\bm{f}^{n+1}}\right) &\text{ \hspace{0.1cm}  on } \partial \Omega_e^{D}.
                \end{array}
              \right.
              \label{eq:ipflux}
\end{equation}
The same interior penalty parameter definition as for the viscous term is employed\cite{Hillewaert13} and the vorticity vector $\bm{\omega}_h \in \mathcal{V}^{{\bm{u}}}_h$ is precomputed via $L^2$-projection\cite{Krank16b}.
\end{itemize}

Finally, these weak operators allow a compact definition of the dual-splitting scheme in element-wise Galerkin form. The explicit convective step reads
\begin{equation}
\frac{\gamma_0 \mathscr{M}_e({\bm{v}_h},\hat{\bm{u}}_h)-\sumop_{i=0}^{J-1}\left(\alpha_i \mathscr{M}_e({\bm{v}_h},{\bm{u}}_h^{n-i})\right)}{\Delta t}
= -\sumop_{i=0}^{J-1}\beta_i \mathscr{F}^c_e\left({\bm{v}_h},\bm{u}_h^{n-i}\right) + \mathscr{M}_e\left({\bm{v}_h},\bm{f}_h^{n+1}\right).
\label{eq:convstepspat}
\end{equation}
The pressure Poisson equation is 
\begin{equation}
-\mathscr{L}_e\left({q}_h, p_h^{n+1}\right)
= -\frac{\gamma_0}{\Delta t}\mathscr{C}_e\left(q_h,\hat{\bm{u}}_h\right)
\label{eq:poissonspat}
\end{equation}
and as boundary condition for $\hat{\bm{u}}_h$ on $\partial \Omega^D$ we employ the interior value $\{\!\{ \hat{\bm{u}}_h \}\!\}=\hat{\bm{u}}_h^-$.
The element-wise projection step is obtained by adding the div-div penalty term~\eqref{eq:div-div} to the left-hand side, resulting in
\begin{equation}
\mathscr{M}_e\left({\bm{v}_h},\hat{\hat{\bm{u}}}_h\right) + \mathscr{D}_e\left({\bm{v}}_h,\hat{\hat{\bm{u}}}_h\right) \\
= \mathscr{M}_e\left({\bm{v}_h},\hat{\bm{u}}_h\right) - \mathscr{P}_e\left(\bm{v}_h, p_h^{n+1}\right)
\label{eq:projectionspat}
\end{equation}
where boundary conditions on the pressure variable are again applied using the internal value $\{\!\{p_h^{n+1}\}\!\}=p_h^{n+1,-}$ on $\partial \Omega^D$.
The viscous step may be written as
\begin{equation}
\frac{\gamma_0}{\Delta t}\left( \mathscr{M}_e\left({\bm{v}_h},{\bm{u}}_h^{n+1}\right)- \mathscr{M}_e\left({\bm{v}_h},\hat{\hat{\bm{u}}}_h\right)\right) = \mathscr{F}^{\nu}_e\left(\bar{\bm{v}}_h,\widetilde{\bm{v}}_h,\bar{\bm{u}}_h^{n+1},\widetilde{\bm{u}}_h^{n+1}\right).
\label{eq:helmholtzspat}
\end{equation}
Boundary conditions for all remaining quantities on no-slip walls are specified using the so-called mirror principle\cite{Hesthaven07}, defining the exterior velocity contribution with $\bm{u}_h^+=-\bm{u}_h^-+2\bm{g}_{\bm{u}}=-\bm{u}_h^-$ on $\partial \Omega^D$.

\subsection{Coercivity analysis}
\label{sec:coercivity}
We sketch a coercivity analysis of the viscous term in order to investigate the stability of our numerical scheme. The Lax--Milgram theorem ensures solvability of the variational Helmholtz problem~\eqref{eq:helmholtzspat} if there is a constant $C \geq 0$ such that
\begin{equation}
-\mathscr{F}^{\nu}_e\left(\bar{\bm{v}},\widetilde{\bm{v}},\bar{\bm{v}},\widetilde{\bm{v}}\right) \geq C \left(\Vert \bar{\bm{v}}\Vert^2+\Vert\widetilde{\bm{v}}\Vert^2\right), \text{ for all } \bar{\bm{v}} \in \mathcal{V}^{\bar{\bm{u}}}, \widetilde{\bm{v}} \in \mathcal{V}^{\widetilde{\bm{u}}}
\end{equation}
holds. Herein, we disregard the mass terms, as they always yield a positive contribution on the left-hand side of the inequality and the Helmholtz equation degenerates to an $L^2$ projection if $-\mathscr{F}^{\nu}_e \to 0$.

Due to their skew-symmetric construction, the face terms on $\partial \widetilde{\Omega}_e$ in Equation~\eqref{eq:viscous_multiscale} cancel each other when the solution and weighting functions are equal. Solely the interior penalty face terms remain, which have an entirely positive contribution. Therefore, the following inequality holds for enriched cells,
\begin{equation}
-\mathscr{F}^{\nu}_e\left(\bar{\bm{v}},\widetilde{\bm{v}},\bar{\bm{v}},\widetilde{\bm{v}}\right) \geq \left( \bm{\epsilon}\left(\bar{\bm{v}}\right), 2 \nu \bm{\epsilon}\left(\bar{\bm{v}}\right)\right)_{\widetilde{\Omega}_e} + \left( \bm{\epsilon}\left(\bar{\bm{v}}\right), 2 {(\nu+\nu_t)} \bm{\epsilon}\left(\widetilde{\bm{v}}\right) \right)_{\widetilde{\Omega}_e} + \left( \bm{\epsilon}\left(\widetilde{\bm{v}}\right), 2 (\nu+\nu_t) \bm{\epsilon}\left(\widetilde{\bm{v}}\right) \right)_{\widetilde{\Omega}_e}
\label{eq:coercivity_ieq}
\end{equation}
leaving only the volume terms for further consideration. The first and the last volume term are each symmetric and thus positive while the nonsymmetric second term is problematic as it can yield a negative contribution. In the following, we derive an estimate for this term in order to guarantee that the sum of all terms in the bilinear form is bounded from below by $C \left(\Vert \bar{\bm{v}}\Vert^2+\Vert\widetilde{\bm{v}}\Vert^2\right)$. 

Young's inequality
\begin{equation}
2 a b \geq -\frac{a^2}{\varepsilon} - \varepsilon b^2 
\end{equation}
for any $\varepsilon > 0$ applied to the second volume term in~\eqref{eq:coercivity_ieq} results in:
\begin{equation}
\left( \bm{\epsilon}(\bar{\bm{v}}), 2 \widehat{(\nu+\nu_t)} \bm{\epsilon}(\widetilde{\bm{v}}) \right)_{\Omega_e} \geq
\left( \bm{\epsilon}(\bar{\bm{v}}), - \frac{\widehat{(\nu+\nu_t)}}{\varepsilon} \bm{\epsilon}(\bar{\bm{v}}) \right)_{\Omega_e} +
\left( \bm{\epsilon}\left(\widetilde{\bm{v}}\right), - \varepsilon\widehat{(\nu+\nu_t)} \bm{\epsilon}\left(\widetilde{\bm{v}}\right) \right)_{\Omega_e}.
\label{eq:youngs_volume}
\end{equation}
Herein, we have introduced the modified viscosity term $\widehat{(\nu+\nu_t)}$.
Inserting this relation in inequality~\eqref{eq:coercivity_ieq} yields an estimate for $-\mathscr{F}^{\nu}_e$ based on the two symmetric terms,
\begin{equation}
-\mathscr{F}^{\nu}_e\left(\bar{\bm{v}},\widetilde{\bm{v}},\bar{\bm{v}},\widetilde{\bm{v}}\right) \geq \left( \bm{\epsilon}(\bar{\bm{v}}), (2 \nu - \frac{\widehat{(\nu+\nu_t)}}{\varepsilon}) \bm{\epsilon}(\bar{\bm{v}})\right)_{\Omega_e}
 + \left( \bm{\epsilon}(\widetilde{\bm{v}}), \left(2 (\nu+\nu_t)- \varepsilon\widehat{(\nu+\nu_t)} \right) \bm{\epsilon}(\widetilde{\bm{v}}) \right)_{\Omega_e},
\end{equation}
allowing the conclusion that the Lax--Milgram theorem is fulfilled under the following conditions:
\begin{equation}
2 \nu - \frac{\widehat{(\nu+\nu_t)}}{\varepsilon} \geq 0 ,  \hspace{2cm}
2 (\nu+\nu_t)- \varepsilon\widehat{(\nu+\nu_t)} \geq 0.
\end{equation}
Upon rearrangement, these inequalities directly result in the condition for coercivity of the viscous multiscale term:
\begin{equation}
\widehat{(\nu+\nu_t)} \leq 2\sqrt{\nu(\nu+\nu_t)}.
\end{equation}
This means that the amount of viscous dissipation introduced in the LES scale by the RANS solution has to be limited if ${\nu+\nu_t} > 4 \nu$ for reasons of stability. We apply the modified eddy viscosity
\begin{equation}
\widehat{\nu+\nu_t} = \mathrm{min}(\nu+\nu_t,2\sqrt{\nu(\nu+\nu_t)})
\label{eq:minvt}
\end{equation}
in the volume term of the row ($\bar{\bm{v}}, \widetilde{\bm{u}}$) in~\eqref{eq:viscous_multiscale}, which comes along with a minor limitation of the application range of the wall model. This aspect will be investigated in detail in Section~\ref{sec:channel}. We anticipate at this point that the width of the first off-wall cell should not exceed a $y^+$-range of approximately 120 wall units in the statistical data. The relation~\eqref{eq:minvt} guarantees the stability of the scheme in outliers due to turbulent fluctuations. We emphasize that this behavior is related to the particular discrete method used for the viscous multiscale term in the present work and we encourage mathematicians to develop alternative formulations in order to weaken or remove this limitation.

For completeness, we remark on the relations for the case with the viscous LES term according to Equation~\eqref{eq:generic_vms_vtilde}, ie, without the modification introduced in Equation~\eqref{eq:generic_vms_vtilde_mod} and therefore including all terms in the rows ($\widetilde{\bm{v}}, \bar{\bm{u}}$) and ($\bar{\bm{v}},\widetilde{\bm{u}}$) in $\widetilde{\Omega}_e$ as well as on $\partial \widetilde{\Omega}_e$ (in Equation~\eqref{eq:viscous_multiscale}). The nonsymmetric term on the right hand side of inequality~\eqref{eq:coercivity_ieq} would get a material factor of $2 {(2\nu+\nu_t)}$ instead of $2 {(\nu+\nu_t)}$, resulting in the condition $\widehat{(\nu+\nu_t)} \leq 2\sqrt{\nu(\nu+\nu_t)}-\nu$ for the material parameter of the volume term in the row ($\bar{\bm{v}},\widetilde{\bm{u}}$). This condition limits the growth of the eddy viscosity variable if ${\nu+\nu_t} > \nu$, which would represent a too strong restriction in practical applications.

\begin{table}[t]
\caption{Overview of simulation cases for the turbulent channel flow: Investigation of several Reynolds numbers (Figure~\ref{fig:ch_reydep_u}), comparison of the new approach to delayed-detached-eddy simulation (DDES) including function enrichment\cite{Krank17c} (Figure~\ref{fig:ch_comparison_ddes}), grid stretching factors and mesh aspect ratios (Figure~\ref{fig:ch_mesh_u}), transition from wall-resolved to wall-modeled LES (Figure~\ref{fig:ch_wallres_u}), weighting of the enrichment function with a constant or linear basis (Figure~\ref{fig:ch_l0l1_u}), and the performance comparing the wall model with wall-resolved LES (Figure~\ref{fig:ch_performance_u}). The number of cells per direction $i$ is denoted $N_{ie}$ and the normalized dimensions of the first off-wall cell is defined as $\Delta  y_{1e}^+=\Delta y_{1e} u_{\tau}/\nu$ in wall-normal direction and $\Delta  \{x,z\}_{e}^+=\Delta \{x,z\}_{e} u_{\tau}/\nu$ in wall-parallel direction. The number of LES grid points per direction $i$ is $N_i=(k+1)N_{ie}$ with $k=4$ in all cases.}
\label{tab:ch_flows}
\begin{tabular*}{\textwidth}{c| @{\extracolsep{\fill}} l l l l l l l l}
\hline
\multicolumn{1}{l}{Fig.}  & Case   & $N_{1e} {\times} N_{2e} {\times} N_{3e}   $ &$Re_{\tau}$ & $l$  & $\gamma$ & $\Delta  y_{1e}^+   $ & $\max(\Delta x_e^+,\Delta z_e^+)$ & $\frac{\Delta y_{1e}^+(k+1)}{\max(\Delta x_e^+,\Delta z_e^+)}$
\\ \hline \noalign{\smallskip}
\multirow{4}{*}{\ref{fig:ch_reydep_u}} 
&\ $ch395\_N12{\times}8{\times}12\_k4l0\_\gamma0.8$  & $12 {\times} 8 {\times} 12$   & $395$  & $0$ & $0.8$ & $76$& $207$ & $1.84$\\
& \ $ch950\_N12{\times}8{\times}12\_k4l0\_\gamma1.6$  & $12 {\times} 8 {\times} 12$   & $950$  & $0$ & $1.6$ & $91$& $497$ &$0.92$\\
& \ $ch2000\_N24{\times}8{\times}24\_k4l0\_\gamma2.2$  & $24 {\times} 8 {\times} 24$   & $2{,}000$  & $0$ & $2.2$ & $96$& $524$ & $0.92$\\
& \ $ch5200\_N48{\times}16{\times}48\_k4l0\_\gamma2.05  $  & $48 {\times} 16 {\times} 48$   & $5{,}200$  & $0$ & $2.05$& $114$ &$681$ &$0.84$\\ 
\noalign{\smallskip}
\multirow{2}{*}{\ref{fig:ch_comparison_ddes}} 
& \ $ch950\_N16^3\_k4l0\_\gamma0.001$  & $16 {\times} 16 {\times} 16$   & $950$  & $0$ & $0.001   $ & $119$& $373$ & $1.60$\\
& \ $ch950\_N16^3\_k4l0\_\gamma0.001\_DDES$\cite{Krank17c}  & $16 {\times} 16 {\times} 16$   & $950$  & $0$ & $0.001   $ & $119$& $373$ & -\\
\noalign{\smallskip}
\multirow{6}{*}{\ref{fig:ch_mesh_u}} 
& \ $ch950\_N16^3\_k4l0\_\gamma0.001$  & $16 {\times} 16 {\times} 16$   & $950$  & $0$ & $0.001   $ & $119$& $373$ & $1.60$\\
& \ $ch950\_N16^3\_k4l0\_\gamma0.8$   & $16 {\times} 16 {\times} 16$   & $950$  & $0$ & $0.8$ & $85$& $373$ & $1.14$\\
& \ $ch950\_N16^3\_k4l0\_\gamma1.2$   & $16 {\times} 16 {\times} 16$   & $950$  & $0$ & $1.2$ & $59$& $373$ & $0.79$\\
& \ $ch950\_N16^3\_k4l0\_\gamma1.5$   & $16 {\times} 16 {\times} 16$   & $950$  & $0$ & $1.5$ &  $42$& $373$ & $0.56$\\
& \ $ch950\_N16^3\_k4l0\_\gamma1.0$   & $16 {\times} 16 {\times} 16$   & $950$  & $0$ & $1.0$ &  $72$& $373$ & $0.97$\\
&\ $ch950\_N16{\times}16{\times}32\_k4l0\_\gamma1.0$  & $16 {\times} 16 {\times} 32$   & $950$  & $0$ & $1.0$ & $72$& $373$ &$0.97$\\
\noalign{\smallskip}
\multirow{5}{*}{\ref{fig:ch_wallres_u}} 
& \ $ch950\_N24^3\_k4\_\gamma2.25$  & $24 {\times} 24 {\times} 24$   & $950$  & - & $2.25   $ & $10$& $249$ & $0.20$\\
& \ $ch950\_N24^3\_k4l0\_\gamma1.75$  & $24 {\times} 24 {\times} 24$   & $950$  & $0$ & $1.75   $ & $19$& $249$ & $0.38$\\
& \ $ch950\_N24^3\_k4l0\_\gamma1.5$  & $24 {\times} 24 {\times} 24$   & $950$  & $0$ & $1.5   $ & $27$& $249$ & $0.54$\\
& \ $ch950\_N24^3\_k4l0\_\gamma1.25$  & $24 {\times} 24 {\times} 24$   & $950$  & $0$ & $1.25   $ & $36$& $249$ & $0.72$\\
& \ $ch950\_N24^3\_k4l0\_\gamma1.0$  & $24 {\times} 24 {\times} 24$   & $950$  & $0$ & $1.0   $ & $46$& $249$ & $0.92$\\
\noalign{\smallskip}
\multirow{2}{*}{\ref{fig:ch_l0l1_u}} 
& \ $ch950\_N16^3\_k4l0\_\gamma1.2$   & $16 {\times} 16 {\times} 16$   & $950$  & $0$ & $1.2$ & $59$& $373$ &$0.79$\\
& \ $ch950\_N16^3\_k4l1\_\gamma1.2$   & $16 {\times} 16 {\times} 16$   & $950$  & $1$ & $1.2$ & $59$& $373$ &$0.79$\\
\noalign{\smallskip}
\multirow{3}{*}{\ref{fig:ch_performance_u}} 
& \ $ch950\_N12{\times}8{\times}12\_k4l0\_\gamma1.6$  & $12 {\times} 8 {\times} 12$   & $950$  & $0$ & $1.6$ & $91$& $497$ &$0.92$\\
& \ $ch950\_N24^3\_k4\_\gamma2.25$  & $24 {\times} 24 {\times} 24$   & $950$  & - & $2.25   $ & $10$& $249$ & $0.20$\\
& \ $ch950\_N32^3\_k4\_\gamma2.0$  & $32 {\times} 32 {\times} 32$   & $950$  & - & $2.0$ & $10$& $187$ & $0.27$\\ \noalign{\smallskip} 
\hline
\end{tabular*}
\end{table}

\begin{figure}[htb]
\centering
\begin{minipage}[b]{1.0\linewidth}
\centering
\includegraphics[trim= 10mm 40mm 10mm 40mm,clip,width=0.497\textwidth]{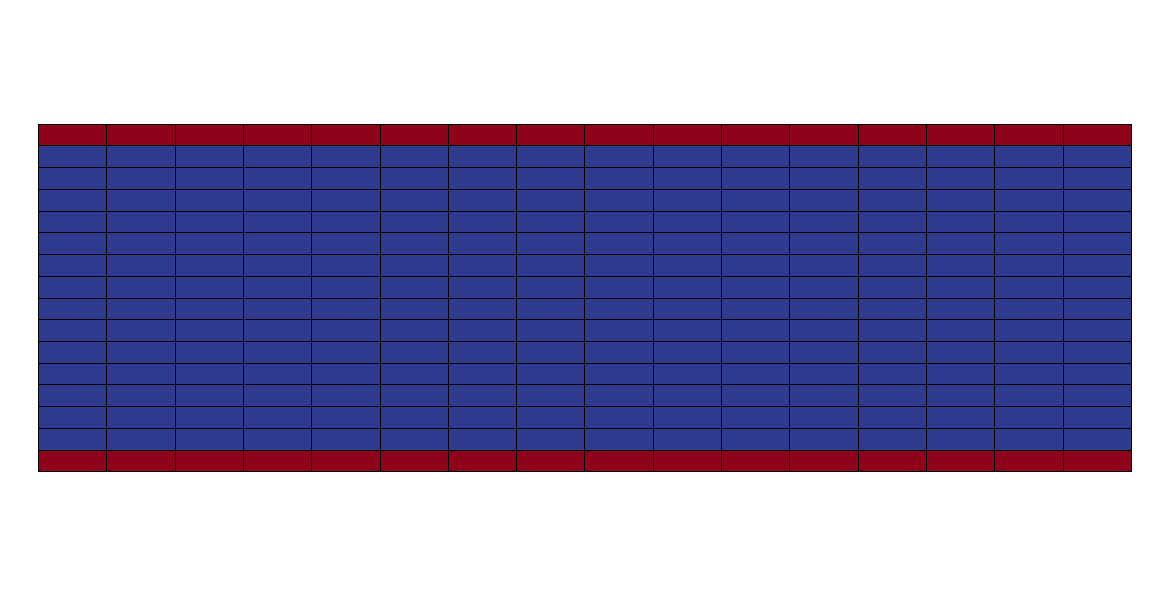}
\caption{Grid of the case $ch950\_N16^3\_k4l0\_\gamma0.001$. The solution is represented by a polynomial of degree four in each cell plus the enrichment shape functions in the wall-layer (red cells).}
\label{fig:ch_mesh_picture}
\end{minipage}
\end{figure}
\begin{figure}[htb]
\begin{minipage}[b]{0.497\linewidth}
\centering
\includegraphics[trim= 10mm 40mm 10mm 40mm,clip,width=1.\textwidth]{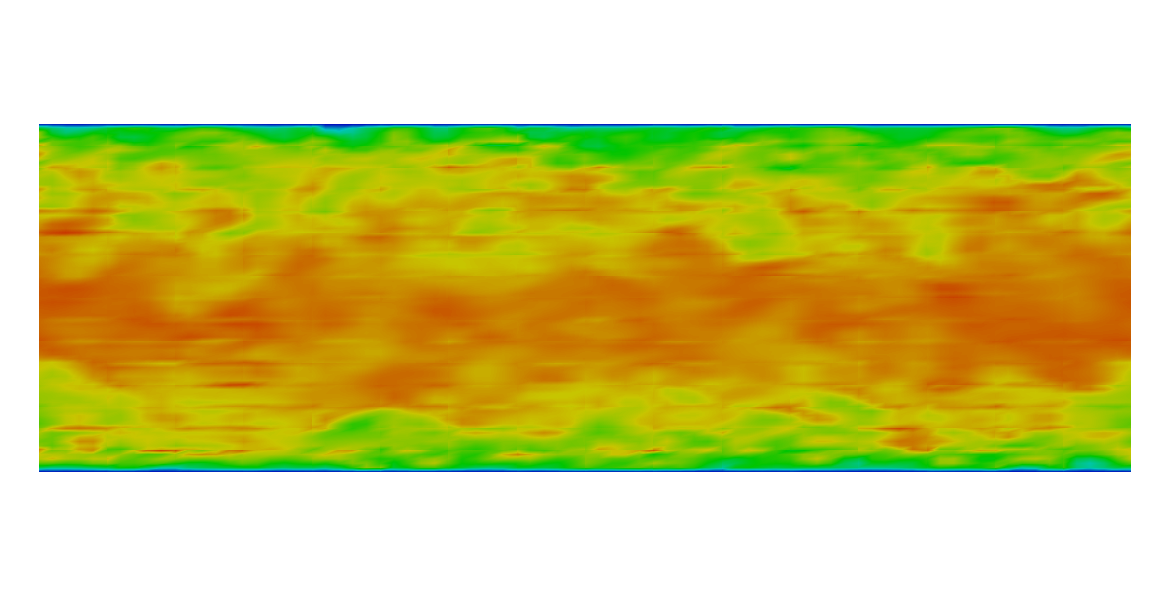}
\end{minipage}
\begin{minipage}[b]{0.497\linewidth}
\centering
\includegraphics[trim= 10mm 40mm 10mm 40mm,clip,width=1.\textwidth]{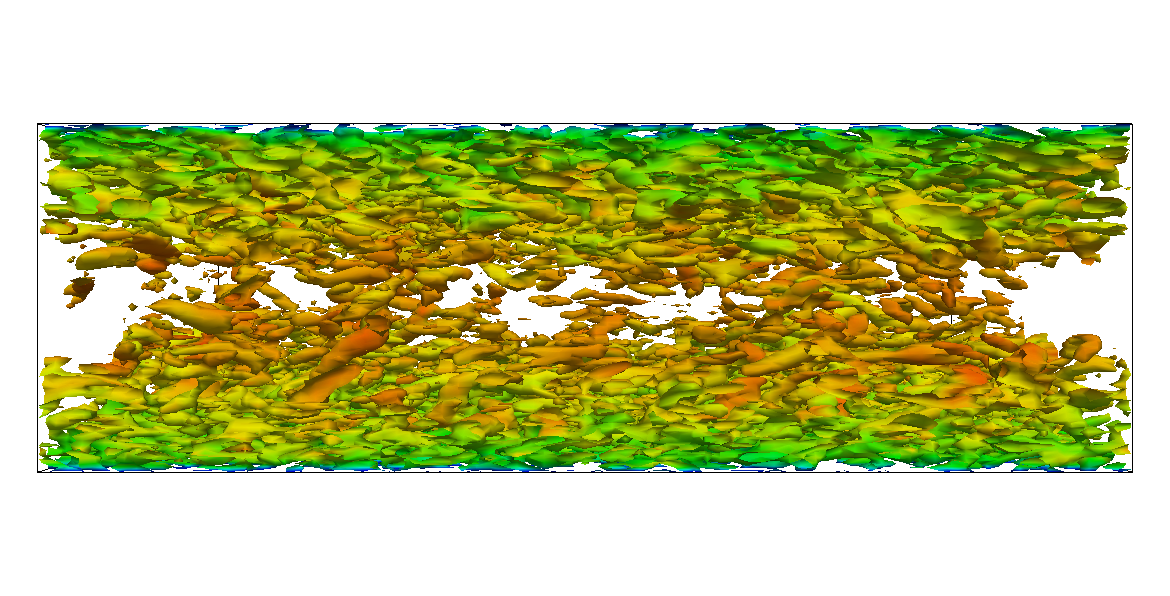}
\end{minipage}
\begin{minipage}[b]{0.497\linewidth}
\centering
\includegraphics[trim= 10mm 40mm 10mm 40mm,clip,width=1.\textwidth]{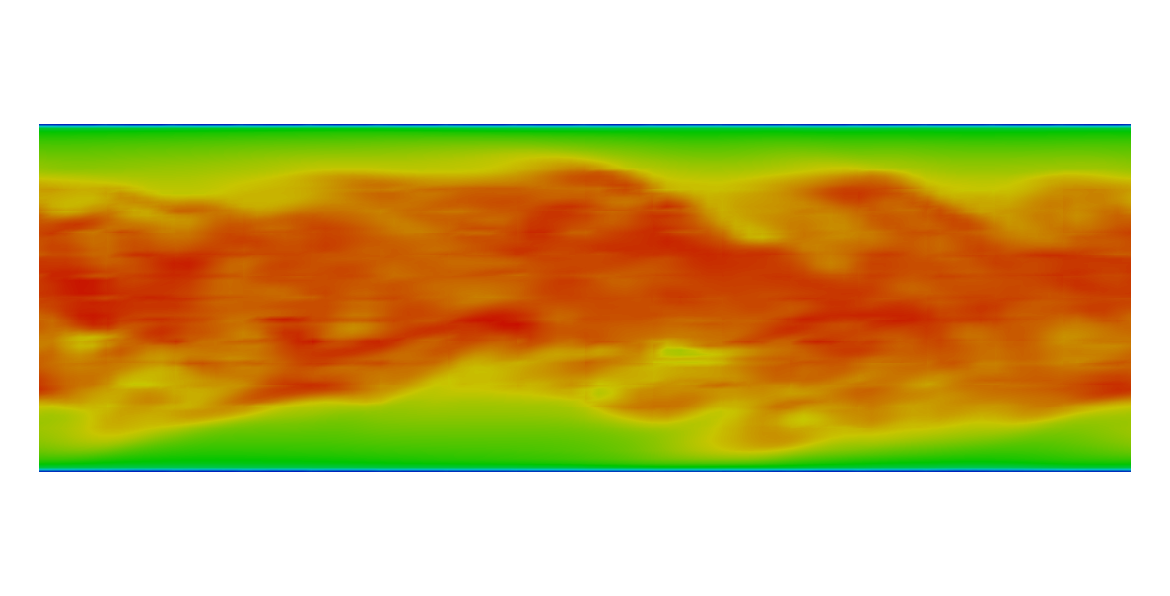}
\end{minipage}
\begin{minipage}[b]{0.497\linewidth}
\centering
\includegraphics[trim= 10mm 40mm 10mm 40mm,clip,width=1.\textwidth]{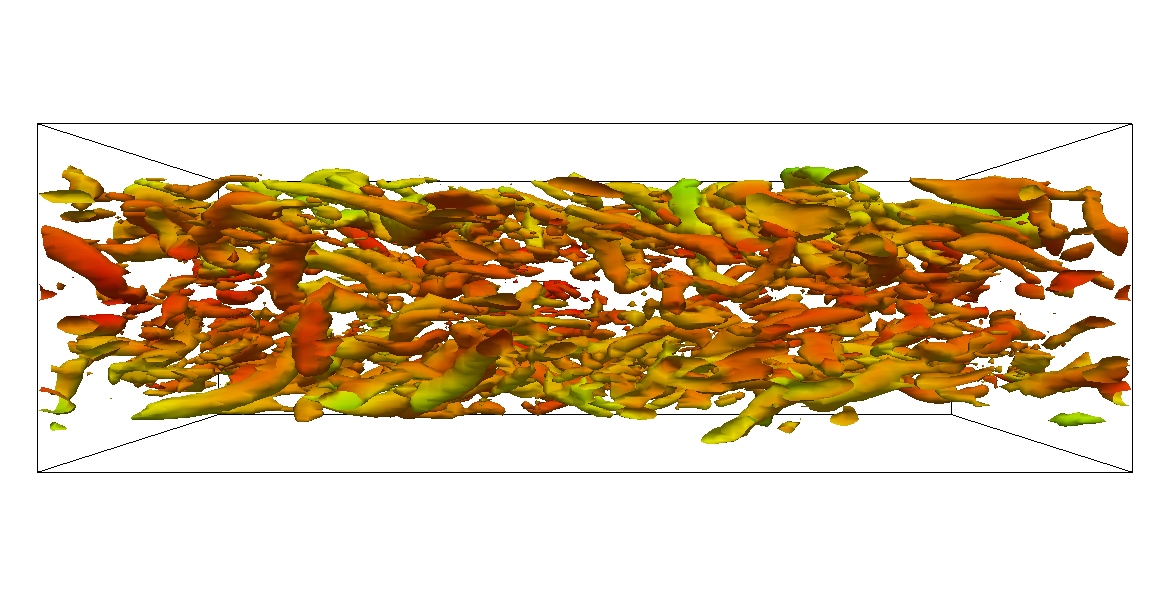}
\end{minipage}
\centering
\caption{Qualitative comparison of the present multiscale wall model including enrichment (top) with DDES using enrichment\cite{Krank17c} (bottom) on the same mesh. Velocity magnitude (left) and visualization of turbulent eddies via the Q-criterion colored by velocity magnitude (right) of the cases $ch950\_N16^3\_k4l0\_\gamma0.001$ and $ch950\_N16^3\_k4l0\_\gamma0.001\_DDES$. All cases use the same color scale and the Q-criterion shows the same iso-value. Red indicates high and blue low values.}
\label{fig:ch_u}
\end{figure}

\subsection{Implementation}
\label{sec:implementation}
The Galerkin formulations of the dual splitting scheme, Equations~\eqref{eq:convstepspat}--\eqref{eq:helmholtzspat} are integrated by numerical quadrature, yielding a matrix formulation for each sub-step. The matrix forms are similar to the ones described in earlier studies\cite{Krank16c,Krank16b}. The numerical evaluation and integration is performed using the high-performance sum factorization kernels by Kronbichler and Kormann\cite{Kronbichler12} available in the deal.II finite element library\cite{Arndt17}. Herein, the nonenriched cells are evaluated with Gaussian quadrature rules of appropriate order such that all integrals are evaluated exactly on affine cells ($k+1$ quadrature points for linear terms and $\left \lfloor \frac{3k}{2}\right \rfloor +1$ points for nonlinear terms). Enriched cells pose higher requirements on the accuracy of the quadrature rules due to the presence of the nonpolynomial shape functions. Therefore, $8^3$ to $12^3$ quadrature points have been used for the examples in the present paper, where more points have been employed if the enriched cells span a wider $y^+$-range. Face integrals are evaluated with the same number of quadrature points per direction if at least one of the two adjacent cells is enriched. In an earlier work\cite{Krank16c}, we have shown how an object-oriented programming language with template capabilities (eg C++) may be used to implement the present boundary layer enrichment with minimal computational overhead in nonenriched cells and zero extra cost in nonenriched simulations in one single implementation of the Navier--Stokes solver.

All sub-steps are implemented in a matrix-free manner, including the iterative solvers for the global Poisson and Helmholtz problems as well as the local projection solver according to Reference\cite{Krank16b}; further details on the implementation of the multigrid solver of the Poisson problem are given in Reference\cite{Kronbichler16}. In particular, we have developed an algorithmic toolbox for implementing enrichments in such a matrix-free context at very little additional cost. For example, these algorithms allow a much faster application of the inverse mass matrix than the approach via LU factorization chosen in Krank et al\cite{Krank16c}. Furthermore, most of the additional cost of the higher quadrature rules used in enriched cells can be circumvented. Since a detailed description and performance analysis of these algorithms is beyond the scope of this article, it is the subject of a separate publication\cite{Kronbichler17}; see also the monograph~\cite{Krank18}.

\begin{figure}[t]
\centering
\begin{minipage}[b]{0.497\linewidth}
\centering
\includegraphics[trim= 5mm 3mm 8mm 8mm,clip,width=1.\textwidth]{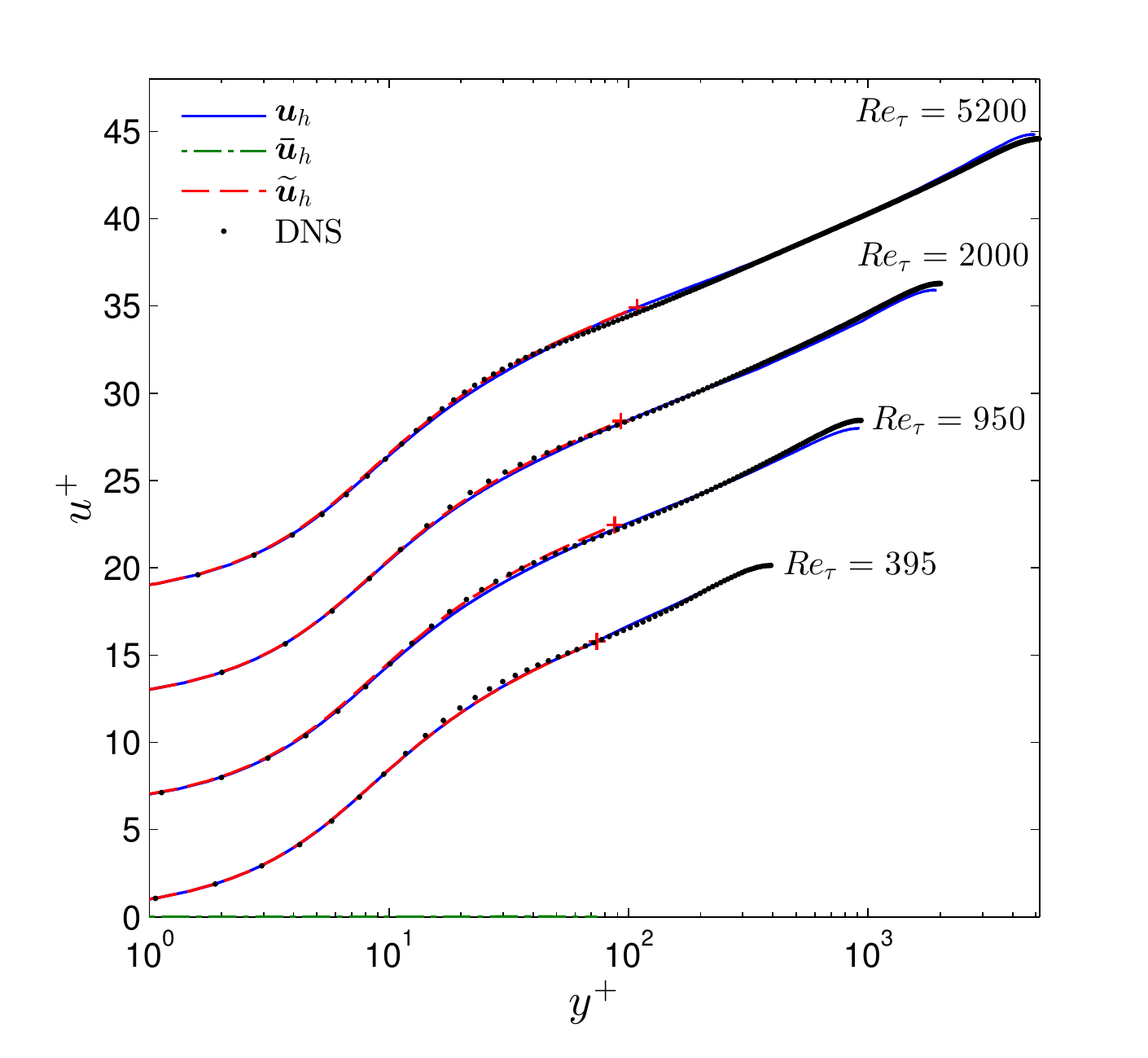}
\end{minipage}
\begin{minipage}[b]{0.497\linewidth}
\centering
\includegraphics[trim= 5mm 3mm 8mm 8mm,clip,width=1.\textwidth]{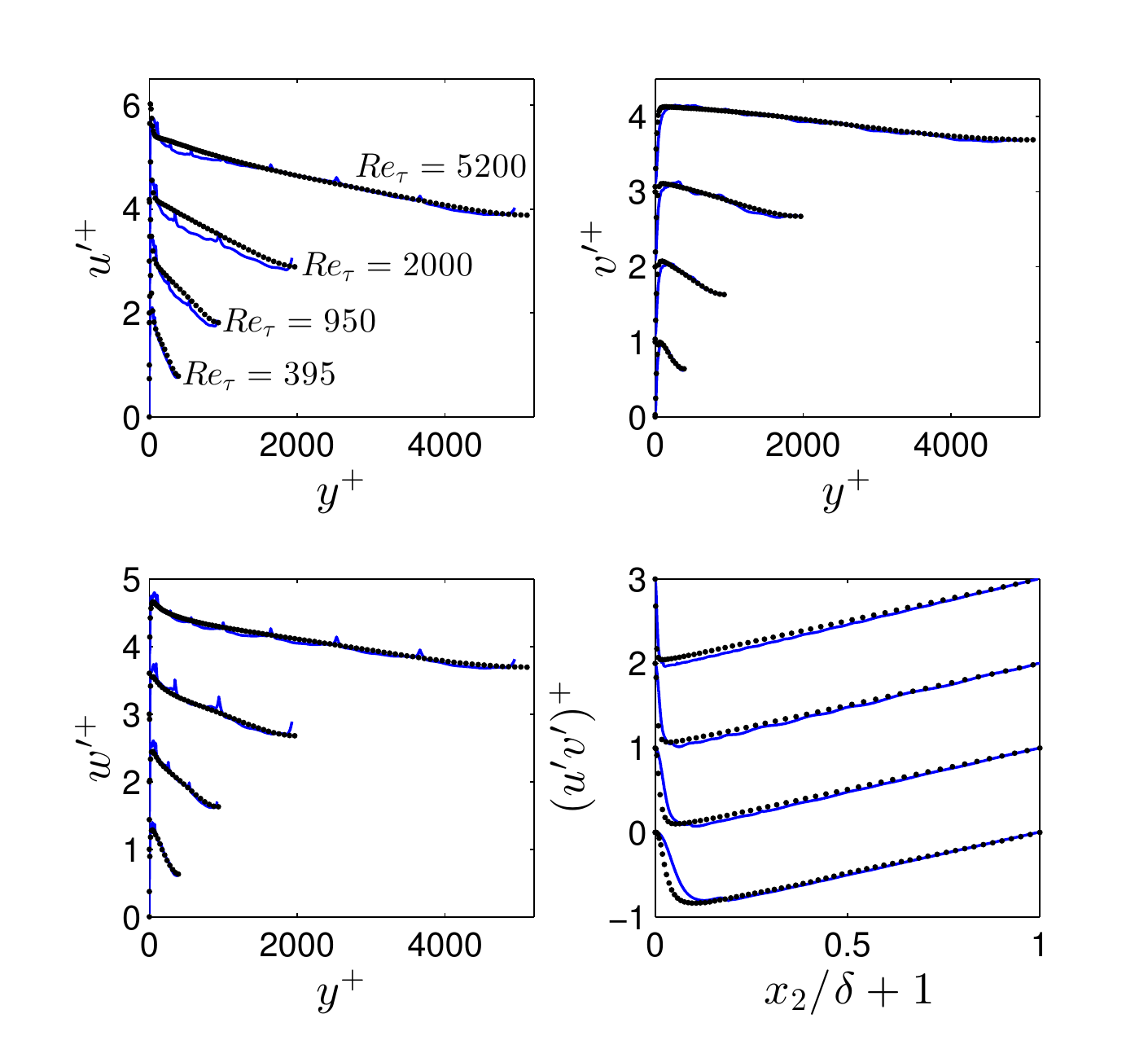}
\end{minipage}
\caption{Wall-modeled LES of turbulent channel flow at several Reynolds numbers: Mean velocity (left) and RMS-velocities as well as Reynolds shear stress (right). All quantities are normalized according to $u^+=\langle u_1\rangle/u_{\tau}$, $u^{\prime+}= \sqrt{\langle u_1^{\prime2}\rangle}/u_{\tau}$, $v^{\prime+}= \sqrt{\langle u_2^{2}\rangle}/u_{\tau}$, $w^{\prime+}=\sqrt{\langle u_3^{2}\rangle}/u_{\tau}$, and $(u^{\prime}v^{\prime})^{+}=\langle u_1 u_2\rangle/u_{\tau}^2$. The polynomial part is inside the wall-layer only displayed for $Re_{\tau}=395$.}
\label{fig:ch_reydep_u}
\end{figure}

\begin{figure}[t]
\centering
\begin{minipage}[b]{0.497\linewidth}
\centering
\includegraphics[trim= 5mm 3mm 8.0mm 8mm,clip,width=1.\textwidth]{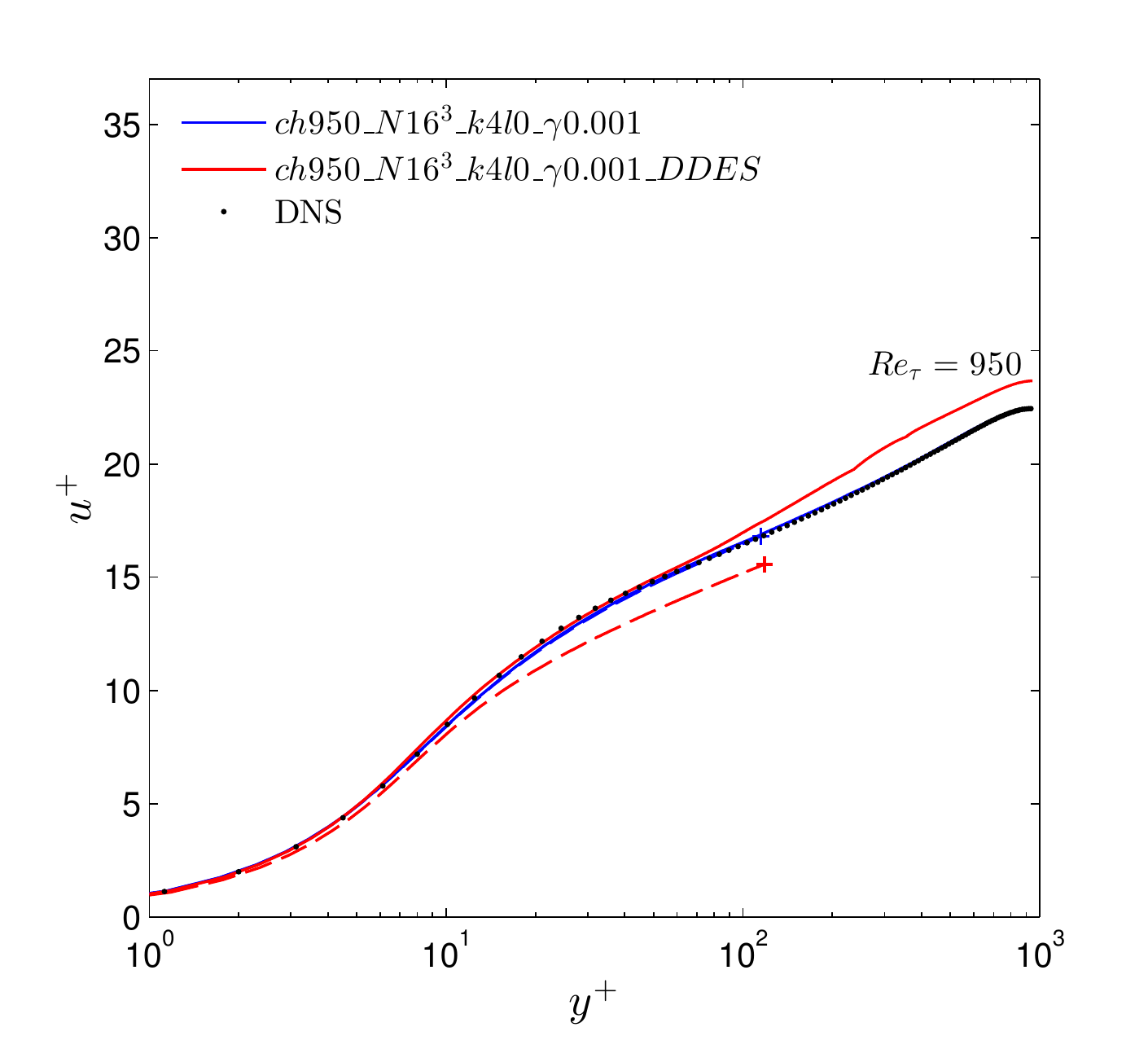}
\end{minipage}
\begin{minipage}[b]{0.497\linewidth}
\centering
\includegraphics[trim= 5.0mm 3mm 8mm 8mm,clip,width=1.\textwidth]{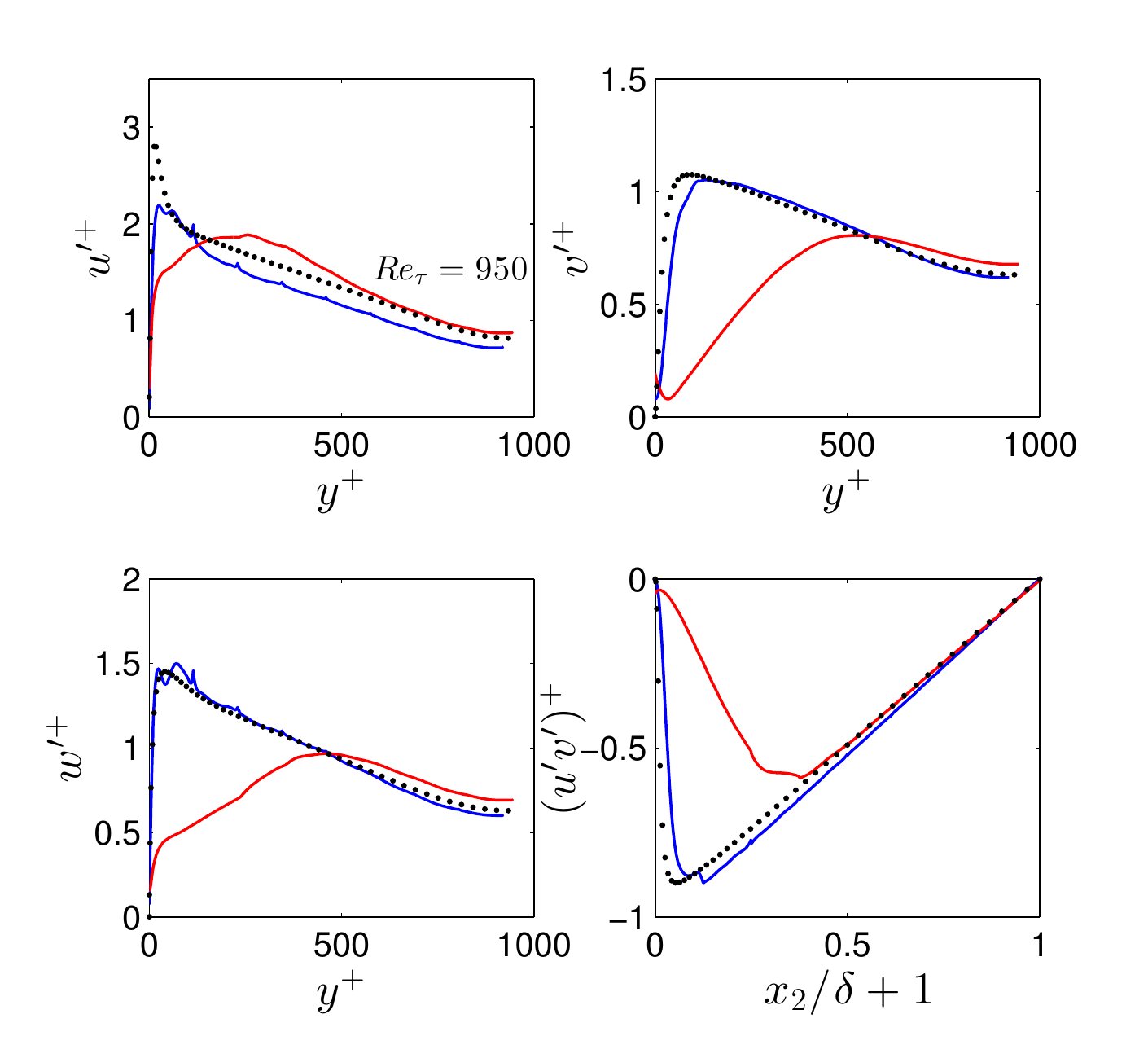}
\end{minipage}
\caption{Quantitative comparison of the present multiscale wall model to DDES using function enrichment\cite{Krank17c} employing the same grid: Mean velocity (left) and RMS-velocities as well as Reynolds shear stress (right). All quantities are normalized according to $u^+=\langle u_1\rangle/u_{\tau}$, $u^{\prime+}= \sqrt{\langle u_1^{\prime2}\rangle}/u_{\tau}$, $v^{\prime+}= \sqrt{\langle u_2^{2}\rangle}/u_{\tau}$, $w^{\prime+}=\sqrt{\langle u_3^{2}\rangle}/u_{\tau}$, and $(u^{\prime}v^{\prime})^{+}=\langle u_1 u_2\rangle/u_{\tau}^2$. The full solution is displayed as solid line and the enrichment component as dashed line.}
\label{fig:ch_comparison_ddes}
\end{figure}

\section{Numerical examples}
\label{sec:examples}

The wall modeling approach is validated with turbulent channel flow (Section~\ref{sec:channel}) as well as flow past periodic hills (Section~\ref{sec:ph}). These two benchmark examples provide insight in the performance of the wall model both in attached and separated boundary layers. We follow our earlier recommendations of using a polynomial degree of $k=4$ for the discretization of the LES scale, since this choice yields a good compromise between accuracy and time-to-solution within the present solver\cite{Krank16c,Krank16b,Krank17c,Krank17b}. In addition, a polynomial of $4^{\mathrm{th}}$ degree inside the wall layer is capable of resolving sufficient turbulence within a single layer of cells. The influence of the degree of the polynomial weighting of the enrichment function, $l$, is investigated and guidelines for the use of the wall model are formulated.

\begin{figure}[t]
\centering
\begin{minipage}[b]{0.497\linewidth}
\centering
\includegraphics[trim= 5mm 3mm 8.0mm 8mm,clip,width=1.\textwidth]{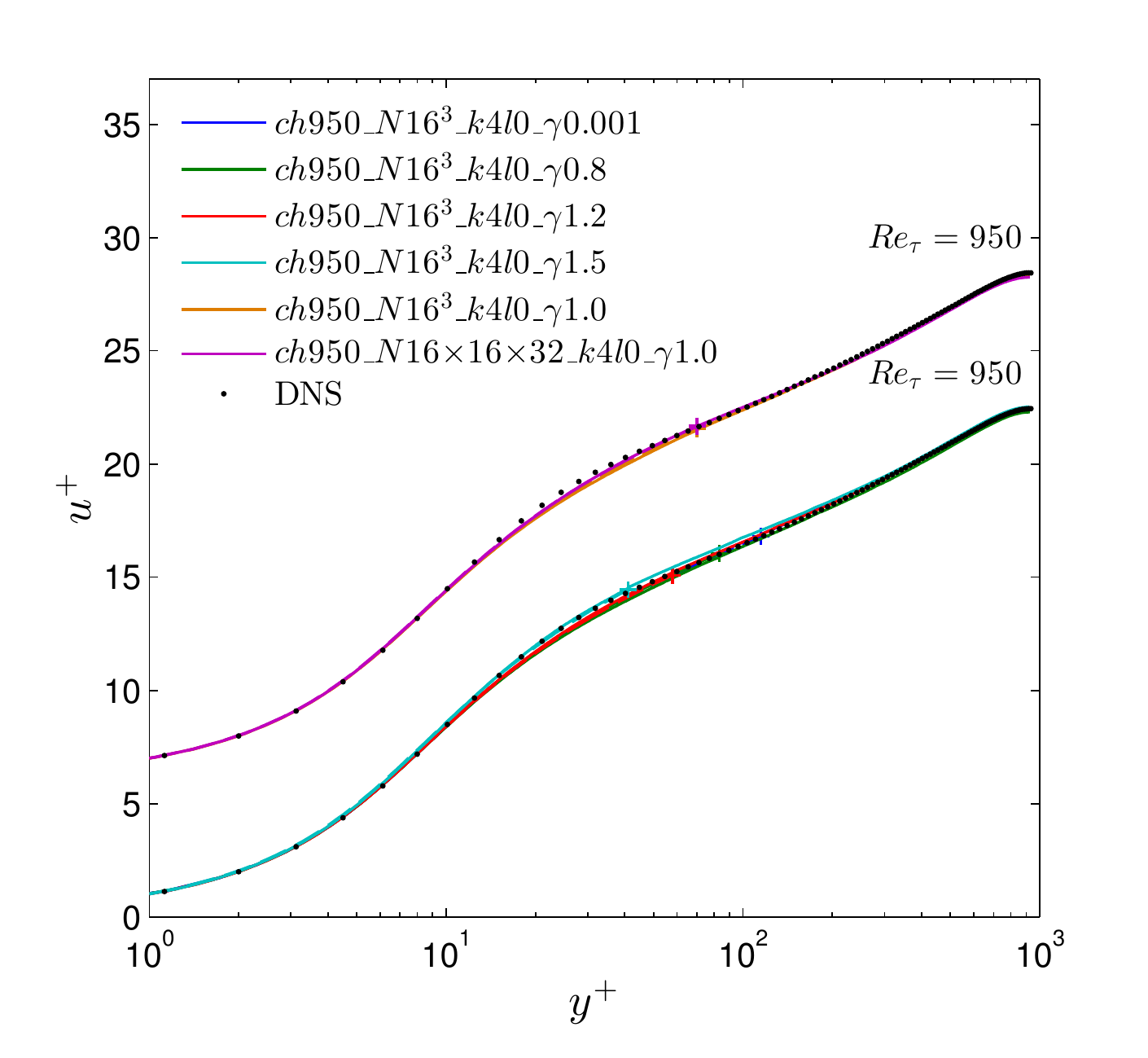}
\end{minipage}
\begin{minipage}[b]{0.497\linewidth}
\centering
\includegraphics[trim= 5.0mm 3mm 8mm 8mm,clip,width=1.\textwidth]{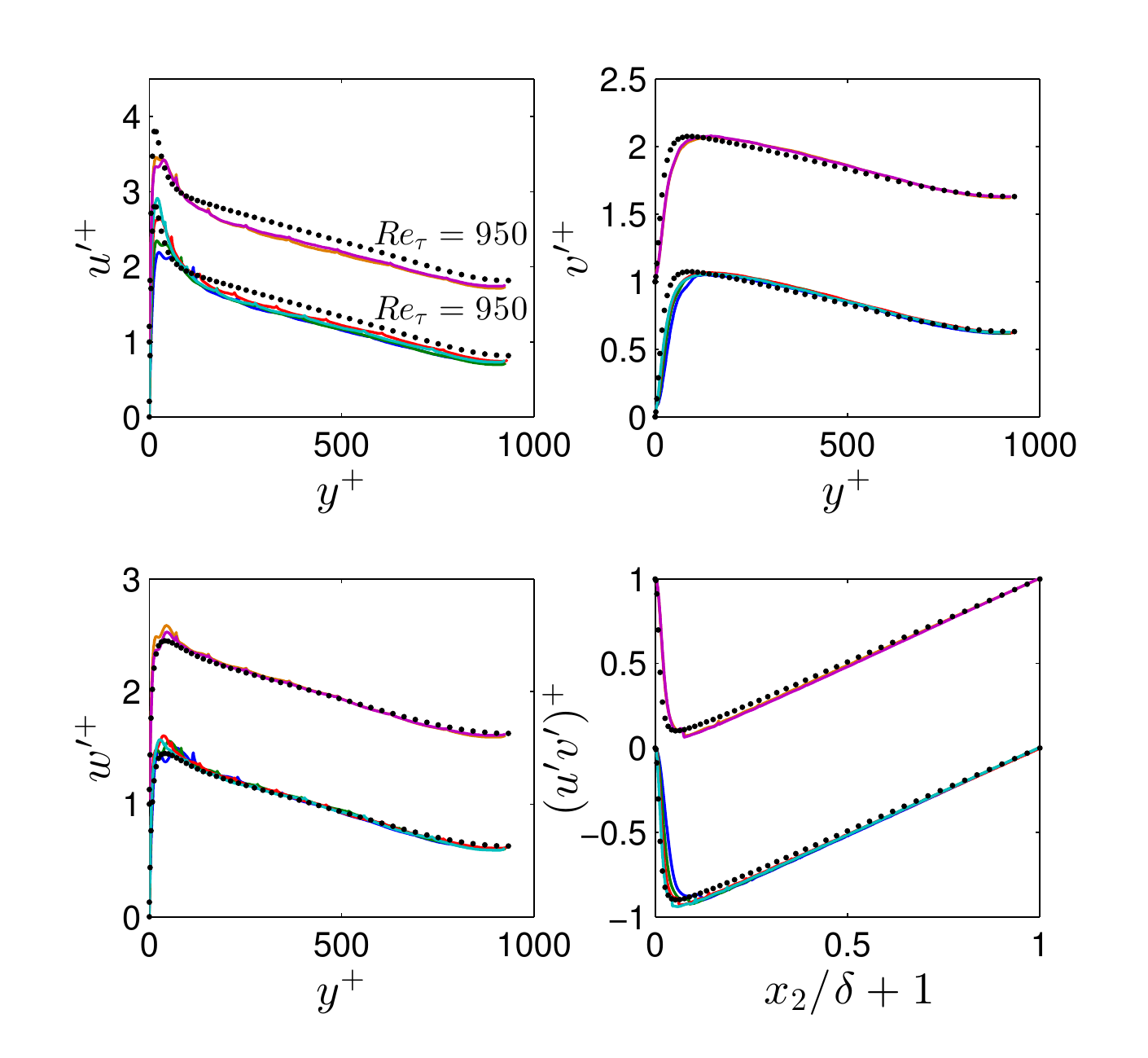}
\end{minipage}
\caption{Grid stretching and element aspect ratios: Mean velocity (left) and RMS-velocities as well as Reynolds shear stress (right). All quantities are normalized according to $u^+=\langle u_1\rangle/u_{\tau}$, $u^{\prime+}= \sqrt{\langle u_1^{\prime2}\rangle}/u_{\tau}$, $v^{\prime+}= \sqrt{\langle u_2^{2}\rangle}/u_{\tau}$, $w^{\prime+}=\sqrt{\langle u_3^{2}\rangle}/u_{\tau}$, and $(u^{\prime}v^{\prime})^{+}=\langle u_1 u_2\rangle/u_{\tau}^2$. The full solution is displayed as solid line and the enrichment component as dashed line.}
\label{fig:ch_mesh_u}
\end{figure}

\subsection{Turbulent channel flow}
\label{sec:channel}
As a first validation example, we consider turbulent flow in a rectangular channel of the dimensions $2\pi\delta{\times}2\delta{\times}\pi \delta$ in streamwise, vertical, and spanwise direction, respectively, with the channel half-width $\delta$. Periodic boundary conditions are applied in the streamwise and spanwise direction and no-slip boundary conditions at the top and bottom wall. The grid is graded towards the walls by the hyperbolic mapping given as $f$: $[0,1] \to [-\delta, \delta]$:
\begin{equation}
x_2 \mapsto f(x_2)=\delta \frac{\tanh(\gamma (2x_2-1))}{\tanh(\gamma)}
\end{equation}
with the mesh stretching parameter $\gamma$. The wall model is active in the first off-wall element layer and a typical mesh is shown in Figure~\ref{fig:ch_mesh_picture}. The initial conditions are solely applied on the polynomial degrees of freedom of the solution and the RANS component develops in the course of the simulation once $y^+>30$ at any quadrature point of a wall-layer cell according to Section~\ref{sec:vandriest}. The flow is driven by a constant pressure gradient derived from the nominal flow quantities and the results are normalized using the numerical wall shear stress in the friction velocity $u_{\tau}=\sqrt{\tau_w/\rho}$. We consider the friction Reynolds numbers $Re_{\tau}=u_{\tau}\delta/\nu$ in accordance to DNS reference data at $Re_{\tau}=395$\cite{Moser99}, $Re_{\tau}=950$\cite{Alamo03}, $Re_{\tau}=2{,}000$\cite{Hoyas06}, and $Re_{\tau}=5{,}200$\cite{Lee15}. One snapshot of the turbulent flow is visualized in Figure~\ref{fig:ch_u} via contours of the velocity magnitude and turbulent vortex structures are made visible using iso-surfaces of the Q-criterion. All simulation parameters and discretization cases are summarized in Table~\ref{tab:ch_flows}. Statistics were averaged in a time interval corresponding to approximately 60--80 flow-through times based on a fixed time interval.

The results of a first application of the wall model including all four Reynolds numbers are depicted in Figure~\ref{fig:ch_reydep_u}. Overall, excellent agreement is observed both for the mean velocity, the RMS velocities and the Reynolds shear stress. Furthermore, it is apparent that the results do not depend on the Reynolds number. Regarding the mean velocity, the enrichment represents the whole mean velocity and the time-averaged LES solution is almost zero in the wall layer. Therefore, the LES result is only visualized for the lowest Reynolds number and not considered in the remainder of this section. The Reynolds shear stress is computed based on the resolved fluctuations, which explains the small gap to the reference data in the near-wall zone. In particular the $u^{\prime+}$ and $w^{\prime+}$ curves exhibit small ticks at the element interfaces, which is a typical result for coarse meshes in DG and has been observed in an earlier study as well\cite{Krank16b}. These ticks vanish with increasing resolution\cite{Krank16b}.

Since the primary incentive for the development of the present multiscale wall model was the deficiencies of DES in the hybrid RANS/LES transition region, we directly compare the two simulation methodologies. To this end, we consider DDES including wall modeling via function enrichment\cite{Krank17c} for the spatial discretization, such that the exact same meshes can be used for both models. Figure~\ref{fig:ch_u} shows a qualitative comparison of two simulations at $Re_{\tau}=950$; they deviate drastically. DDES operates in RANS mode in the inner layer up to $y/\delta\approx0.05$ and the eddy viscosity acts on the polynomial velocity component as well, such that no vortices are resolved in the inner layer. This stands in contrast to the present multiscale wall model, which computes turbulent motions in the inner boundary layer as well.  In DDES, turbulent eddies evolve outside of the inner layer, but the flow generally behaves differently due to the use of the LES eddy viscosity subgrid model given through the Spalart--Allmaras model in LES mode. These two simulations are further compared quantitatively through velocity statistics in Figure~\ref{fig:ch_comparison_ddes}. As it is expected, DDES overpredicts the mean velocity in the outer layer as a result of a log-layer mismatch caused by the RANS--LES transition. Such a log-layer mismatch is not observed with the present model. The turbulent stresses show that the transition in the DDES model extends until approximately $y/\delta=0.4$, whereas the present multiscale wall model makes much better use of the eddy-resolving capability of the mesh and resolves the important turbulent motions beyond $y/\delta=0.05$.

The wall model including its application range and robustness with regard to grid dependence is analyzed in a systematic manner in the following. In Figure~\ref{fig:ch_mesh_u}, five grid stretching factors are investigated using the same number of cells in each spatial direction. In addition, the number of cells is doubled in the $x_3$-direction in one case. The results show that the wall model generally exhibits an excellent robustness regarding the choice of the mesh. The dependence of the cell aspect ratio may be analyzed by considering the measure ${\Delta y_{1e}^+(k+1)}/{\max(\Delta x_e^+,\Delta z_e^+)}$, which quantifies the available number of grid points in the wall-parallel direction to resolve a turbulent motion of the size of $\Delta y_{1e}^+$; this quantity is included in Table~\ref{tab:ch_flows} for all simulation cases. Remember that we require sufficient turbulence to be resolved at a distance from the wall $\Delta y_{1e}^+$ such that the RANS model can be `switched off', so the cell aspect ratio should not exceed a certain limit. It is noted that other wall modeling approaches, such as wall stress models, in principle have the same requirement if not even more stringent. The factor ${\Delta y_{1e}^+(k+1)}/{\max(\Delta x_e^+,\Delta z_e^+)}$ varies in the range $0.56$ to $1.60$. The simulation case with the highest aspect ratio, $ch950\_N16^3\_k4l0\_\gamma1.5$, shows a minor overprediction of the mean velocity where the wall model ends, allowing the conclusion that ${\Delta y_{1e}^+(k+1)}/{\max(\Delta x_e^+,\Delta z_e^+)}$ should not go below the limit $\sim 0.5$. According to Jim{\'e}nez~\cite{Jimenez11}, the streamwise size of the energetic eddies in the log-layer is approximately $5y$ so they are well-resolved with $2.5$ grid points at the height $\Delta y_{1e}^+$, where the RANS layer ends.

The coercivity analysis in Section~\ref{sec:coercivity} requires the clipping of the eddy viscosity beyond $\widehat{\nu+\nu_t} > 4 \nu$ in the nonsymmetric volume term of the viscous operator. A normalization of this relation may be obtained by division with the viscosity $\widehat{\nu}_t^+=\widehat{\nu+\nu_t}/\nu > 4 $. This relation is universal in $y^+$-units. It is thus possible to investigate the clipping of $\widehat{\nu}_t^+$ for one flow configuration and to transfer the conclusions to other simulations. Given that the maximum values of $\widehat{\nu}_t^+$ grow with the width of the wall layer, it is sufficient to find the upper limit of the application range (cf. Equation~\eqref{eq:minvt}). In Figure~\ref{fig:ch_mesh_u}, $\widehat{\nu}_t^+$ is clipped on approximately $11\%$ of the quadrature points of all wall-layer cells for the case $ch950\_N16^3\_k4l0\_\gamma0.001$. Although the results are excellent for this case, we do not recommend the application of the wall model significantly beyond this $y^+$-range in order to guarantee the reliability of the results. In conclusion, the wall model should not extend beyond approximately $120$ wall units in the statistical data. We stress at this point that stability of the numerical method is guaranteed irrespective of the application of the wall model and this $y^+$-limit is only due to the result quality, which may degrade for thicker wall layers.

\begin{figure}[t]
\centering
\begin{minipage}[b]{0.497\linewidth}
\centering
\includegraphics[trim= 5mm 3mm 8.0mm 8mm,clip,width=1.\textwidth]{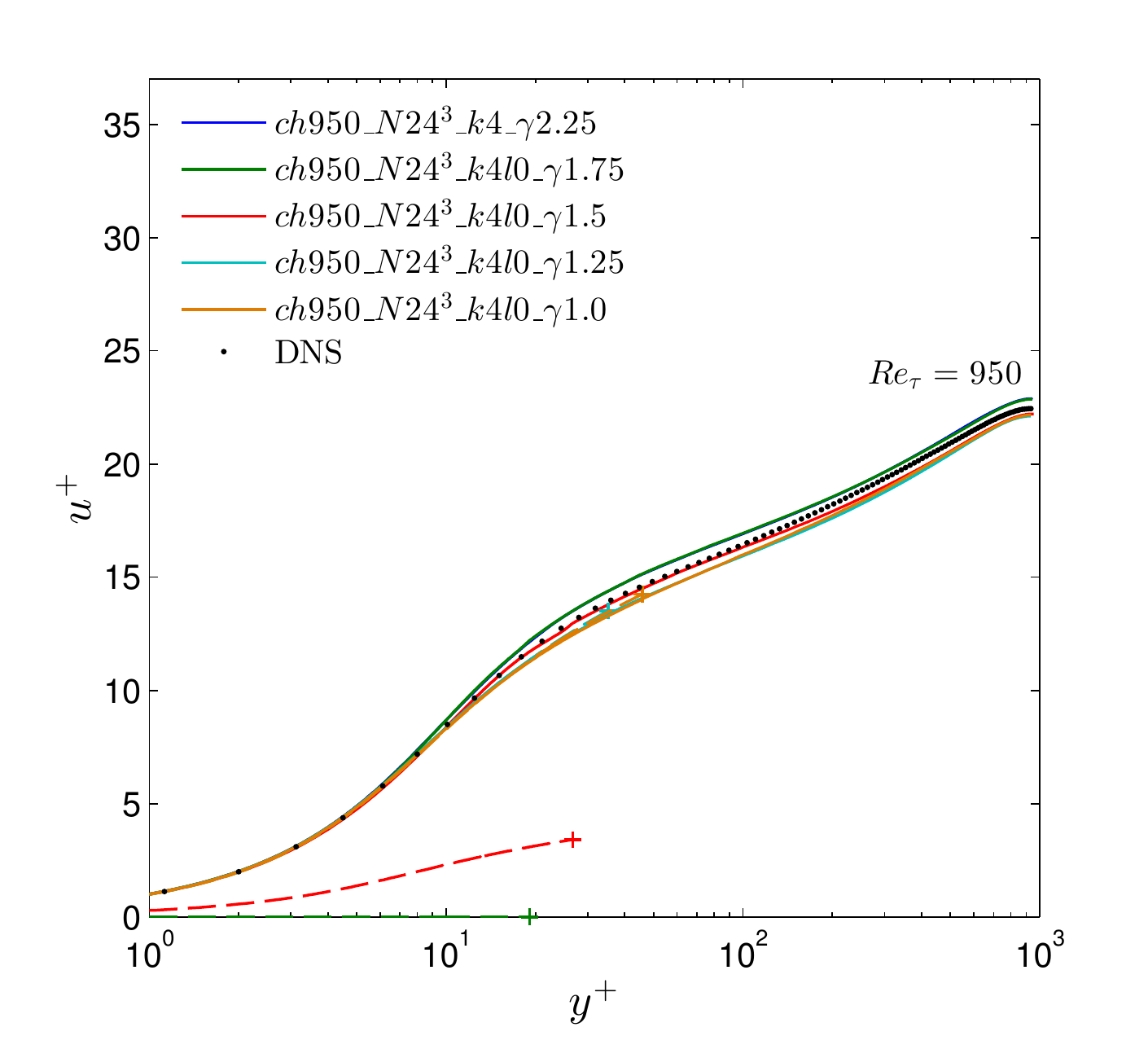}
\end{minipage}
\begin{minipage}[b]{0.497\linewidth}
\centering
\includegraphics[trim= 5.0mm 3mm 8mm 8mm,clip,width=1.\textwidth]{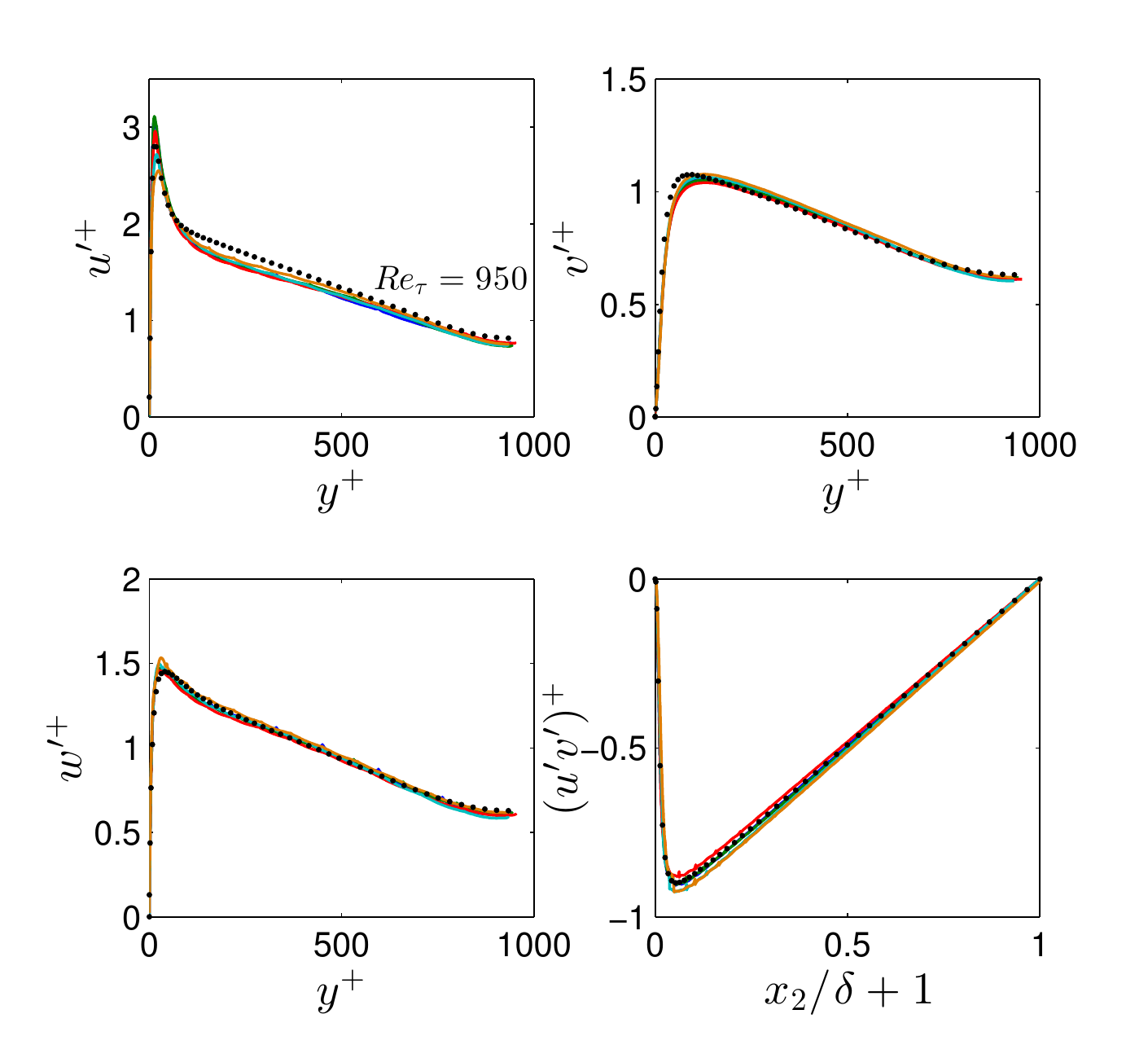}
\end{minipage}
\caption{Transition from wall-resolved to wall-modeled LES: Mean velocity (left) and RMS-velocities as well as Reynolds shear stress (right). All quantities are normalized according to $u^+=\langle u_1\rangle/u_{\tau}$, $u^{\prime+}= \sqrt{\langle u_1^{\prime2}\rangle}/u_{\tau}$, $v^{\prime+}= \sqrt{\langle u_2^{2}\rangle}/u_{\tau}$, $w^{\prime+}=\sqrt{\langle u_3^{2}\rangle}/u_{\tau}$, and $(u^{\prime}v^{\prime})^{+}=\langle u_1 u_2\rangle/u_{\tau}^2$. The full solution is displayed as solid line and the enrichment component as dashed line.}
\label{fig:ch_wallres_u}
\end{figure}

The transition from wall-modeled to wall-resolved LES is investigated in Figure~\ref{fig:ch_wallres_u}. At least $24^3$ cells are required for sufficiently resolving the energetic scales in the case without wall model, based on resolution guidelines by Chapman\cite{Chapman79} and our experience with the present scheme\cite{Krank16b} (with $k+1$ grid points per cell, we have $\Delta  y_{1}^+\approx\Delta  y_{1e}^+/(k+1)\approx 2$, $\Delta  z_e^+\approx\Delta  z^+/(k+1)\approx 20$). In Figure~\ref{fig:ch_wallres_u}, the grid stretching of the wall resolved LES is successively reduced until the wall model is fully active. The wall-resolved simulation slightly overpredicts the mean velocity and the turbulent fluctuations agree well with the reference. Regarding the case $ch950\_N24^3\_k4l0\_\gamma1.75$, the wall model is taken into account in the algorithm, but the model is switched off during the entire simulation due to the condition described in Section~\ref{sec:vandriest}. The results in Figure~\ref{fig:ch_wallres_u} are equivalent to the wall-resolved case. For the case $ch950\_N24^3\_k4l0\_\gamma1.5$ the wall model is activated dynamically in a temporally varying fraction of the cells and the enrichment shape functions constitute a small part of the mean velocity. The mean velocity profile agrees well with the DNS, even better than the wall-resolved case. For smaller grid stretching factors, the wall model is fully switched on after the initial laminar-turbulent transition and the enrichment solution is equivalent to the mean velocity profile. The mean velocity in these simulation cases is slightly underpredicted compared to the reference data and the turbulent fluctuations agree well with the reference data. From this numerical test it may be concluded that the wall model is well capable of handling the full range from wall-resolved to wall-modeled LES.

\begin{figure}[t]
\centering
\begin{minipage}[b]{0.497\linewidth}
\centering
\includegraphics[trim= 5mm 3mm 8.0mm 8mm,clip,width=1.\textwidth]{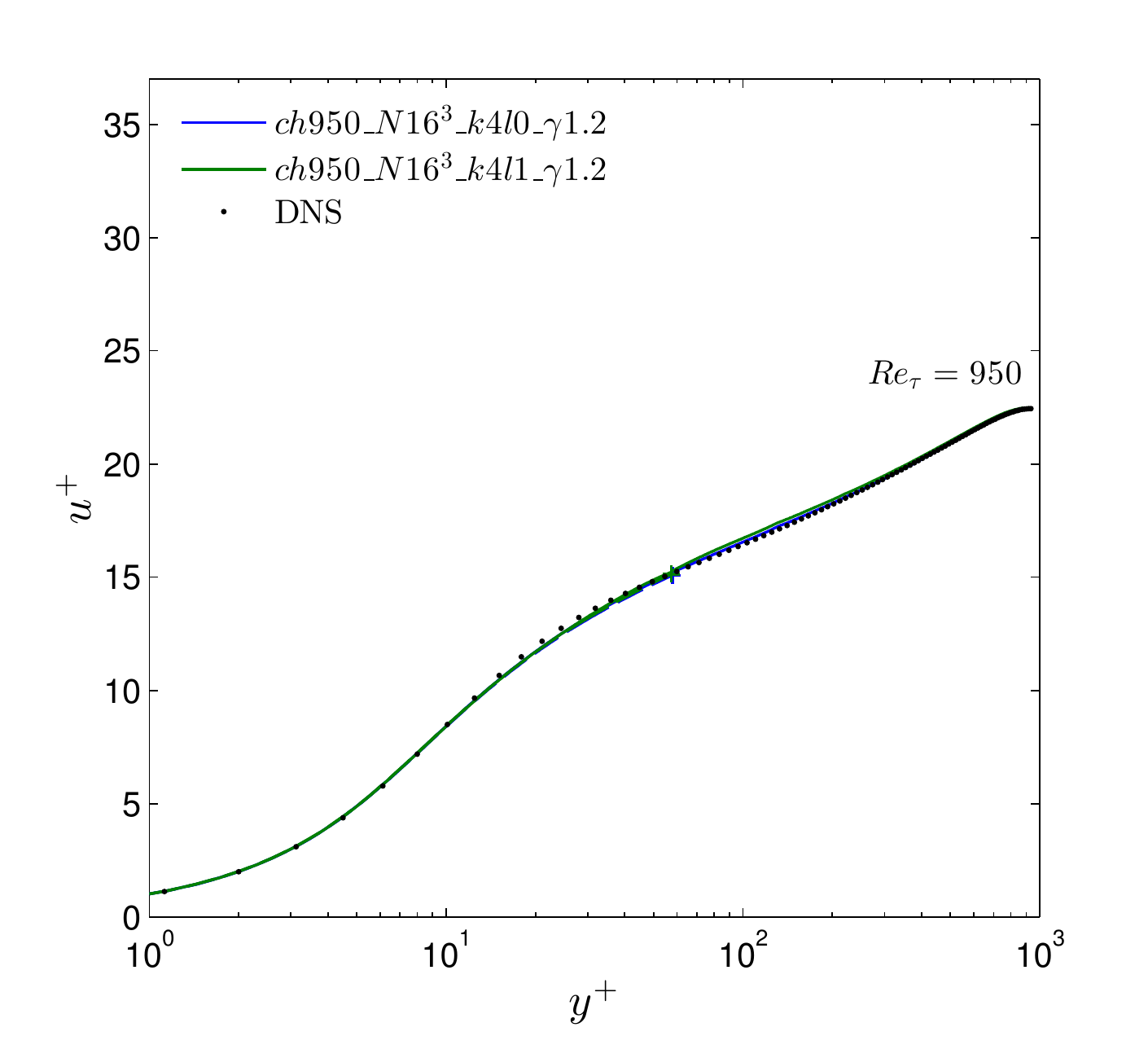}
\end{minipage}
\begin{minipage}[b]{0.497\linewidth}
\centering
\includegraphics[trim= 5.0mm 3mm 8mm 8mm,clip,width=1.\textwidth]{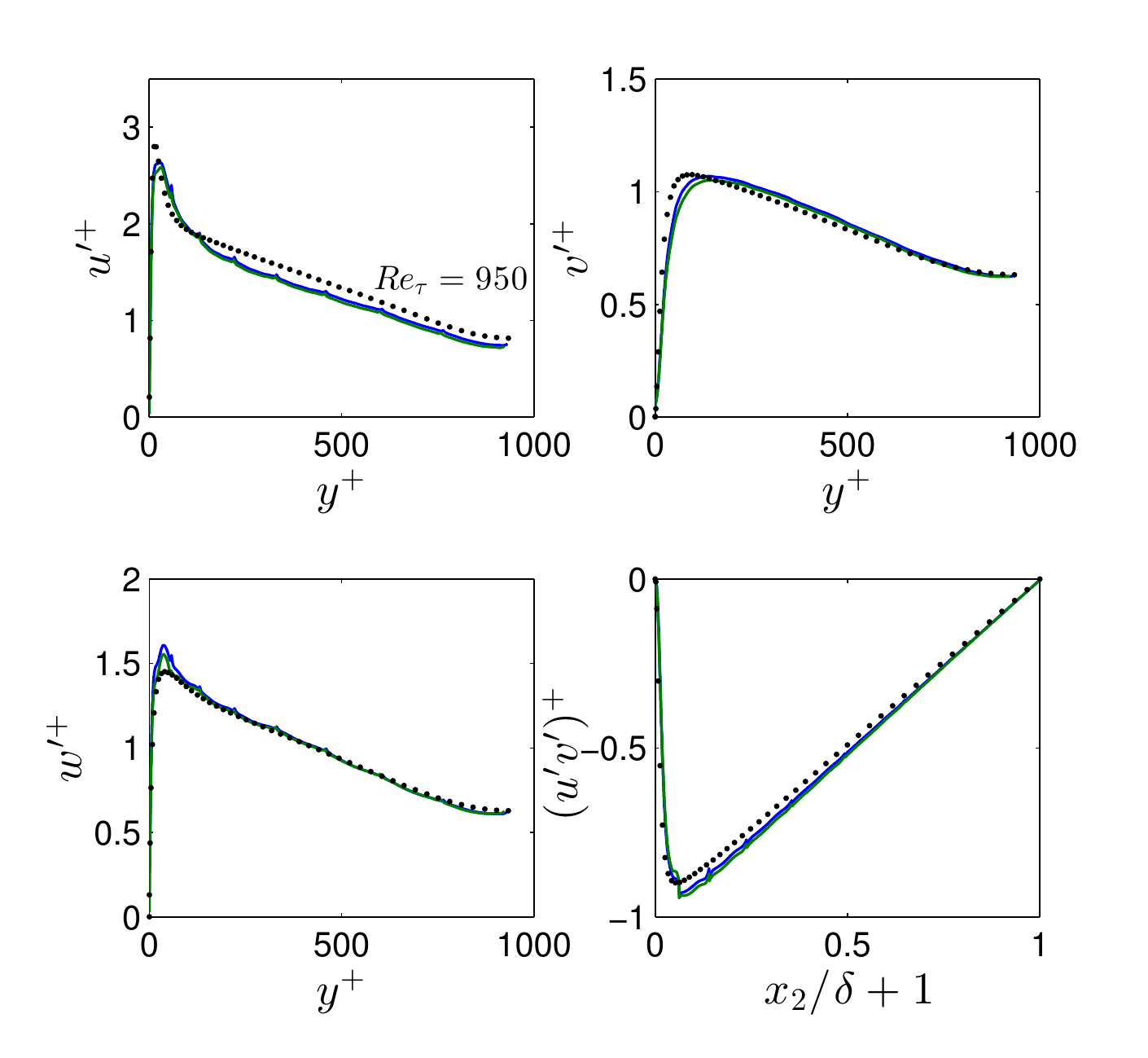}
\end{minipage}
\caption{Comparison of constant and linear shape functions for the weighting of the enrichment function: Mean velocity (left) and RMS-velocities as well as Reynolds shear stress (right). All quantities are normalized according to $u^+=\langle u_1\rangle/u_{\tau}$, $u^{\prime+}= \sqrt{\langle u_1^{\prime2}\rangle}/u_{\tau}$, $v^{\prime+}= \sqrt{\langle u_2^{2}\rangle}/u_{\tau}$, $w^{\prime+}=\sqrt{\langle u_3^{2}\rangle}/u_{\tau}$, and $(u^{\prime}v^{\prime})^{+}=\langle u_1 u_2\rangle/u_{\tau}^2$. The full solution is displayed as solid line and the enrichment component as dashed line.}
\label{fig:ch_l0l1_u}
\end{figure}

Until this point, all simulation cases have used a constant weighting of the enrichment function ($l=0$) and we put forward reasons for this choice. Figure~\ref{fig:ch_l0l1_u} compares two simulation cases, one with $l=0$ and one with $l=1$, where the latter represents a linear weighting of van Driest's law. The lines are on top of each other, including the enrichment solution and the fluctuations. This suggests that, for attached boundary layers, there is no need for a weighting of the enrichment with more than a constant factor. Furthermore, the enrichment using linear functions requires additional degrees of freedom ($24$ instead of $3$ per cell) and, most importantly, results in approximately three times the number of linear iterations in the GMRES solver of the Helmholtz equation due to worse condition numbers. Regarding nonequilibrium boundary layers, a linear weighting may yield slightly better results due to the additional flexibility within the RANS solution, but the higher computational cost is not justified.

\begin{figure}[t]
\centering
\begin{minipage}[b]{0.497\linewidth}
\centering
\includegraphics[trim= 5mm 3mm 8.0mm 8mm,clip,width=1.\textwidth]{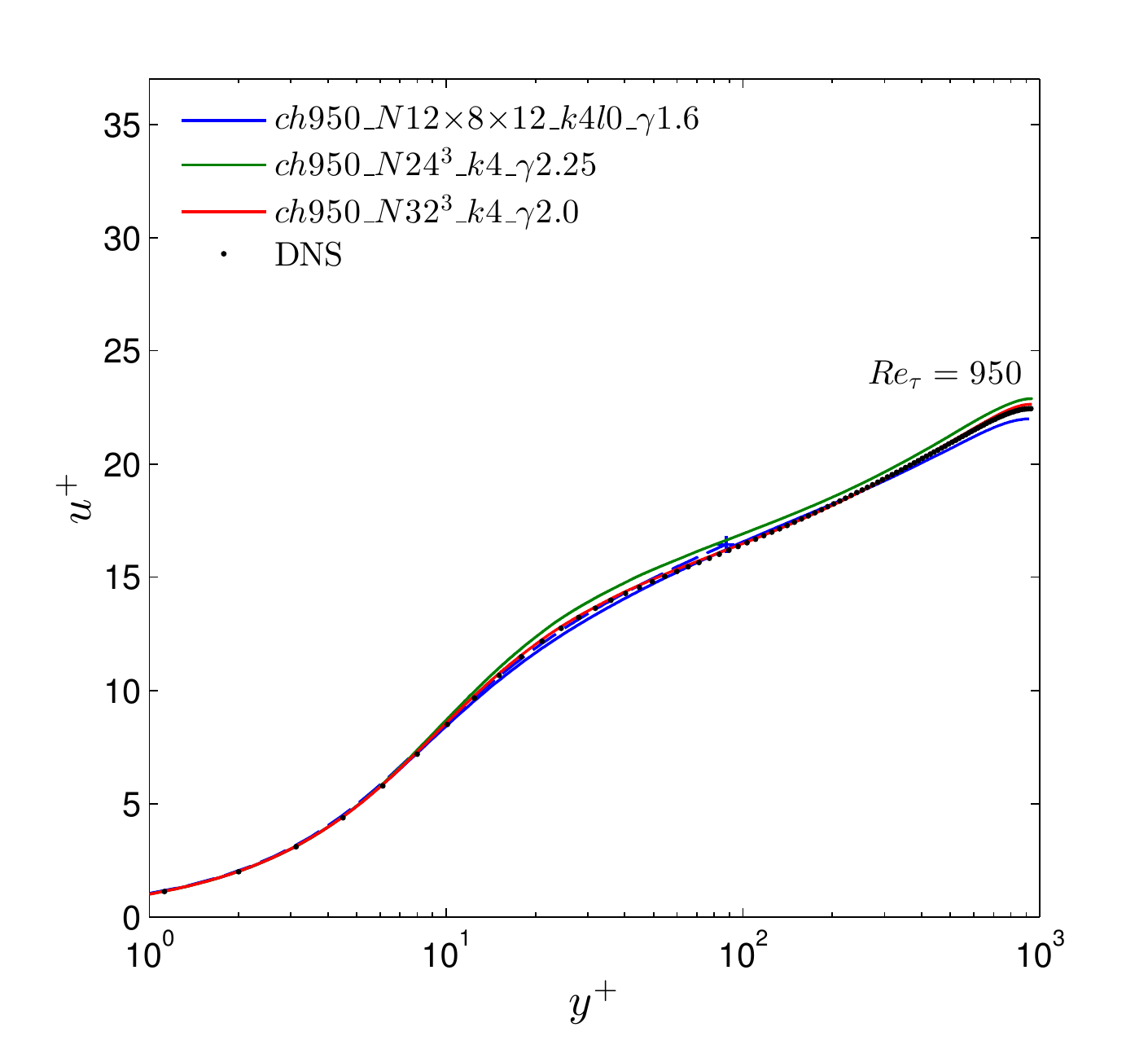}
\end{minipage}
\begin{minipage}[b]{0.497\linewidth}
\centering
\includegraphics[trim= 5.0mm 3mm 8mm 8mm,clip,width=1.\textwidth]{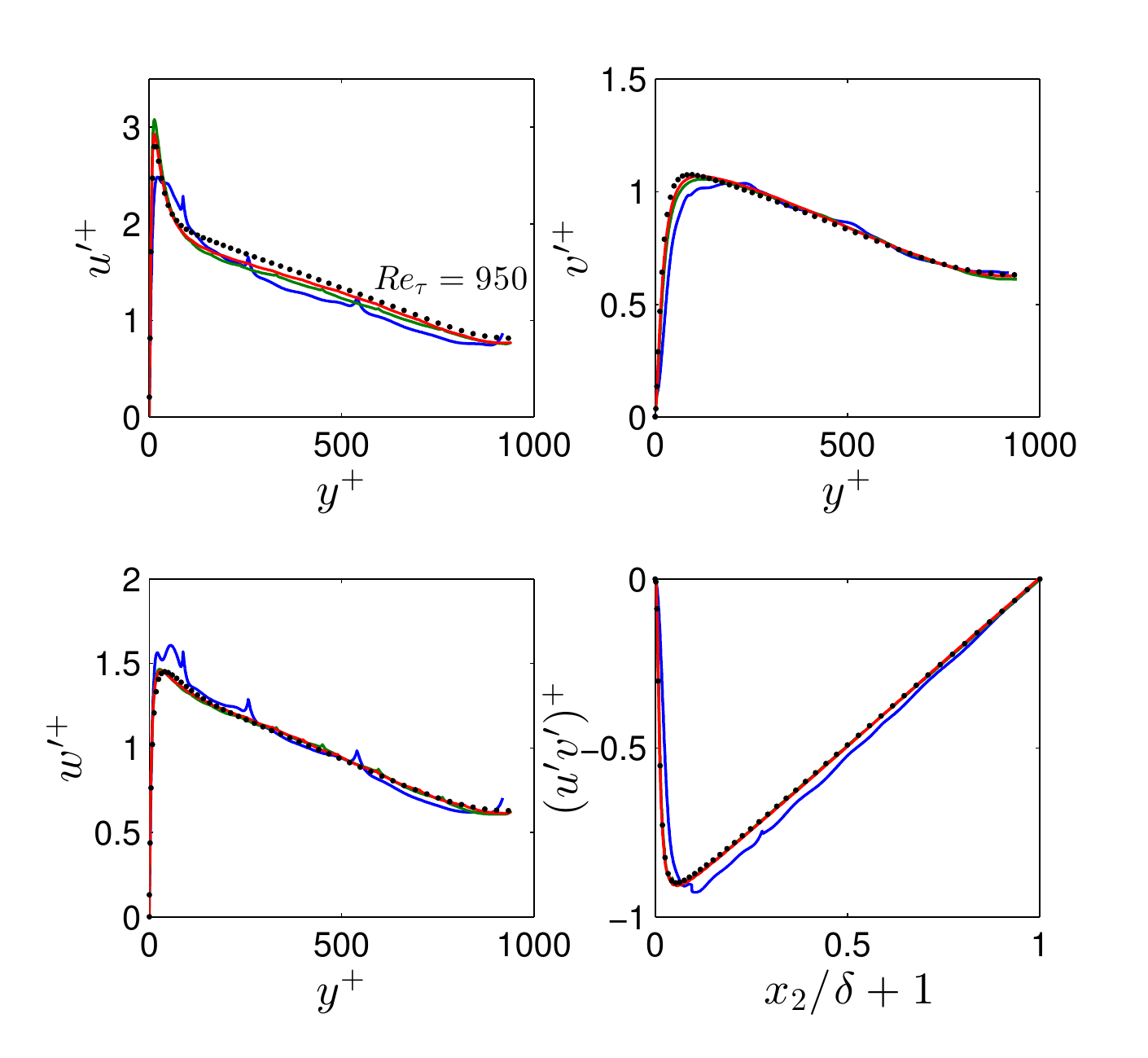}
\end{minipage}
\caption{Performance comparison of wall-resolved and wall-modeled cases: Mean velocity (left) and RMS-velocities as well as Reynolds shear stress (right). All quantities are normalized according to $u^+=\langle u_1\rangle/u_{\tau}$, $u^{\prime+}= \sqrt{\langle u_1^{\prime2}\rangle}/u_{\tau}$, $v^{\prime+}= \sqrt{\langle u_2^{2}\rangle}/u_{\tau}$, $w^{\prime+}=\sqrt{\langle u_3^{2}\rangle}/u_{\tau}$, and $(u^{\prime}v^{\prime})^{+}=\langle u_1 u_2\rangle/u_{\tau}^2$. The full solution is displayed as solid line and the enrichment component as dashed line. The wall-modeled simulation reduces the computational cost by a factor of $77$ compared to the coarser wall-resolved case and by a factor of $158$ compared to the finer wall-resolved case.}
\label{fig:ch_performance_u}
\end{figure}

We conclude the present section of the turbulent channel flow with a performance comparison of wall-resolved and wall-modeled LES. On the one hand, the wall model allows much coarser meshes near the wall, which also comes along with significantly larger time steps due to the explicit time integration, and the lower grid anisotropy yields fewer iterations in the expensive pressure Poisson solver. On the other hand, the wall model requires additional computational effort due to the larger number of quadrature points in the enriched cells and supplementary algorithmic steps. Figure~\ref{fig:ch_performance_u} compares the results of a wall-modeled case with the above coarse wall-resolved simulation as well as an additional finer wall-resolved case. The wall-modeled simulation exhibits better agreement with the DNS than the coarse wall-resolved simulation, but worse than the fine wall-resolved case. The computational cost is measured in number of processor cores times wall clock time of the entire simulation and the result is normalized by the wall-modeled simulation cost. The wall-modeled case ($ch950\_N12{\times}8{\times}12\_k4l0\_\gamma1.6$) gives a cost of 1, the coarse LES simulation ($ch950\_N24^3\_k4\_\gamma2.25$) yields a factor of $77$, and the fine wall-resolved calculation ($ch950\_N32^3\_k4\_\gamma2.0$) a cost of $158$. Comparing the coarse wall-resolved calculation and the wall-modeled calculation in more detail, the use of the wall model reduces the number of time steps by a factor of $5.7$ (for the same simulation time) and reduces the number of cells by a factor of 12. In addition, the wall-resolved simulation requires approximately 22 Poisson solver iterations instead of 9 in the wall-modeled case to yield the same relative accuracy in the iterative solver due to the higher mesh stretching, which overcompensates the extra cost of the wall model. These results allow the conclusion that the wall model promises a speed-up compared to wall-resolved LES by a factor of approximately two orders of magnitude and we expect an even larger benefit for higher Reynolds numbers.

The wall model has been thoroughly investigated regarding its application limits, mesh dependence and performance. The major results are that the model can compute the whole range from wall-resolved to wall-modeled LES and is extremely robust regarding different meshes as well as aspect ratios. However, the wall model should not be used beyond $y^+=120$ to guarantee accurate results. Given this limitation, the wall model has accelerated the simulations of turbulent channel flow by a factor of approximately two orders of magnitude.

\subsection{Flow over periodic hills}
\label{sec:ph}
\begin{table}[t]
\caption{Simulation cases and resolutions of the periodic hill flow. The cases at $Re_H=10{,}595$ and $19{,}000$ use $32 {\times} 16 {\times} 16$ cells with varying grid stretching in vertical direction and, at $Re_H=37{,}000$, a finer mesh with $64 {\times} 32 {\times} 32$ grid cells is additionally considered. The polynomial degrees are $k=4$ and $l=0$ for all simulation cases, and the number of grid points per direction is $k+1$ in each cell. The addition $NWM$ stands for `no wall model'. The separation and reattachment lengths $x_{1,\mathrm{sep}}$ and $x_{1,\mathrm{reatt}}$ correspond to the zero-crossings of the skin friction.}
\label{tab:ph_flows}
\begin{tabular*}{\textwidth}{l @{\extracolsep{\fill}} l l l l l l l}
\hline
Case   & $N_{e1} {\times} N_{e2} {\times} N_{e3}$ &$N_{1} {\times} N_{2} {\times} N_{3}$ &$Re_{H}$ & $\mathrm{max}(\Delta y^+_{1e})$ & $x_{1,\mathrm{sep}}/H$ & $x_{1,\mathrm{reatt}}/H$
\\ \hline \noalign{\smallskip} 
$ph10595\_stretch1$ \ & $32 {\times} 16 {\times} 16$ & $160 {\times} 80 {\times} 80$ &  $10{,}595$ & $86$ & $0.32$ & $4.23$\\
$ph10595\_stretch2$ \ & $32 {\times} 16 {\times} 16$ & $160 {\times} 80 {\times} 80$ &  $10{,}595$ & $73$  & $0.23$ & $4.24$\\
$ph10595\_stretch3$ \ & $32 {\times} 16 {\times} 16$ & $160 {\times} 80 {\times} 80$ &  $10{,}595$ & $60$  & $0.19$ & $3.98$\\
$ph10595\_stretch2\_NWM$ \ & $32 {\times} 16 {\times} 16$ & $160 {\times} 80 {\times} 80$ &  $10{,}595$ & -  & $0.29$ & $3.99$\\
$KKW\_DNS$\cite{Krank17b} \ & -  & $896 {\times} 448 {\times} 448$ & $10{,}595$ & -  & $0.19$ & $4.51$ \\
\noalign{\smallskip} 
$ph19000\_stretch1$ \ & $32 {\times} 16 {\times} 16$ & $160 {\times} 80 {\times} 80$ &  $19{,}000$ & $138$  & $0.42$ & $2.85$\\
$ph19000\_stretch2$ \ & $32 {\times} 16 {\times} 16$ & $160 {\times} 80 {\times} 80$ &  $19{,}000$ & $113$  & $0.35$ & $3.60$\\
$ph19000\_stretch3$ \ & $32 {\times} 16 {\times} 16$ & $160 {\times} 80 {\times} 80$ &  $19{,}000$ & $95$  & $0.25$ & $3.64$\\
$ph19000\_stretch2\_NWM$ \ & $32 {\times} 16 {\times} 16$ & $160 {\times} 80 {\times} 80$ & $19{,}000$ &  - & $0.32$ & $1.99$\\
$RM\_EXP$\cite{Rapp11} & - & - & $19{,}000$ & - & - & $3.94$ \\ \noalign{\smallskip}  
 $ph37000\_stretch3$ \ & $32 {\times} 16 {\times} 16$ & $160 {\times} 80 {\times} 80$ &  $37{,}000$ & $156$  & $0.37$ & $2.78$\\
$ph37000\_stretch3\_fine$ \ & $64 {\times} 32 {\times} 32$ & $320 {\times} 160 {\times} 160$ &  $37{,}000$ & $78$  & $0.28$ & $3.38$\\
 $ph37000\_stretch3\_NWM$ \ & $32 {\times} 16 {\times} 16$ & $160 {\times} 80 {\times} 80$ &  $37{,}000$ & -  & - & -\\
 $RM\_EXP$\cite{Rapp11} & - & - & $37{,}000$ & - & - & $3.76$\\
 $CM\_WMLES$\cite{Wiart17} & - & $128 {\times} 64 {\times} 64$ & $37{,}000$ & - & - & $2.3$\\
 $CM\_WMLES\_fine$\cite{Wiart17} & - & $256 {\times} 128 {\times} 128$ & $37{,}000$ & - &  - & $2.8$\\ 
\hline
\end{tabular*}
\end{table}

\begin{figure}[t]
\centering
\begin{minipage}[b]{0.497\linewidth}
\centering
\includegraphics[trim= 10mm 0mm 10mm 0mm,clip,width=1.\textwidth]{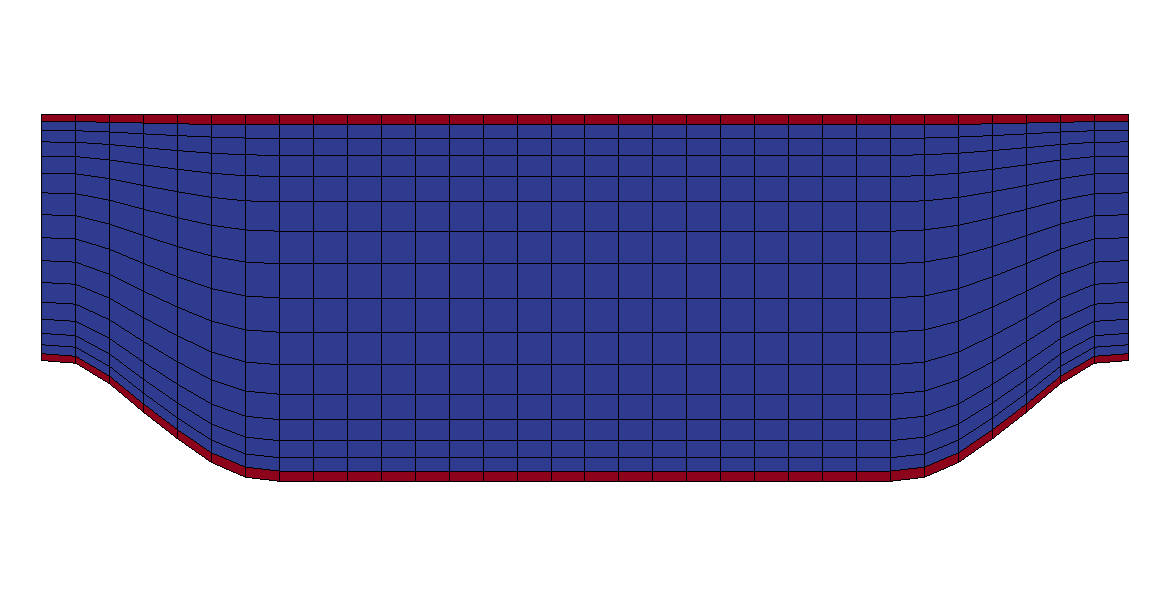}
\end{minipage}
\begin{minipage}[b]{0.497\linewidth}
\centering
\includegraphics[trim= 10mm 0mm 10mm 0mm,clip,width=1.\textwidth]{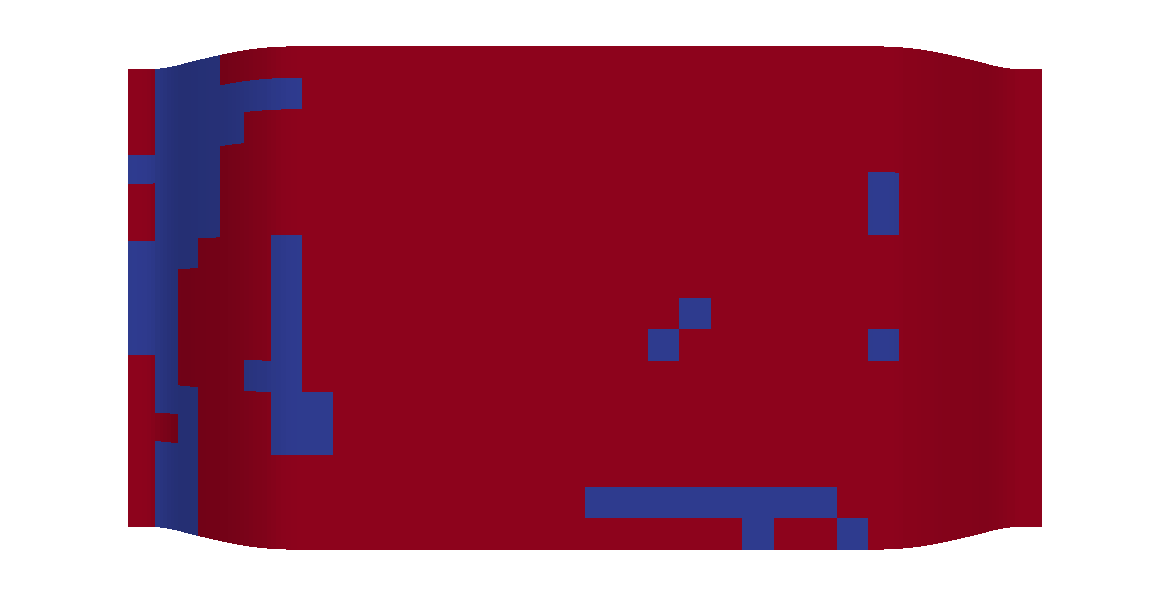}
\end{minipage}
\caption{Grid of the case $ph10595\_stretch2$ (left). The wall-model cells are used in one layer near the wall. However, the enrichment is in these cells switched off dynamically if the polynomial is sufficient to resolve the turbulent structures. Therefore, the enrichment is in some cells at the lower wall inactive in the instantaneous flow field (right, computational domain shown from below). Enriched cells are depicted red and standard nonenriched cells blue.}
\label{fig:grid}
\end{figure}
As a second benchmark example, we consider flow over periodic hills at the Reynolds numbers $Re_H=10{,}595$, $Re_H=19{,}000$, and $Re_H=37{,}000$ based on the hill height $H$ and the bulk velocity $u_b$. This test case is a popular benchmark for wall models due to its simplicity regarding simulation setup and boundary conditions on the one hand and complexity with respect to turbulence phenomena and modeling on the other hand. For example, common RANS models fail to predict the separation from the curved hill crest\cite{Krank16c,Jakirlic15}. In addition, the nonequilibrium boundary layer within the recirculation bubble necessitates the consideration of all terms in the Navier--Stokes equations in order to yield accurate results. As an example, the simulations using an equilibrium wall model within a high-order DG method do not converge to the reference data\cite{Wiart17}. In contrast, the results obtained with wall modeling via function enrichment for LES in a standard FEM code\cite{Krank16} are very promising, since all terms of the Navier--Stokes equations are taken into consideration.

For this validation case, several reference databases exist. The lower Reynolds number has been computed by LES using multiple codes\cite{Frohlich05,Breuer09}, experiments have been carried out\cite{Rapp11,Breuer09}, and the database has recently been extended by a fully resolved DNS using the standard version of the present solver\cite{Krank16b} at a higher polynomial degree ($7^{\mathrm{th}}$ order of accuracy) and 180 million grid points\cite{Krank17b}. This reference data may be downloaded from the public repository\cite{Krank18b}. Therein, it was found that the DNS is in closer agreement with the experiments than the LES data and predicts the reattachment length at $4.51H$, slightly shorter than the previous LES references. Except for the spatial discretization and temporal order of accuracy, we use the same simulation setup as for the DNS here. Reference data for the higher Reynolds numbers is available via experiments\cite{Rapp11}.

\begin{figure}[t]
\centering
\begin{minipage}[b]{0.33\linewidth}
\centering
\includegraphics[trim= 0mm 0mm 9mm 4mm,clip,width=1\linewidth]{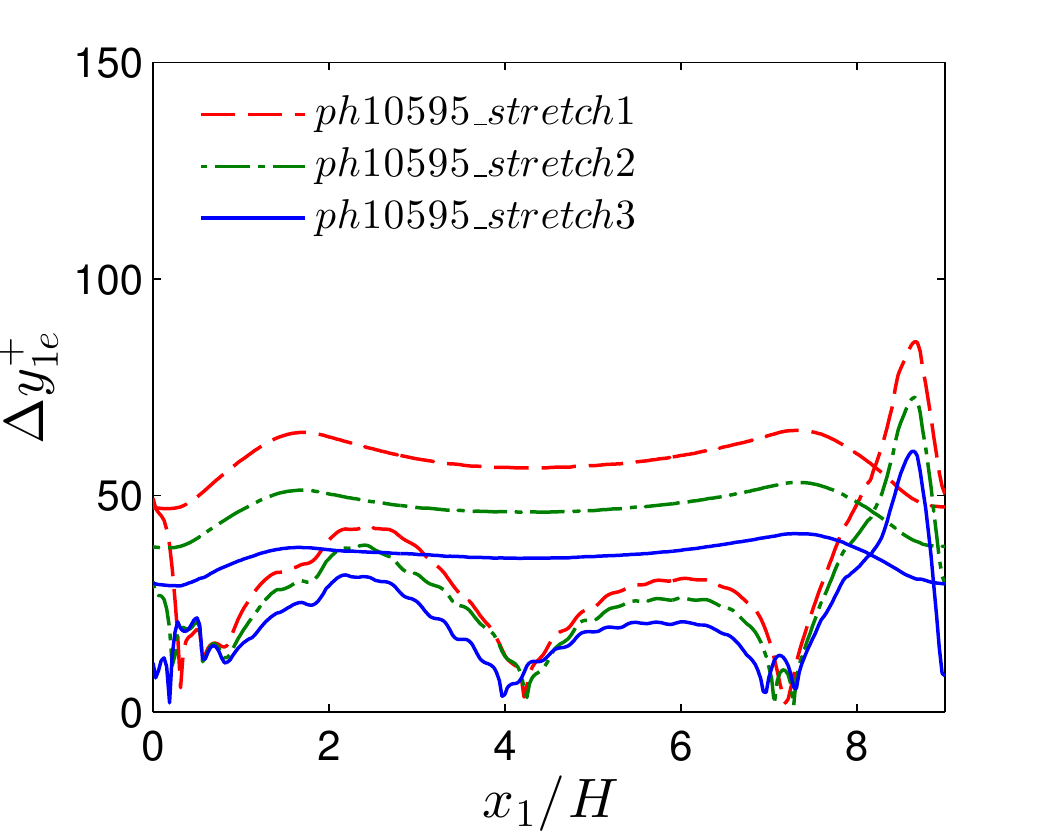}
\end{minipage}
\begin{minipage}[b]{0.33\linewidth}
\centering
\includegraphics[trim= 0mm 0mm 9mm 4mm,clip,width=1\linewidth]{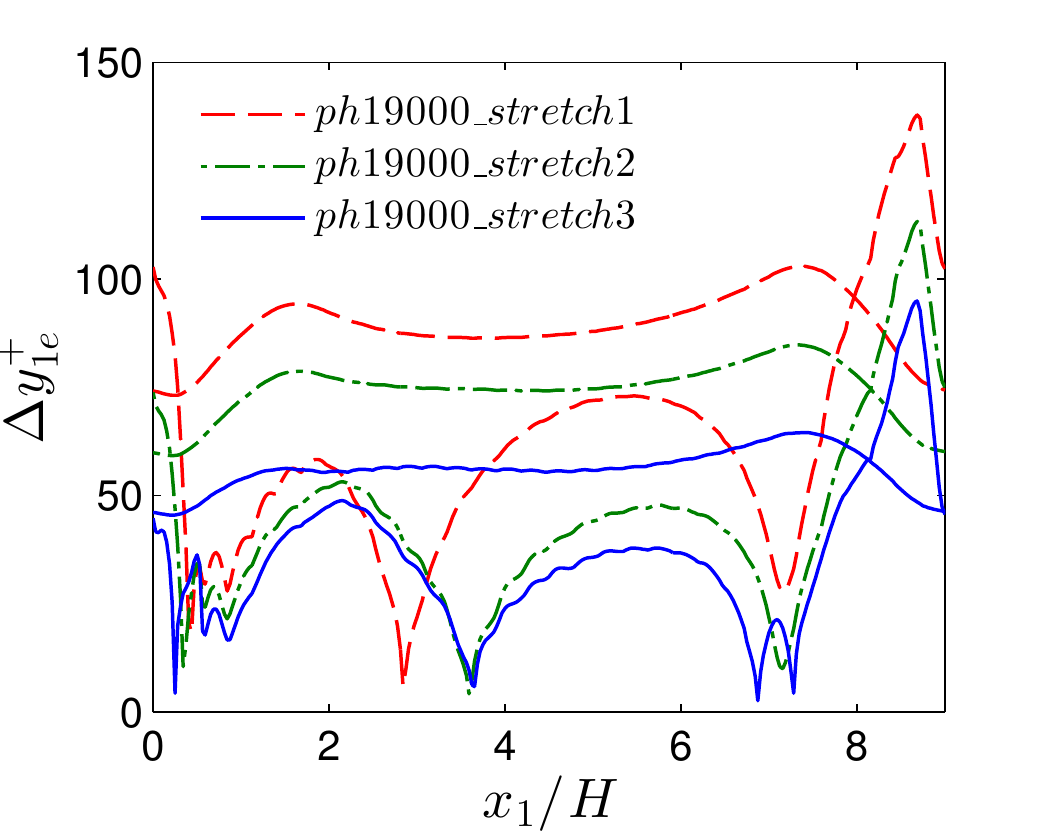}
\end{minipage}
\begin{minipage}[b]{0.33\linewidth}
\centering
\includegraphics[trim= 0mm 0mm 9mm 4mm,clip,width=1\linewidth]{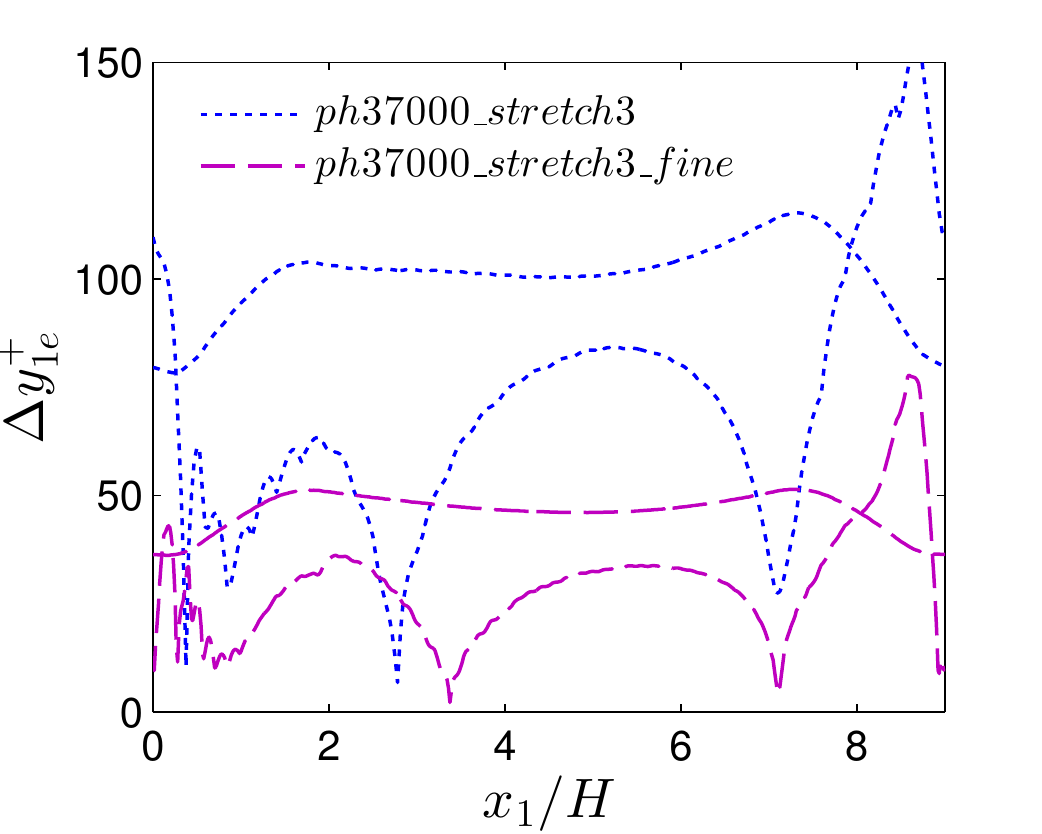}
\end{minipage}
\centering
\caption{Width of wall-layer (width of first off-wall cell) for $Re_H=10{,}595$, $Re_H=19{,}000$ (top right), and $Re_H=37{,}000$ (from left to right). The shallower curves correspond to the upper wall.}
\label{fig:ph_yp1}
\end{figure}

\begin{figure}[t]
\begin{minipage}[b]{0.497\linewidth}
\centering
\includegraphics[trim= 10mm 0mm 10mm 0mm,clip,width=1.\textwidth]{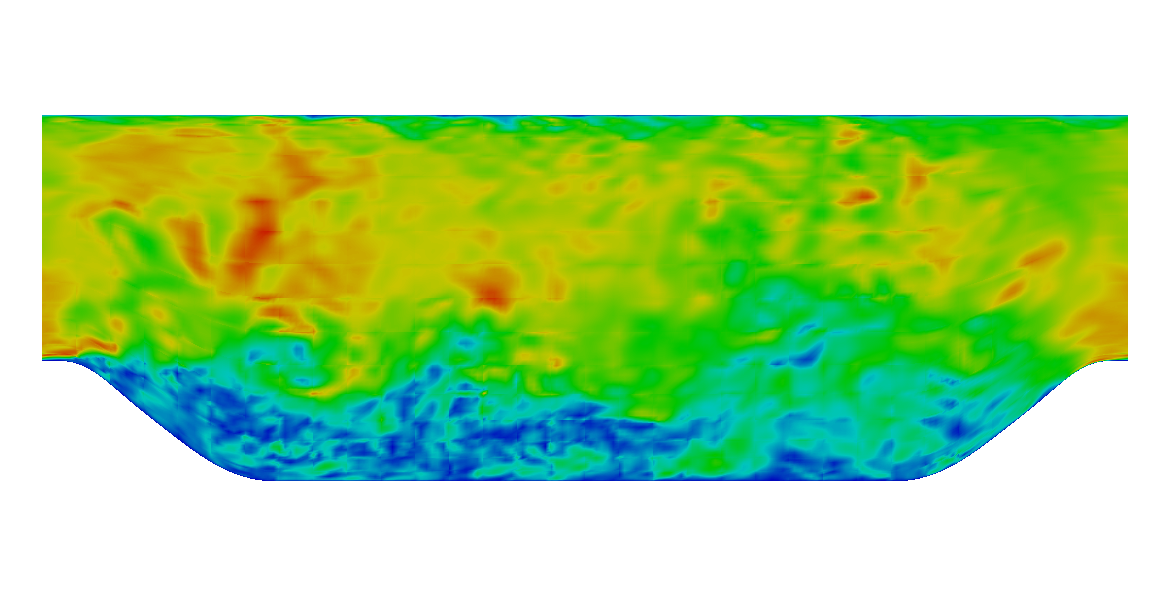}
\end{minipage}
\begin{minipage}[b]{0.497\linewidth}
\centering
\includegraphics[trim= 10mm 0mm 10mm 0mm,clip,width=1.\textwidth]{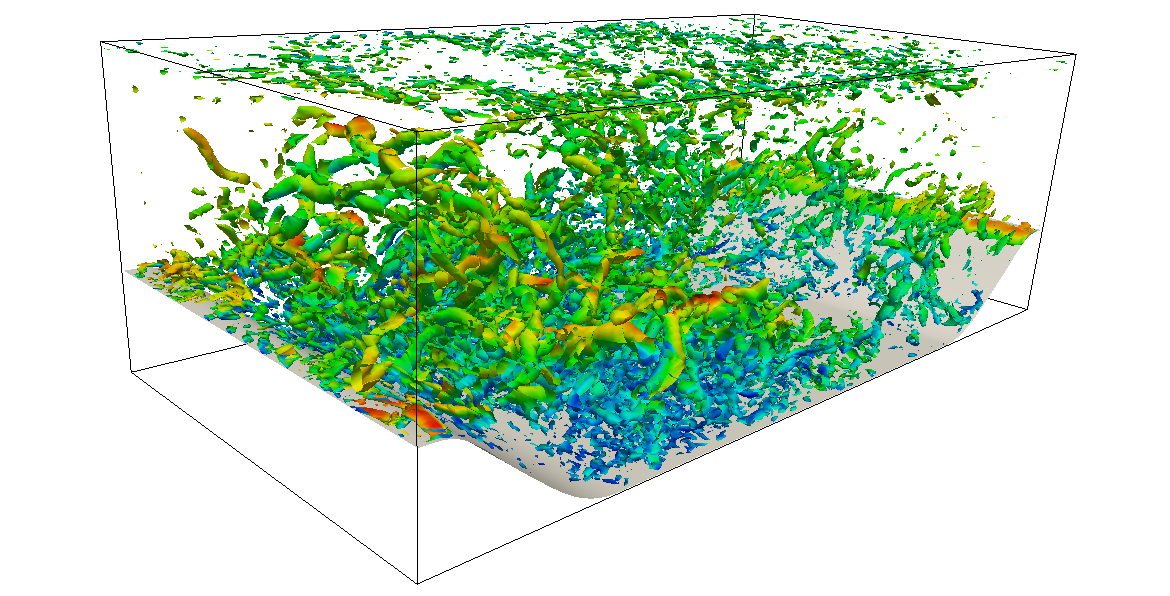}
\end{minipage}
\centering
\caption{Velocity magnitude (left) and visualization of turbulent eddies via the Q-criterion colored by velocity magnitude (right) of the case $ph10595\_stretch2$. Red indicates high and blue low values.}
\label{fig:ph_u}
\end{figure}

The computational domain for the present simulations is of the size $9H{\times}3.036H{\times}4.5H$ in streamwise, vertical, and spanwise direction, respectively, and the lower wall is smoothly curved in the surrounding of the hill. Since the baseline solver used in this study has already been investigated in detail in the context of wall-resolved LES\cite{Krank17b} using four polynomial degrees and three mesh refinement levels, we focus on the performance of the wall model in the present work. For most of the simulation cases, a mesh consisting of $32{\times}16{\times}16$ cells of degree $k=4$ is selected, resulting in $160{\times}80{\times}80$ nodes, and the enrichment with $l=0$ is included in the first off-wall element layer, see Figure~\ref{fig:grid}. A single simulation case is presented, in which the number of cells is doubled in each spatial direction, considering the highest Reynolds number. The grid is graded towards the walls and we investigate three grid stretching factors in order to show the influence of the width of the wall layer. According to Figure~\ref{fig:ph_yp1}, the wall layer and thus the first off-wall cells span a range in $y^+$-units up to $86$ for the lowest, $138$ for the medium, and $156$ for the highest Reynolds number, with peaks near the hill top and minima near the separation and reattachment points. The mesh is mapped onto the exact hill geometry using the same polynomial ansatz as for the LES scale via manifold techniques available in the deal.II library\cite{Arndt17}. The enrichment degrees of freedom constitute equal or less than $0.1\%$ of the overall number of degrees of freedom. Further, the enrichment and thus the multiscale wall model are switched off dynamically if the boundary layer is sufficiently resolved. The wall model is therefore effectively not taken into account behind the hill crest and in the region of the flow reattachment. More cells are switched off in the lower Reynolds number cases; the active cells in one snapshot are depicted in Figure~\ref{fig:grid}. The statistical quantities are averaged during a simulation time corresponding to 61 flow-through times. The velocity field of the fully turbulent flow and turbulent vortex structures are visualized in Figure~\ref{fig:ph_u}. All discretization cases are summarized in Table~\ref{tab:ph_flows}, including the labels employed in the subsequent figures. The investigations of the wall model include a comparison to one simulation for each Reynolds number, which uses the exact same mesh but without the wall model and is labeled with the addition $NWM$ (no wall model) in Table~\ref{tab:ph_flows}.

\begin{figure}[t]
\centering
\includegraphics[trim= 9mm 0mm 9mm 4mm,clip,scale=0.7]{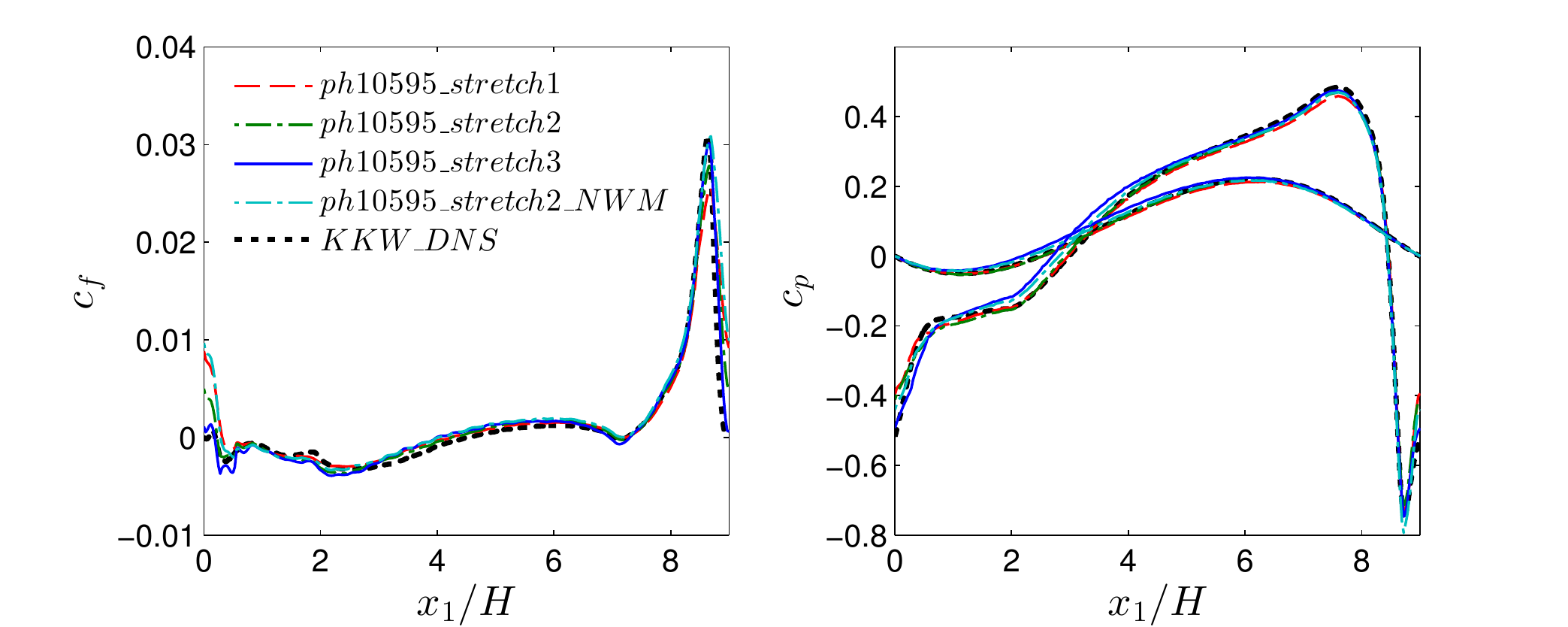}
\caption{Skin friction coefficient at the lower wall (left) and pressure coefficient at the lower and upper boundary (right). The shallower pressure coefficient curves correspond to the upper wall.}
\label{fig:ph10595_cfcp}
\end{figure}

We begin the discussion of the results with the friction and pressure coefficients for the lowest Reynolds number. They are defined as
\begin{equation*}
c_f=\frac{\tau_w}{\frac{1}{2} \rho u_b^2},\hspace{1cm}
c_p=\frac{\rho(p-p_{\mathrm{ref}})}{\frac{1}{2} \rho u_b^2}
\label{eq:cfcpdef}
\end{equation*}
and for the reference pressure $p_{\mathrm{ref}}$ we take the value at $x_1=0$ at the upper wall. The pressure coefficient is multiplied by the density since the kinematic pressure is used in this work. The curves are compared to the DNS reference data in Figure~\ref{fig:ph10595_cfcp}. Overall, good agreement of all curves with the DNS is observed and the largest error occurs near the hill top. Even the case without wall modeling is in acceptable agreement with the reference data, although the peak in the skin friction is predicted less distinct in that case. Also, the separation and reattachment lengths in Table~\ref{tab:ph_flows}, given as the zero-crossings of the skin friction, are in good agreement with the reference. The reattachment lengths of the cases $ph10595\_stretch1$ and $ph10595\_stretch2$ at $\sim 4.25H$ are slightly underpredicted in comparison to the DNS reference of $4.51H$. Regarding the case $ph10595\_stretch3$, the reattachment length is predicted even shorter as $3.98H$. With respect to the latter, a possible source of error could be the coarser mesh in the shear layer within the LES region, resulting in a more dissipative behavior of the numerical scheme and thus affecting the length of the recirulation zone.

\begin{figure}[t]
\centering
\includegraphics[trim= 0mm 10mm 0mm 0mm,clip,scale=0.7]{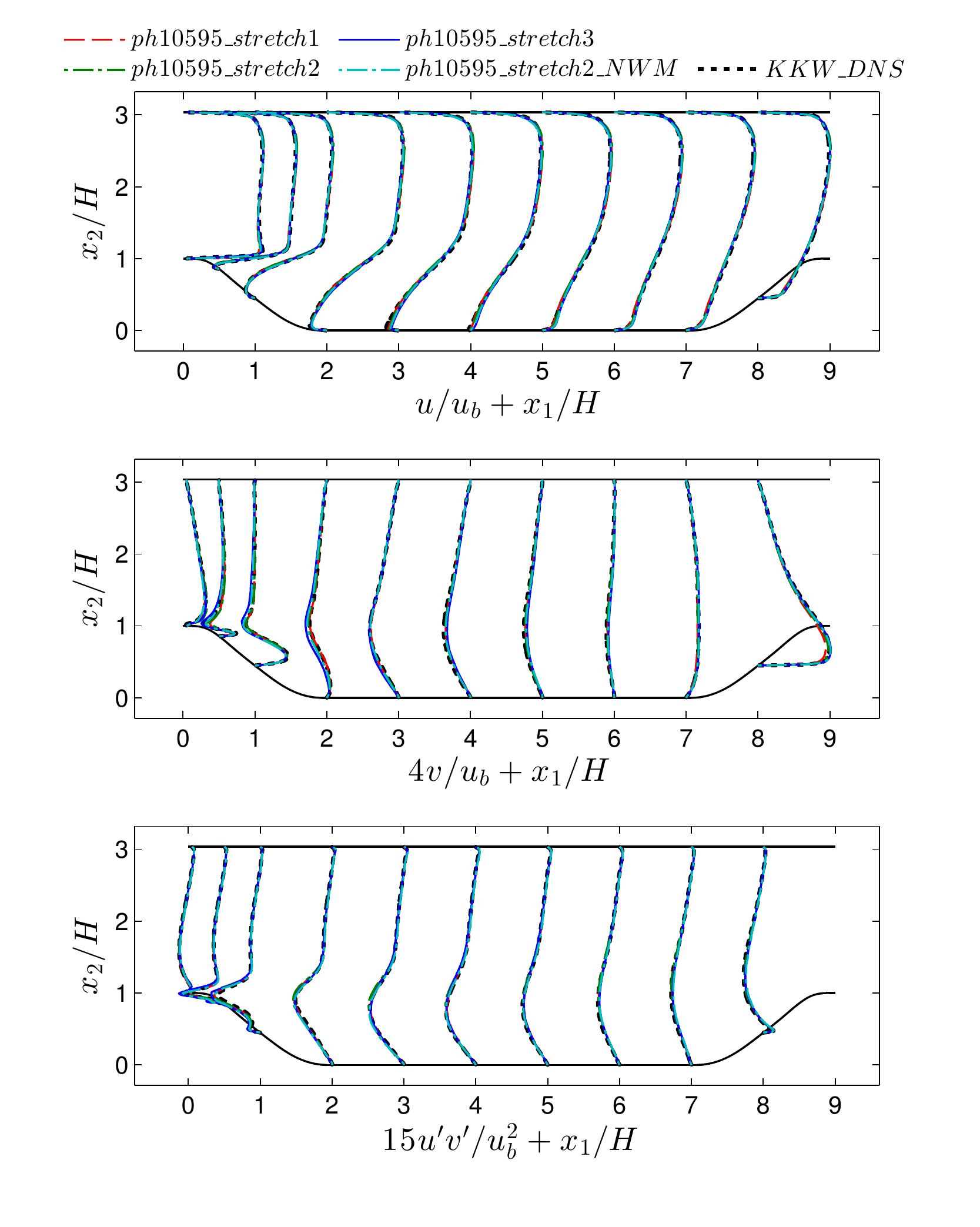}
\caption{Streamwise $u=\langle u_1\rangle$ and vertical $v=\langle u_2\rangle$ mean velocity as well as Reynolds shear stress $u'v'=\langle u_1 u_2\rangle - \langle u_1 \rangle \langle u_2\rangle$ of the periodic hill flow at $Re_H=10{,}595$.}
\label{fig:ph10595_um}
\end{figure}

The time-averaged streamwise velocity, vertical velocity, and Reynolds shear stress are depicted in Figure~\ref{fig:ph10595_um} at ten stations. The curves essentially lie on top of each other, including the wall-modeled simulations, the underresolved case, and the DNS. From this fact we conclude that the wall model shows an outstanding performance for nonequilibrium boundary layers, which confirms the results obtained for wall modeling via function enrichment within the continuous FEM\cite{Krank16} and in DES\cite{Krank17c} as well as the robustness of the method with respect to mesh aspect ratios observed in the previous section.

\begin{figure}[t]
\centering
\includegraphics[trim= 0mm 10mm 0mm 0mm,clip,scale=0.7]{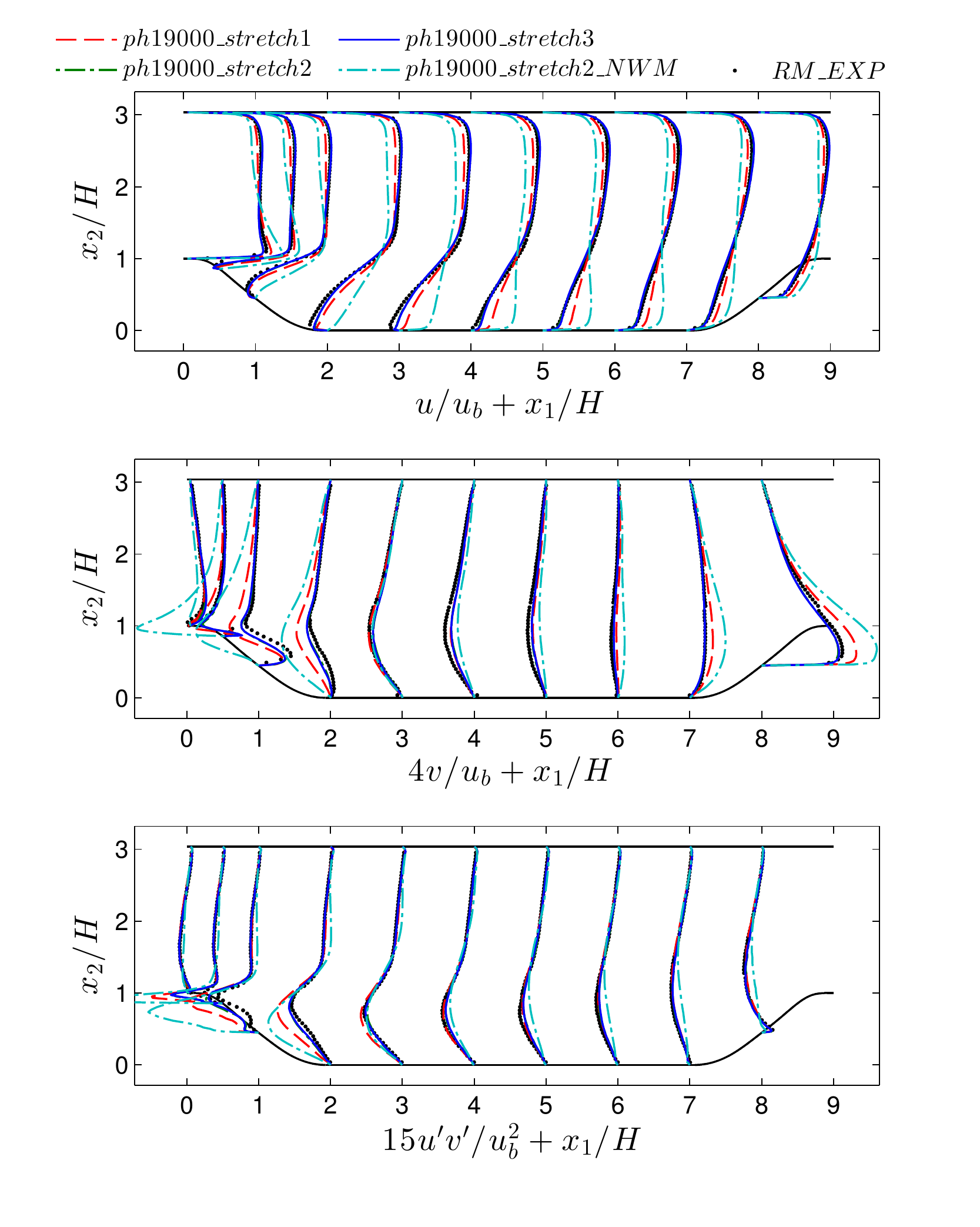}
\caption{Streamwise $u=\langle u_1\rangle$ and vertical $v=\langle u_2\rangle$ mean velocity as well as Reynolds shear stress $u'v'=\langle u_1 u_2\rangle - \langle u_1 \rangle \langle u_2\rangle$ of the periodic hill flow at $Re_H=19{,}000$.}
\label{fig:ph19000_um}
\end{figure}

\begin{figure}[htb]
\centering
\includegraphics[trim= 0mm 137mm 0mm 0mm,clip,scale=0.7]{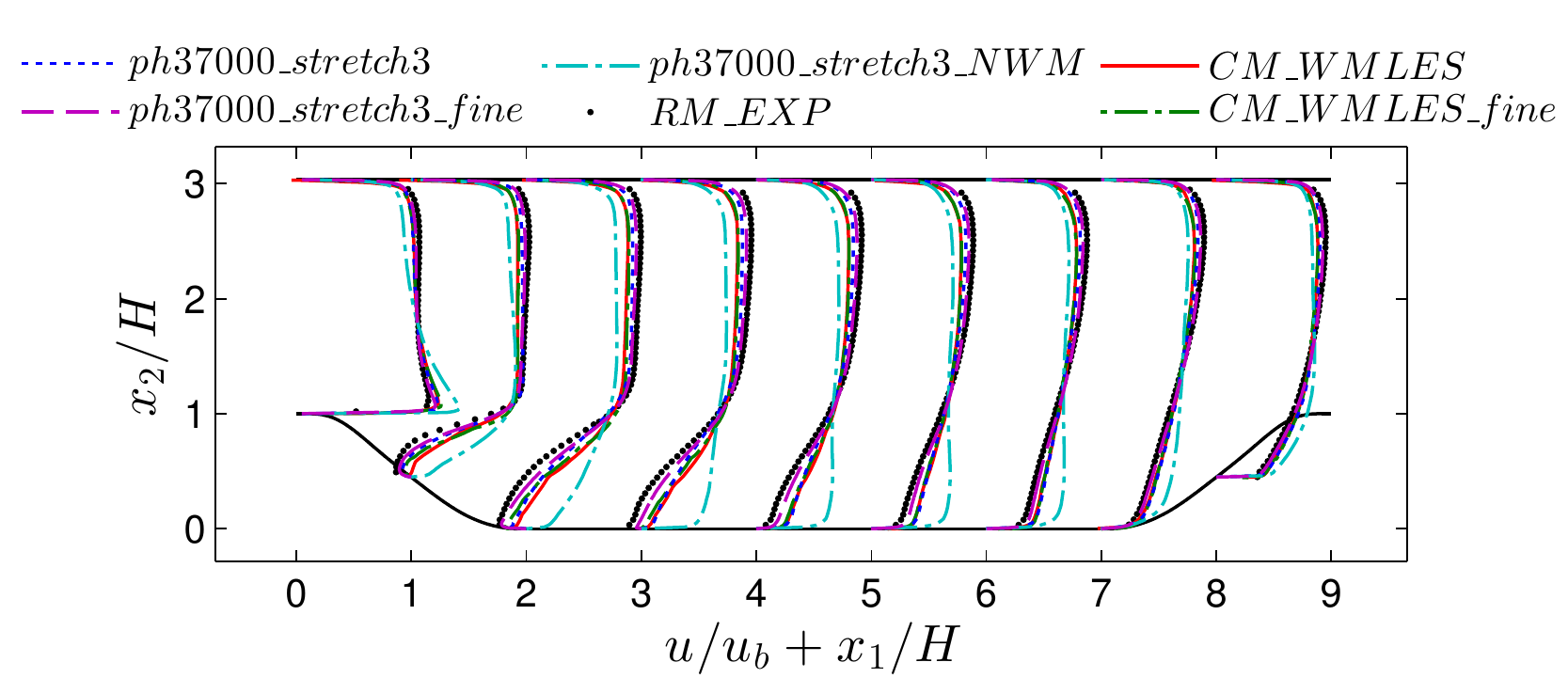}
\caption{Streamwise $u=\langle u_1\rangle$ mean velocity of the periodic hill flow at $Re_H=37{,}000$.}
\label{fig:ph37000_um}
\end{figure}

The fact that even the underresolved case without the wall model yields a good agreement with the reference data motivates an application of the same meshes to a higher Reynolds number. The mean streamwise velocity, vertical velocity, and Reynolds shear stress for $Re_H=19{,}000$ are shown in Figure~\ref{fig:ph19000_um}. The wall model exhibits results of similar quality as for the lower Reynolds number and the sensitivity regarding the grid stretching is slightly higher. However, the results for the case without wall model do not agree with the reference data at all. In particular, the reattachment length  with $1.99H$ instead of $3.94H$ is significantly underpredicted.

Finally, we assess whether the wall model gives any added value in separated flow conditions in comparison to the simpler wall stress model by Carton de Wiart and Murman within a high-order DG method\cite{Wiart17}. Two meshes are employed, the grid with the highest mesh stretching considered previously and a refined variant with twice as many cells in each spatial direction. These meshes are quite similar to the baseline and fine grid used by Wiart and Murman\cite{Wiart17}, labeled as $CM\_WMLES$ and $CM\_WMLES\_fine$, respectively, in the following. Slightly more points are used herein, the approach in the reference\cite{Wiart17} exhibits an order of accuracy of 8 in comparison to the $4^{\mathrm{th}}$ order accuracy of the present method, however. A full data set including reference data is only available for the mean streamwise velocity, which is presented in Figure~\ref{fig:ph37000_um}. The coarser case using wall modeling via function enrichment exhibits solutions of similar quality as the coarse case with the equilibrium wall model. A possible explanation for this agreement may be that some of the relevant scales in the separation region may not be sufficiently resolved, which would also explain the delayed separation length of the case $ph37000\_stretch3$ of $0.37H$.  Comparing the refined simulation cases, the equilibrium model does not show a significant improvement in comparison to the coarse cases while the results of the present wall model are almost converged to the experimental reference data. An investigation of the reattachment lengths confirms these observations, as the simulation $ph37000\_stretch3\_fine$ shows the best agreement considering this quantity. This superiority of the function-enrichment-based wall model in separated flow conditions is due to the full consistency of the method and the consideration of all terms of the Navier--Stokes equations. The case without wall modeling does not show a separation region at all as the near-wall region is significantly underresolved.

In summary, the present multiscale wall model enables an accurate computation of separated boundary layers, is robust with regard to mesh sensitivities and exhibits much more accurate results compared to an equilibrium wall stress model.

\section{Concluding remarks}
\label{sec:conclusion}
In this work, we have developed a new turbulence modeling approach for the wall modeling concept of function enrichment within the high-order discontinuous Galerkin method. Based on a rigorous derivation of the modeling terms using Germano's framework of additive filtering, a RANS model is applied in a thin layer near the wall. Unlike existing wall modeling approaches, the RANS and an underresolved LES solution overlap inside the near-wall layer. This composition allows the LES solution to develop within the wall layer such that the typical issue of the RANS--LES transition is avoided. The method is also capable of resolving the velocity gradient and thus the wall shear stress in the viscous sublayer with coarse meshes, where the first off-wall cell extends up to $120$ wall units. This is achieved by using the recently proposed technique of function enrichment\cite{Krank16,Krank16c,Krank17c,Krank18} inside the inner boundary layer. The wall model has exhibited outstanding characteristics in attached and separated boundary layers, does not show a log-layer mismatch and has reduced the cost of a benchmark simulation by approximately two orders of magnitude.

The turbulence model used in the original publication on wall modeling via function enrichment within a standard FEM flow solver\cite{Krank16} consists of a residual-based approach, which is supported by a structural LES model (multifractal subgrid scales) in the bulk flow. Therein, the idea was that the residual-based method provides numerical dissipation, which is both appropriate to stabilize the scheme and to model the unresolved energetic scales at the wall. Although the data has not been directly compared in this article, we argue based on personal experience with both models that the present methodology with a clean separation of the RANS and LES scales including turbulence modeling (the concept of \emph{divide et impera}, as noted in the introduction) yields a more reliable and robust approach. Moreover, the present data has been compared with results of a detached-eddy simulation hybrid RANS/LES approach that uses function enrichment in the near-wall area\cite{Krank17c}. That method showed problems in the hybrid RANS/LES transition region, whereas the present method does not.
 
\section*{Acknowledgments}

\subsection*{Financial disclosure}

The research presented in this paper was partly funded by the German Research Foundation (DFG) under the project ``High-order discontinuous Galerkin for the EXA-scale'' (ExaDG) within the priority program ``Software for Exascale Computing'' (SPPEXA), grant agreement no. KR4661/2-1 and WA1521/18-1.

\subsection*{Conflict of interest}

The authors declare no potential conflict of interests.


\appendix 

\section{Numerical evaluation of van Driest's law}
\label{sec:vdriest_num}
\begin{table}[t]
\center
\caption{Location of support points $y_i^+$ and corresponding values $\psi_i$ for the efficient evaluation of the enrichment function.}
\label{tab:psi}
\begin{tabular*}{0.5\textwidth}{l @{\extracolsep{\fill}} l l}
\hline
$i$     & $y_i^+$ & $\psi_i$
\\ \hline \noalign{\smallskip}
$0$ & $0.0$ & $0.0$\\
$1$ & $5.0$ & $4.88298776233176$\\
$2$ & $11.0$ & $8.91824406645381$\\
$3$ & $24.0$ & $12.3978516813118$\\
$4$ & $59.0$ & $15.1875389926298$\\
$5$ & $144.0$ & $17.4177125619900$\\
$6$ & $361.0$ & $19.6484300823042$\\
$7$ & $946.0$ & $21.9930107788854$\\
$8$ & $2517.0$ & $24.3778307011372$ \\
\hline
\end{tabular*}
\end{table}

The enrichment function $\psi$ given as van Driest's wall-law in Equation~\eqref{eq:psi} has to be computed on each quadrature point during evaluation of the weak forms. The wall function necessitates the evaluation of an integral in the interval $[0,y^+]$. In order to avoid computing the integral over the entire interval each time the wall function is evaluated, we tabulate the wall function $\psi_i$ at several discrete values of the wall coordinate $y^+_i$. When the integral is evaluated at run time, solely a small slice of the interval has to be computed, namely $[y^+_i,y^+]$ with $i$ such that $y^+_i \leq y^+ < y_{i+1}^+$:
\begin{equation}
\psi=\psi_i + \int_{y_i^+}^{y^+}\frac{2\ \mathrm{d}y^+}{1+\sqrt{1+\left(2\kappa y^+(1-\mathrm{exp}(-y^+/A^+))\right)^2}}
\label{eq:psinum}
\end{equation}
This remaining integral is evaluated employing a nine-point Gaussian quadrature rule (with a polynomial order of accuracy of 17) irrespective of the width of the interval in our code. The fixed selection allows for efficient SIMD instructions that include quadrature points in several elements each with different parameters. The tabulated quantities have been computed using 32 digit precision in Matlab and are listed in Table~\ref{tab:psi} for $\kappa=0.41$ and $A^+=26$ and yield a guaranteed relative accuracy for $\psi$ of $1e-12$ up to $y^+=5{,}000$. The rapid change of the $y_i^+$ values shows that the near-wall area necessitates a denser distribution of the support points, while the integrand varies more slowly beyond $y^+>1{,}000$, such that only a few quadrature nodes are sufficient. As the wall function is only computed once per time step on each quadrature point and stored, the evaluation cost is negligible. Aiming at fine-tuning the approach, one may consider more support points $y_i^+$ and a Gaussian rule with fewer support points.

Besides the enrichment function itself, the spatial derivative $\mathrm{d}\psi / \mathrm{d} y^+$ is also required for evaluating the weak forms. This derivative is given by the integrand in Equation~\eqref{eq:psinum}. The derivatives with respect to the spatial location vector $\bm{x}$ are computed starting from Equation~\eqref{eq:space} by successive application of the chain rule as it is detailed Krank and Wall\cite{Krank16}.

\clearpage

\end{document}